\documentclass[aerospace,article,submit,pdftex,moreauthors]{mdpi} 
\firstpage{1} 
\pubvolume{1}
\issuenum{1}
\articlenumber{0}
\pubyear{2026}
\copyrightyear{2026}
\datereceived{ } 
\daterevised{ } 
\dateaccepted{ } 
\datepublished{ } 

\usepackage{subcaption}
\usepackage{amssymb}
\usepackage{amsmath}
\usepackage{bm}
\usepackage{soul}
\usepackage{pdfpages}
\usepackage{changepage}
\definecolor{mycolor}{rgb}{0, 0.69019607843, 0.31372549019}
\definecolor{mycolor1}{rgb}{0.39215686274, 0.56078431372, 1}
\definecolor{mycolor2}{rgb}{0.86274509803, 0.14901960784, 0.49803921568}
\definecolor{mycolor3}{rgb}{1, 0.69019607843, 0}
\prependpdf{preprint}
\Title{Operational Modal Analysis of Aeronautical Structures via Tangential Interpolation}

\Author{Gabriele Dessena $^{1,}$*\orcidA{}, Marco Civera $^{2}$\orcidB{} and Oscar E. Bonilla-Manrique $^{3}$\orcidC{}}

\AuthorNames{Firstname Lastname, Firstname Lastname and Firstname Lastname}

\address{%
$^{1}$ \quad Department of Aerospace Engineering, Universidad Carlos III de Madrid, Avenida de la Universidad 30, Leganés, 28911, Madrid, Spain; \href{mailto:gdessena@ing.uc3m.es}{GDessena@ing.uc3m.es} (G.D.)\\
$^{2}$ \quad Department of Structural, Geotechnical and Building Engineering, Politecnico di Torino, 10129, Turin, Italy; \href{mailto:Marco.Civera@polito.it}{Marco.Civera@polito.it} (M.C.)\\
$^{3}$ \quad Electronic Technology Department, Universidad Carlos III de Madrid, Avenida de la Universidad 30, Leganés, 28911, Madrid, Spain; \href{mailto:obonilla@ing.uc3m.es}{OBonilla@ing.uc3m.es} (OE.BM.)}

\corres{Correspondence: \href{mailto:gdessena@ing.uc3m.es}{GDessena@ing.uc3m.es} (G.D.)}

\firstnote{This is an extended version of the paper (\textit{Tangential interpolation for the operational modal analysis of aeronautical structures} by Gabriele Dessena, Marco Civera, and Oscar E. Bonilla-Manrique) accepted in Proceedings of the 15\textsuperscript{th} EASN International Conference on Innovation in Aviation \& Space Towards Sustainability Today and Tomorrow, Madrid, Spain, 14–17 October 2025.}  

\secondnote{{\hl{This is an author-accepted manuscript version of the following article: G. Dessena, M. Civera, and O. E. Bonilla-Manrique, ‘Operational Modal Analysis of Aeronautical Structures via Tangential Interpolation’, \textit{Aerospace}, vol. 13, no. 4, p. 378, Apr. 2026, doi: 10.3390/aerospace13040378. \textbf{\textit{Available online in open access at}}} [\url{https://doi.org/10.3390/aerospace13040378}]}}

\abstract{Over the last decades, progress in modal analysis has enabled increasingly routine use of modal parameters for applications such as structural health monitoring and finite element model updating. For output-only identification, or Operational Modal Analysis (OMA), widely adopted approaches include Stochastic Subspace Identification (SSI) methods and the Natural Excitation Technique combined with the Eigensystem Realization Algorithm (NExT-ERA). Nevertheless, SSI-based techniques may become cumbersome on large systems, while NExT-ERA fitting can struggle when measurements are contaminated by noise. To alleviate these, this work investigates an OMA frequency-domain formulation for aeronautical structures by coupling the Loewner Framework (LF) with NExT, yielding the proposed NExT-LF method. The method exploits the computational efficiency of LF{, due to the effectiveness of tangential interpolation, }together with the impulse response function retrieval enabled by NExT. NExT-LF is assessed on two experimental benchmarks: the eXperimental BeaRDS 2 high-aspect-ratio wing main spar and an Airbus Helicopters H135 bearingless main rotor blade. The identified modal parameters are compared against available experimental references and results obtained via SSI with Canonical Variate Analysis and NExT-ERA. The results show that the modes identified by NExT-LF correlate well with benchmark data, particularly for high-amplitude tests and in the low-frequency range.}


\begin{document}

\section{Introduction}
Modal analysis is a key tool for characterising the structural dynamics of engineering systems \cite{Ewins2000}. It can be performed in an input-controlled experimental setting or by measuring a structure \textit{in situ}, either during operation or at rest (Ambient Vibration Testing -- AVT), without the application of an external controlled input. The former is referred to as Experimental Modal Analysis (EMA), {while} the latter is known as Operational Modal Analysis (OMA) \cite{Rainieri2014}. In OMA, which is the focus of this work, the objective is to extract modal parameters \cite{Reynders2012} -- natural frequencies (\(\omega_n\)), damping ratios (\(\zeta_n\)), and mode shapes (\(\boldsymbol{\phi}_n\)) -- to describe the dynamic behaviour of a structure for purposes including Finite Element Model (FEM) updating \cite{ZanottiFragonara2017} and Structural Health Monitoring (SHM) \cite{Cadoret2025}. In aeronautical and aerospace engineering, OMA is employed in a {wide} range of applications. For instance, \cite{Jelicic2014} uses OMA results to track flutter stability in real time for a wind-tunnel wing model, while \cite{Sinske2017} reports in-flight aeroelastic identification for the German Aerospace Center (DLR) HALO research aircraft. OMA has also been applied during {launch vehicles} rollout operations, as reported for Artemis~I in \cite{Akers2025}, and to extract modal characteristics from in-flight data for launch vehicles and missiles in \cite{Eugeni2018} and \cite{James1993}, respectively. Rotorcraft applications are also well established: \cite{Ameri2013} applies OMA to Ground Vibration Testing (GVT) data from a Westland Lynx Mark~7 helicopter airframe, while \cite{Sibille2023} and \cite{Agneni2010} identify time-expired helicopter blades from an Airbus Helicopters H135 and an Agusta-Bell AB~204, respectively.

Traditional OMA techniques include the Natural Excitation Technique (NExT) paired with the Eigensystem Realization Algorithm (ERA), commonly {denoted} NExT-ERA \cite{GeorgeIII1993}, Stochastic Subspace Identification with Canonical Variate Analysis (SSI-CVA) \cite{VanOverschee1996}, and Frequency Domain Decomposition (FDD) \cite{Brincker2001}. NExT-ERA and SSI-CVA are time-domain approaches, whereas FDD operates directly in the frequency domain. Recent developments have aimed {to improve the} robustness to measurement noise and {support} consistent mode selection. For example, \cite{OConnell2024,OConnell2025} propose a robust COVariance-driven SSI (SSI-COV) formulation {that incorporates} a probabilistic criterion to reduce the likelihood that outliers are misinterpreted as physical modes. With a related objective, \cite{Amador2024} introduces a subspace-based Poly-reference Complex Frequency approach to reduce uncertainties associated with the null space of the system matrices used in the identification process, by adopting a subspace-driven parametric formulation and estimating modal properties via eigensystem realisations across selected subspaces. Another SSI-based development couples SSI-COV with Fuzzy C-Means (SSI-FCM) clustering to reject spurious modes in stabilisation diagrams, and has been demonstrated on {applications to }a train bogie and a five-storey experimental structure in \cite{Lv2026}. Furthermore, the Multi-Reference Transmissibility Complex Frequency method is proposed in \cite{Tarinejad2026} to address bias errors related to the application of the Fast Fourier Transform (FFT) to non-stationary signals, although validation is limited to numerical systems. Finally, NExT combined with Dynamic Mode Decomposition (DMD) is introduced {for OMA} on aeronautical structures in \cite{Wu2026}, {based} on the DMD formulation for experimental modal analysis in \cite{Saito2020, Wu2025}, which in turn extend the original DMD framework for extracting coherent features in fluid flows \cite{Schmid2010}.

In parallel, automated OMA frameworks (AOMA) have been proposed to streamline the identification process, including machine learning-based strategies \cite{Mugnaini2022}, hierarchical clustering \cite{Neu2017}, k-means clustering \cite{DeAlmeidaCardoso2018}, and Density-Based Spatial Clustering of Applications with Noise (DBSCAN) \cite{Sibille2023}. {T}he interested reader is referred to \cite{Mostafaei2025} for a recent review{ on AOMA applications}.

Within this broader context, the frequency-domain Loewner Framework (LF) has been considered for EMA \cite{Dessena2022} and subsequently extended to output-only identification in civil engineering by pairing it with NExT, yielding the NExT-LF approach \cite{Dessena2024f}. The aim of this coupling is to exploit the computational efficiency of the LF fitting process while retaining the output-only capability provided by NExT. These characteristics make NExT-LF attractive for aeronautical applications, where datasets can be large and measurements may be affected by operational noise. To address this gap, the present work investigates NExT-LF on {applications to }two aeronautically relevant experimental datasets: the eXperimental BeaRDS~2 (XB-2) flexible wing spar model \cite{Dessena2022b} and an Airbus Helicopters H135 Bearingless Main Rotor (BMR) blade \cite{Weber2021}. The two datasets are selected due to their different levels of complexity in terms of system dimensions (1.45~m vs.\ 4.99~m span) and test conditions (shaker excitation vs.\ AVT). The identification accuracy of NExT-LF is evaluated against concurrent identifications obtained via SSI-CVA and NExT-ERA, as well as against available benchmark results. The overall objective is to assess whether the accuracy achieved by NExT-LF is suitable for aeronautical {structures}. 
To achieve this aim, the remainder of the paper is structured as follows:
\begin{itemize}
    \item \Cref{sec:met} (\textit{Methodology}) introduces the LF identification procedure, the output-only enabling method (NExT), and their coupling to {reach} NExT-LF;
    \item \Cref{sec:spar} (\textit{The eXperimental BeaRDS 2 Flexible Wing Spar Model}) presents the XB-2 dataset and discusses the corresponding identification results;
    \item \Cref{sec:h135} (\textit{Airbus Helicopters H135 Bearingless Main Rotor Blade}) reports the blade dataset and the associated modal identification outcomes;
    \item \Cref{sec:disc} (\textit{Discussion}) discusses the main findings, limitations, and practical implications of applying NExT-LF to aeronautical structures;
    \item \Cref{sec:conc} (\textit{Conclusions}) summarises the main conclusions and closes this article.
\end{itemize}
\section{Methodology} \label{sec:met}
\subsection{Loewner Framework}\label{sec:LF}
The LF algorithm has previously been adopted for modelling multi-port electrical systems \cite{Lefteriu2009}, for aeroservoelastic modelling \cite{Quero2019}, and for parametric model order reduction \cite{Vojkovic2023}. Subsequently, the first and second authors {employed} LF to identify modal parameters in SIMO mechanical systems \cite{Dessena2022}. More recent work has extended the LF to {allow} the extraction of modal parameters from {multiple input, multiple output systems} \cite{Dessena2024,Dessena2025a}, output-only systems \cite{Dessena2024f}, and flutter estimation \cite{Dessena2025}. The present study considers this latter output-only formulation.

{To properly introduce LF}, we first define the Loewner matrix ${\bm{\mathbb{L}}}$:
\noindent {\emph{Let ($\mu_j$,${v}_j$), $j=1$,...,$q$ denote a row-wise collection of complex pairs and ($\lambda_i$,${w}_j$), $i=1$,...,$k$ a column-wise collection of complex pairs, with all points $\lambda_i$ and $\mu_j$ mutually distinct. The matrix $\boldsymbol{\bm{\mathbb{L}}}$, commonly termed the {divided-differences matrix }${\bm{\mathbb{L}}}$, is then given by:}}
\begin{equation}
\label{eq:LM}
\boldsymbol{\bm{\mathbb{L}}}=\begin{bmatrix}
\frac{\bm{v}_1-\bm{w}_1}{\mu_1-\lambda_1} & \cdots & \frac{\bm{v}_1-\bm{w}_k}{\mu_1-\lambda_k}\\
\vdots & \ddots & \vdots\\
\frac{\bm{v}_q-\bm{w}_1}{\mu_q-\lambda_1} & \cdots & \frac{\bm{v}_q-\bm{w}_k}{\mu_q-\lambda_k}\\
\end{bmatrix}\:\in \mathbb{C}^{q\times k}
\end{equation}
\emph{If there is a known underlying function $\pmb{\phi}$, then $\bm{w}_i=\pmb{\phi}(\lambda_i)$ and $\bm{v}_j=\pmb{\phi}(\mu_j).$}

Karl Löwner first linked $\boldsymbol{\mathbb{L}}$ to rational interpolation, a setting that is frequently described as Cauchy interpolation \cite{Lowner1934}. Through this link, interpolants may be characterised via the determinants of suitable submatrices of $\boldsymbol{\mathbb{L}}$. Moreover, as shown in \cite{Antoulas2017, Mayo2007}, {rational interpolants can be obtained} directly from $\boldsymbol{\mathbb{L}}$. In the present work, the formulation adopted is the Loewner pencil, defined by the pair of matrices $\boldsymbol{\mathbb{L}}$ and $\boldsymbol{\mathbb{L}}_s$. The latter, $\boldsymbol{\mathbb{L}}_s$, {indicates} the \emph{shifted Loewner matrix}.

To illustrate the working principle of the LF, consider a linear time-invariant dynamical system $\bm{\Sigma}$ characterised by $m$ inputs, $p$ outputs, and $k$ internal variables, expressed in descriptor form as:

\begin{equation}
\bm{\Sigma}:\;\bm{E}\frac{d}{dt}\bm{x}(t)=\bm{A}\bm{x}(t)+\bm{B}\bm{u}(t);\;\;\;
\bm{y}(t)=\bm{C}\bm{x}(t)+\bm{D}\bm{u}(t)
 \label{eq:LTI}
 \end{equation}
\noindent where $\bm{x}(t) \in \mathbb{R}^{k}$ denotes the vector of internal variables, $\bm{u}(t) \in \mathbb{R}^{m}$ is the input signal, and $\bm{y}(t) \in \mathbb{R}^{p}$ is the corresponding output vector. The system is specified by the following constant matrices:

\begin{equation}
    \bm{E},\bm{A}\in \mathbb{R}^{k\times k},\; \bm{B}\in \mathbb{R}^{k\times m}\; \bm{C}\in \mathbb{R}^{p\times k}\; \bm{D}\in \mathbb{R}^{p\times m}
\end{equation}
\noindent The Laplace-domain transfer function $\bm{H}(s)$ corresponding to $\bm{\Sigma}$ may be expressed as {a rational matrix function }$p \times m$, provided that the pencil $\bm{A}-\lambda\bm{E}$ is regular, i.e., $\bm{A}-\lambda\bm{E}$ is non-singular for at least one finite $\lambda \in \mathbb{C}$:

\begin{equation}
    \bm{H}(s)=\bm{C}(s\bm{E}-\bm{A})^{-1}\bm{B}+\bm{D}
    \label{eq:trans}
\end{equation}
\noindent {Consider} the general setting of tangential interpolation, which is commonly described as rational interpolation along tangential directions \cite{Kramer2016}. The associated right interpolation dataset is defined as:

\begin{equation}
\begin{gathered}
   (\lambda_i;\bm{r}_i,\bm{w}_i),\: i = 1,\dots,\rho
    \\
    \begin{matrix}
        \bm{\Lambda}=\text{diag}[\lambda_1,\dotsc,\lambda_k]\in \mathbb{C}^{\rho\times \rho}\\
        \bm{R}=[\bm{r}_1\;\dotsc \bm{r}_k]\in \mathbb{C}^{m\times \rho}\\
        \bm{W} = [\bm{w}_1\;\dotsc\;\bm{w}_k]\in \mathbb{C}^{p\times \rho}
        \end{matrix}\Bigg\}
\end{gathered}
\label{eq:RID}
\end{equation}
\noindent Similarly, the left interpolation data can be outlined:
\begin{equation}
    \begin{gathered} 
        (\mu_j,\bm{l}_j,\bm{v}_j),\: j = 1,\dots,v
        \\ 
        \begin{matrix}
            \bm{M}=\mathrm{diag}[\mu_1,\dotsc,\mu_q]\in \mathbb{C}^{v\times v}\\
            \bm{L}^T=[\bm{l}_1\;\dotsc \bm{l}_v]\in \mathbb{C}^{p\times v}\\
            \bm{V}^T = [\bm{v}_1\;\dotsc\;\bm{v}_q]\in \mathbb{C}^{m\times v}
        \end{matrix}\Bigg\}
    \end{gathered}
\label{eq:LID}
\end{equation}
\noindent The values $\lambda_i$ and $\mu_j$ denote the sampling points at which $\bm{H}(s)$ is queried, which in the present context coincide with the chosen frequency bins. The vectors $\bm{r}_i$ and $\bm{l}_j$ specify the right and left tangential directions and are often drawn at random in practical implementations \cite{Quero2019}. The quantities $\bm{w}_i$ and $\bm{v}_j$ are the corresponding tangential measurements. The rational interpolation task is then satisfied by {enforcing the} consistency between $\bm{w}_i$, $\bm{v}_j$ and the transfer function $\bm{H}$ associated with the realisation $\bm{\Sigma}$ in \cref{eq:LTI}:

\begin{equation}
\bm{H}(\lambda_i)\bm{r}_i=\bm{w}_i,\:j=1.\dots,\rho \;\; \text{and} \;\;
\bm{l}_i\bm{H}(\mu_j)=\bm{v}_j,\:i=1,\dots,v
 \label{eq:LS2}
 \end{equation}
\noindent so that the Loewner pencil complies with \cref{eq:LS2}.
Next, consider a collection of points $Z={z_1,\ldots,z_N}\subset\mathbb{C}$ together with a rational function $\bm{y}(s)$. For each $i=1,\ldots,N$, define the sampled values $\bm{y}_i := \bm{y}(z_i)$, and gather them in the set $Y={\bm{y}_1,\ldots,\bm{y}_N}$. Upon introducing the corresponding left and right partitions of the data, one obtains the following relations:

\begin{equation}
\begin{split}
	Z=\{\lambda_1,\dots,\lambda_\rho\} \cup \{\mu_1,\dots,\mu_v\} \;\;\text{and}\;\; 
	Y=\{\bm{w}_1,\dots,\bm{w}_\rho\} \cup \{\bm{v}_1,\dots,\bm{v}_v\}
\end{split}
\label{eq:ZY}
\end{equation}
\noindent where $N=p+v$. Consequently, the Loewner matrix $\boldsymbol{\mathbb{L}}$ can be written in the form:
\begin{equation}
\label{eq:LM2}
\boldsymbol{\bm{\mathbb{L}}}=\begin{bmatrix}
\frac{\bm{v}_1\bm{r}_1-\bm{l}_1\bm{w}_1}{\mu_1-\lambda_1} & \cdots & \frac{\bm{v}_1\bm{r}\rho-\bm{l}_1\bm{w}\rho}{\mu_1-\lambda\rho}\\
\vdots & \ddots & \vdots\\
\frac{\bm{v}_v\bm{r}_1-\bm{l}_v\bm{w}_1}{\mu_v-\lambda_1}& \cdots & \frac{\bm{v}_v\bm{r}\rho-\bm{l}_v\bm{w}\rho}{\mu_v-\lambda\rho}\\
\end{bmatrix}\:\in \mathbb{C}^{v\times \rho}
\end{equation}

\noindent Noting that $\bm{v}_v\bm{r}_p$ and $\bm{l}_v\bm{w}_p$ are scalar quantities, the Sylvester relation associated with $\boldsymbol{\mathbb{L}}$ can be written as:
\begin{equation}
    \boldsymbol{\bm{\mathbb{L}}}\bm{\Lambda}-\bm{M}\boldsymbol{\bm{\mathbb{L}}}=\bm{L}\bm{W}-\bm{V}\bm{R}
    \label{eq:syl}
\end{equation}

\noindent The \emph{shifted Loewner matrix} $\boldsymbol{\mathbb{L}}_s$ is obtained by constructing the Loewner matrix for the function $s\bm{H}(s)$, namely:

\begin{equation}
\label{eq:LS}
\boldsymbol{\bm{\mathbb{L}}}_s=\begin{bmatrix}
\frac{\mu_1\bm{v}_1\bm{r}_1-\lambda_1\bm{l}_1\bm{w}_1}{\mu_1-\lambda_1} & \cdots & \frac{\mu_1\bm{v}_1\bm{r}_\rho-\lambda_\rho\bm{l}_1\bm{w}_\rho}{\mu_1-\lambda_\rho}\\
\vdots & \ddots & \vdots\\
\frac{\mu_v\bm{v}_v\bm{r}_1-\lambda_1\bm{l}_v\bm{w}_1}{\mu_v-\lambda_1}& \cdots & \frac{\bm{v}_v\bm{r}_\rho-\bm{l}_v\bm{w}_\rho}{\mu_v-\lambda_\rho}\\
\end{bmatrix}\:\in \mathbb{C}^{v\times \rho}
\end{equation}

\noindent In an analogous manner, the corresponding Sylvester relation takes the form:
\begin{equation}
    \boldsymbol{\bm{\mathbb{L}}}_s\Lambda-\bm{M}\boldsymbol{\bm{\mathbb{L}}}_s=\bm{L}\bm{W}\bm{\Lambda}-\bm{M}\bm{V}\bm{R}
    \label{eq:syl2}
\end{equation}

Without loss of generality, $\bm{D}$ may be set to zero, since within the LF framework its contribution does not influence the tangential interpolation conditions \cite{Mayo2007}. Accordingly, \cref{eq:trans} simplifies to:

\begin{equation}
\bm{H}(s)=\bm{C}(s\bm{E}-\bm{A})^{-1}\bm{B}
\label{eq:fin}
\end{equation}

A realisation of minimal order is attainable only when the underlying system is both controllable and observable. Under the assumption that the available samples originate from a system whose transfer function is given by \cref{eq:fin}, the generalised tangential observability matrix $\mathcal{O}_v$ and the generalised tangential controllability matrix $\mathcal{R}_\rho$ are introduced in \cite{Lefteriu2010b}. {Then it} follows that \cref{eq:LM2} and \cref{eq:LS} may be recast in the form:

\begin{align}
    \boldsymbol{\bm{\mathbb{L}}}=-\mathcal{O}_v\bm{E}\mathcal{R}_\rho &&
    \boldsymbol{\bm{\mathbb{L}}}_s=-\mathcal{O}_v\bm{A}\mathcal{R}_\rho
    \label{eq:LLs}
\end{align}
Next, define the Loewner pencil as \emph{regular} by requiring that its generalised eigenvalues do not coincide with the sampling points, i.e., $\mathrm{eig}((\boldsymbol{\bm{\mathbb{L}}},\boldsymbol{\bm{\mathbb{L}}}_s)) \neq (\mu_i, \lambda_i)$:
\begin{align}
    \bm{E}=-\boldsymbol{\bm{\mathbb{L}}},&&\bm{A}=-\boldsymbol{\bm{\mathbb{L}}}_s,&&\bm{B}=\bm{V},&&\bm{C}=\bm{W}
\end{align}
Accordingly, the resulting rational interpolant may be written as:
\begin{equation}
    \bm{H}(s)=\bm{W}(\boldsymbol{\bm{\mathbb{L}}}_s-s\boldsymbol{\bm{\mathbb{L}}})^{-1}\bm{V}
\end{equation}
The preceding derivation is restricted to the minimal-data setting, a situation that is rarely met in practice. However, the LF framework admits a natural extension that can incorporate redundant sampling points in an efficient manner. To that end, we introduce the following assumption:
\begin{equation}
\begin{split}
    \mathrm{rank}[\zeta\boldsymbol{\bm{\mathbb{L}}}-\boldsymbol{\bm{\mathbb{L}}}_s]=\mathrm{rank}[\boldsymbol{\bm{\mathbb{L}}}\:\boldsymbol{\bm{\mathbb{L}}}_s]
   =\mathrm{rank}
    \begin{bmatrix}
    \boldsymbol{\bm{\mathbb{L}}} \boldsymbol{\bm{\mathbb{L}}}_s
    \end{bmatrix}=k,\; 
    \forall \zeta \in \{\lambda_j\}\cup\{\mu_i\}
    \end{split}
    \label{eq:cond1}
\end{equation}
Subsequently, a truncated Singular Value Decomposition (SVD) is computed for the matrix $\zeta\boldsymbol{\mathbb{L}} - \boldsymbol{\mathbb{L}}_s$:
\begin{equation}
    \textrm{svd}(\zeta\boldsymbol{\bm{\mathbb{L}}}-\boldsymbol{\bm{\mathbb{L}}}_s)=\bm{Y}\bm{\Sigma}_l\bm{X}
\label{eq:cond2}
\end{equation}
where $\mathrm{rank}(\zeta\boldsymbol{\bm{\mathbb{L}}}-\boldsymbol{\bm{\mathbb{L}}}_s)=\mathrm{rank}(\bm{\Sigma}_l)=\mathrm{size}(\bm{\Sigma}_l)=k,\bm{Y}\in\mathbb{C}^{v \times k}$ and $\bm{X}\in\mathbb{C}^{k\times \rho}$.
Note that:
\begin{equation}
\begin{split}
-\bm{A}\bm{X}+\bm{E}\bm{X}\bm{\Sigma}_l = \bm{Y}^*\boldsymbol{\bm{\mathbb{L}}}_s\bm{X}^*\bm{X}-\bm{Y}^*\boldsymbol{\bm{\mathbb{L}}}\bm{X}^*\bm{X}\bm{\Sigma}_l=\bm{Y}^*(\boldsymbol{\bm{\mathbb{L}}}_s-\boldsymbol{\bm{\mathbb{L}}}\bm{\Sigma}_l )=\bm{Y}^*\bm{V}\bm{R}=\bm{BR}
\end{split}
\label{eq:cond3}
\end{equation}
In an analogous manner, the identity $-\bm{Y}\bm{A} + \bm{M}\bm{Y}\bm{E} = \bm{L}\bm{C}$ is obtained, where $\bm{X}$ and $\bm{Y}$ act as the generalised controllability and observability matrices of the system $\bm{\Sigma}$, respectively, under the assumption $\bm{D}=0$.

Having verified that both the right and left interpolation conditions are fulfilled, the Loewner realisation accommodating redundant data may be expressed as:
\begin{equation}
\begin{split}
\bm{E}=-\bm{Y}^*\boldsymbol{\bm{\mathbb{L}}}\bm{X},\quad \quad
\bm{A}=-\bm{Y}^*\boldsymbol{\bm{\mathbb{L}}}_s\bm{X,} \quad \quad
\bm{B}=\bm{Y}^*\bm{V},\quad \quad
\bm{C}=\bm{W}\bm{X}
\end{split}
\label{eq:real}
\end{equation}

The expression in \cref{eq:real}, which specifies the Loewner realisation in the presence of redundant data, is adopted as the reference formulation in the remainder of this work. The complete step-by-step procedure is provided in \cite{Mayo2007,Antoulas2017}. Finally, the modal parameters of the system {can} be obtained by performing an eigenanalysis of the matrices $\bm{A}$ and $\bm{C}$ {that appear} in \cref{eq:real}. In practical modal analysis, this procedure is repeated over a range of model orders $k$, selected to capture the minimum set of modes of interest while identifying the highest order for which these modes remain stably estimated.
{At each instance}, the frequency domain data, usually an FRF, $\bm{H}_{real}(s)$ is fed to the LF, which then breaks the data down{, using the frequency bins as reference,} into two {disjoint} subsets, the left and right data, so that the system matrices of the transfer function in \cref{eq:fin} can be retrieved. {It is worth noting that frequency band chosen must include the frequency of interest, as the interpolation of out-of-bounds data is not possible, i.e. all poles lay within the band examined.}

\subsection{Natural Excitation Technique}\label{sec:next}

{The NExT algorithm }enables modal identification from output-only measurements by exploiting the statistical structure of ambient (unmeasured) excitations. In many operational settings, the applied forces cannot be {reliably recorded (e.g., } wind, traffic, or in-service aerodynamic loads), which prevents the direct construction of FRFs from measured inputs and outputs. NExT {bypasses }this limitation by {assuming} unknown excitation as a random process {with a} sufficiently broad frequency content to excite the modes of interest, and by using output correlations as impulse-response-type quantities \cite{GeorgeIII1993}, the {IRFs of the system} can be obtained.

Under the standard assumptions of linearity and time invariance, the output-only problem can be interpreted through a stochastic state-space model in discrete time, an LTI:
\begin{equation}
x_{k+1} = A x_k + w_k, 
\qquad
y_k = C x_k + v_k ,
\label{eq:stoch_ss}
\end{equation}
where $k\in\mathbb{N}$ is the sample index, $x_k\in\mathbb{R}^{n}$ is the state vector, $y_k\in\mathbb{R}^{p}$ is the measured output vector, and $A\in\mathbb{R}^{n\times n}$ and $C\in\mathbb{R}^{p\times n}$ are the state-transition and output matrices, respectively. The terms $w_k$ and $v_k$ represent process and measurement noise. NExT typically assumes that the effective excitation is zero-mean and stationary over the analysed record, and that it can be treated as broadband (often idealised as white noise) within the frequency range of interest. Marked departures from stationarity (e.g.\ strongly time-varying operating conditions) reduce the reliability of correlation estimates and may bias the inferred impulse responses. 

Let $y_i(t)$ and $y_j(t)$ denote two measured responses (e.g. accelerations) acquired at distinct sensor locations. The cross-correlation function is defined as
\begin{equation}
R_{y_i y_j}(\tau) = \lim_{T\to\infty}\frac{1}{T}\int_{0}^{T} y_i(t)\,y_j(t+\tau)\,\mathrm{d}t ,
\label{eq:xcorr_cont}
\end{equation}
where $\tau$ is the time {delay}. For finite records, \eqref{eq:xcorr_cont} is estimated from sampled data. Using $N$ samples $y_i[k]$ and $y_j[k]$, a common unbiased estimator for non-negative lags is
\begin{equation}
\widehat{R}_{y_i y_j}[\ell]
=
\frac{1}{N-\ell}\sum_{k=0}^{N-\ell-1} y_i[k]\,y_j[k+\ell],
\qquad \ell = 0,1,\dots,N-1.
\label{eq:xcorr_disc}
\end{equation}
When the ambient excitation is broadband, and the system is LTI, the correlation functions can be written as a superposition of decaying sinusoids whose damped frequencies and decay rates coincide with those of the structural modes \cite{GeorgeIII1993}. In practice, this means that $\widehat{R}_{y_i y_j}[\ell]$ (with $\ell$ denoting the discrete lag index, i.e.\ $\tau=\ell\Delta t$ for sampling period $\Delta t$) has the same functional form as a free-decay (IRF) for modal identification purposes, up to an unknown scaling and sensor-pair-dependent phase. A convenient representation is:
\begin{equation}
\widehat{R}_{y_i y_j}(\tau)
\approx
\sum_{r=1}^{n_m}
A_{r,ij}\,
\mathrm{e}^{-\zeta_r \omega_r \tau}
\sin\!\bigl(\omega_{d,r}\tau+\theta_{r,ij}\bigr),
\qquad
\omega_{d,r}=\omega_r\sqrt{1-\zeta_r^2},
\label{eq:xcorr_decay}
\end{equation}
where $n_m$ is the number of participating modes, $\omega_r$ and $\zeta_r$ are the natural frequency and damping ratio of {the $r$-th mode}, and $A_{r,ij}$ and $\theta_{r,ij}$ are constants that depend on the {pair of sensors and} the (unknown) spatial distribution of the excitation.

Given $p$ output channels, NExT forms a set of impulse-response-like sequences by selecting a reference channel $j$ and computing $\widehat{R}_{y_i y_j}[\ell]$ for $i=1,\dots,p$. The resulting collection
$\{\widehat{R}_{y_1 y_j},\widehat{R}_{y_2 y_j},\dots,\widehat{R}_{y_p y_j}\}$
is interpreted as a multiple-output IRF associated with the chosen reference.
An equivalent frequency-domain route, not considered in this work, computes cross-spectral densities and then obtains correlations via an inverse Fourier transform. Both approaches are {consistent,} provided th{at th}e stationarity assumptions are satisfied and the same normalisation conventions are used.

The performance of NExT depends on (i) approximate stationarity of the operating condition over the analysed record, (ii) sufficient bandwidth of the effective excitation to activate the modes of interest, and (iii) adequate signal-to-noise ratio across the sensor set. Close modes, strong tonal components (harmonics), or dominant narrow-band excitations can complicate the separation of modal contributions and may require longer records, additional averaging, or complementary processing (e.g. harmonic removal).
\subsection{NExT-LF}
\begin{figure}[h!]
\centering
\includegraphics[width=.9\textwidth]{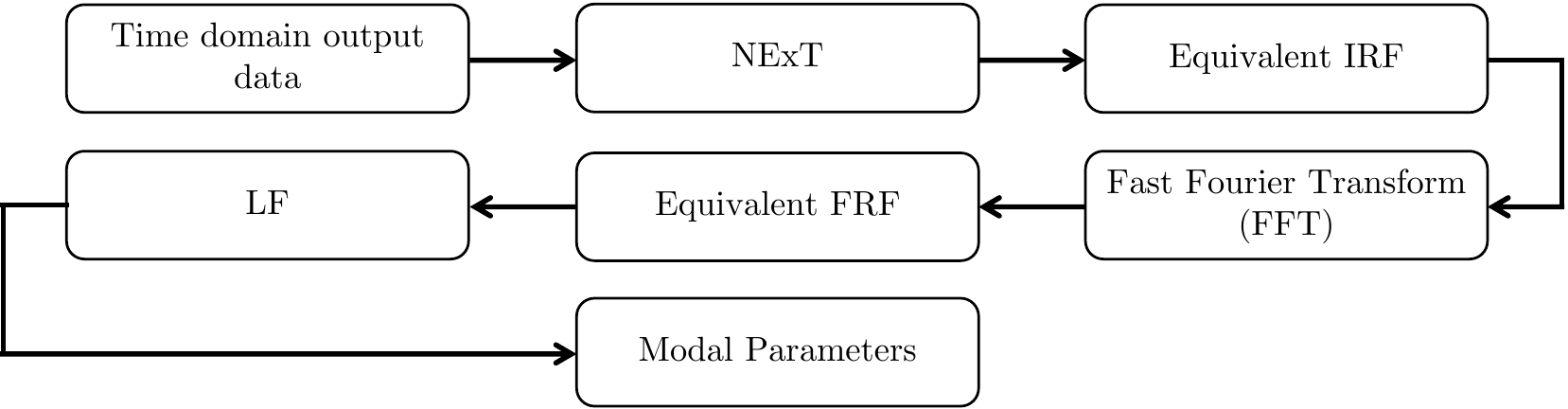}
\caption{NExT-LF workflow.}
\label{fig:fig1}
\end{figure}

{In summary}, NExT can retrieve IRFs from output-only data. IRFs are the time domain counterpart of the FRFs, such that they can be brought into the frequency domain by means of an FTT and then become a suitable input for frequency domain methods, such as the LF. In this work, NExT-LF follows the workflow shown in \cref{fig:fig1}.

\section{The eXperimental BeaRDS 2 Flexible Wing Spar Model}\label{sec:spar}
\begin{figure}[h!]
\centering
\includegraphics[width=.8\textwidth]{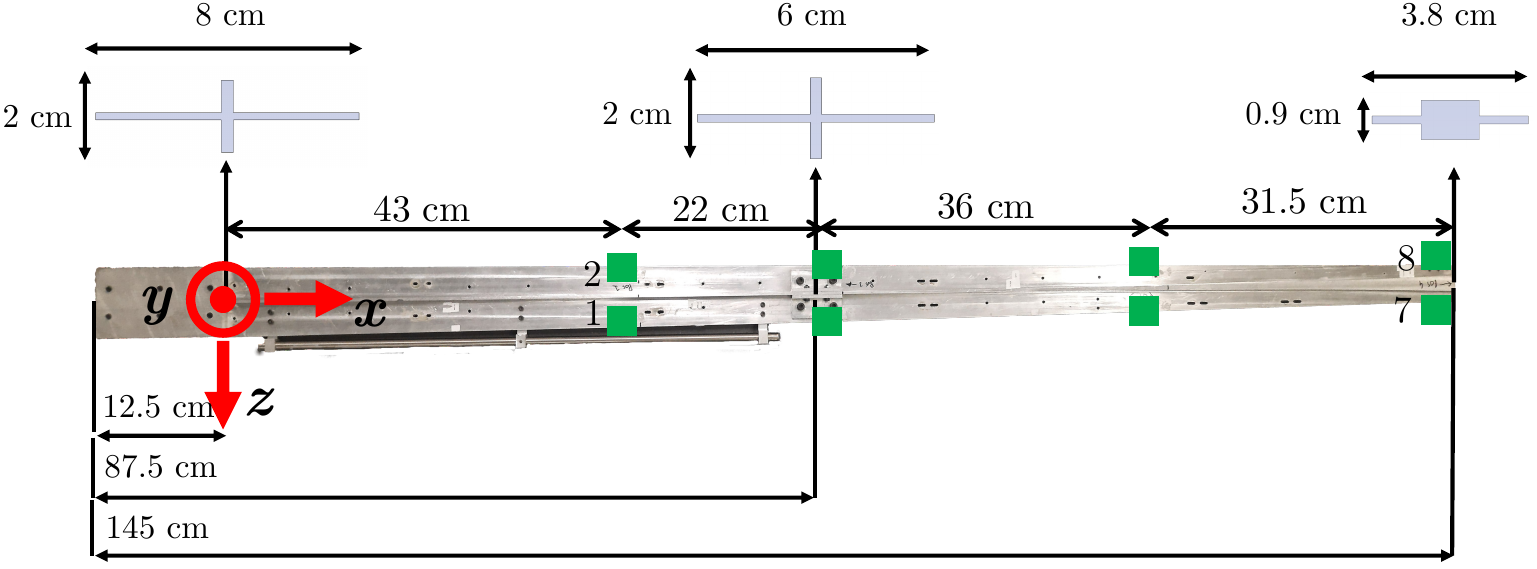}
\caption{XB-2 wing main spar geometrical characteristics and corresponding accelerometer locations (\textcolor{mycolor}{\rule{2.5mm}{2.5mm}}). The odd-numbered accelerometers are positioned on the positive $z$ side, whereas the even-numbered sensors are mounted on the negative $z$ side (retrieved from \cite{Dessena2025c}).}
\label{fig:fig2}
\end{figure}

In the present study, the main spar of the flexible XB-2 wing model is adopted as the experimental platform for the first aeronautically relevant demonstration of NExT-LF. The XB-2 wing is conceived as a dynamically scaled representation of a civil jet airliner wing and is manufactured for testing at Cranfield University 8$\times$6 ft (2.4$\times$1.8 m) wind tunnel within the Beam Reduction Dynamic Scaling (BeaRDS) project \cite{Pontillo2018,Yusuf2019,Pontillo2020}. 
The wing model is dynamically scaled to represent an aircraft broadly comparable to an Airbus A320, but featuring a more slender, higher-aspect-ratio wing with the same wing loading and a cruise condition of M = 0.6, intended to reflect an efficient turboprop configuration. An eXergy-based optimisation yields a full-scale wingspan of 48 m \cite{Hayes2019}. To ensure compatibility with the wind-tunnel facility, a geometric scaling factor of 16 is adopted. The scaling procedure relied on non-dimensionalising the full-scale configuration by matching the governing aeroelastic equations in terms of non-dimensional mass and characteristic frequencies. 
The wing structure consists of three principal components: a spar, a stiffening tube, and an external skin. The aerodynamic planform is based on a NACA 23015 section and features a span of 
1.5 m (corresponding to 1.385 m from the reference origin in \cref{fig:fig1}), a mean aerodynamic chord of 0.172 m, a taper ratio of 0.35, and a leading-edge sweep of 1.49\textsuperscript{o}. The spar and tube together form the wing torque box. The spar, which is the focus of this study, is manufactured from two 6082-T6 aluminium bars that are welded into a single component and locally strengthened using four bolted L-section plates, yielding a mass of 1.225 kg. Its cross-section resembles a Saint George’s cross and tapers along the span. The main geometric and sectional features are reported in \cref{fig:fig2}, together with the accelerometer layout.

The experimental dataset employed in this work is sourced from \cite{Dessena2022b}\footnote{The experimental and benchmark data can be retrieved from \url{https://doi.org/10.17862/cranfield.rd.19077023}}, which also reports benchmark EMA results. The spar is excited for 20 m using a bandwidth-limited random input in the range 2-400 Hz, with an RMS amplitude of 0.305 g. The excitation is applied using a Data Physics Signal Force\texttrademark{} modal shaker operated under closed-loop control via DP760\texttrademark{} software. As shown in \cref{fig:fig2}, the accelerometers are arranged in a 4 by 2 grid to measure vertical acceleration, {thus} capturing both flapwise (out-of-plane bending in the $x$-$z$ plane) and pitching (torsion around the $x$ axis) responses. The signals were sampled at $f_s=$ 5120 Hz using a National Instruments cDAQ-9178 and saved via an in-house LabVIEW programme\footnote{At the time the first author was affiliated with Cranfield University.}. The recorded acceleration time series are then converted to ms\textsuperscript{-2}, band-pass filtered between 3.25 and 85 Hz to suppress low-frequency drift, and band-stop filtered between 49.5 and 50.5 Hz to remove mains-related interference\footnote{In the UK, the mains electricity AC frequency is 50 Hz.}. The upper passband limit is set to 85 Hz, since all modes of interest lie below this frequency. {More} details on the experimental setup and instrumentation used in the test are {available} in \cite{Dessena2022b}{;} additional details on a nonlinear vibration testing campaign on XB-2 and components can be found in \cite{Dessena2022h}.

\begin{figure}[h!]
\centering
\includegraphics[width=.6\textwidth]{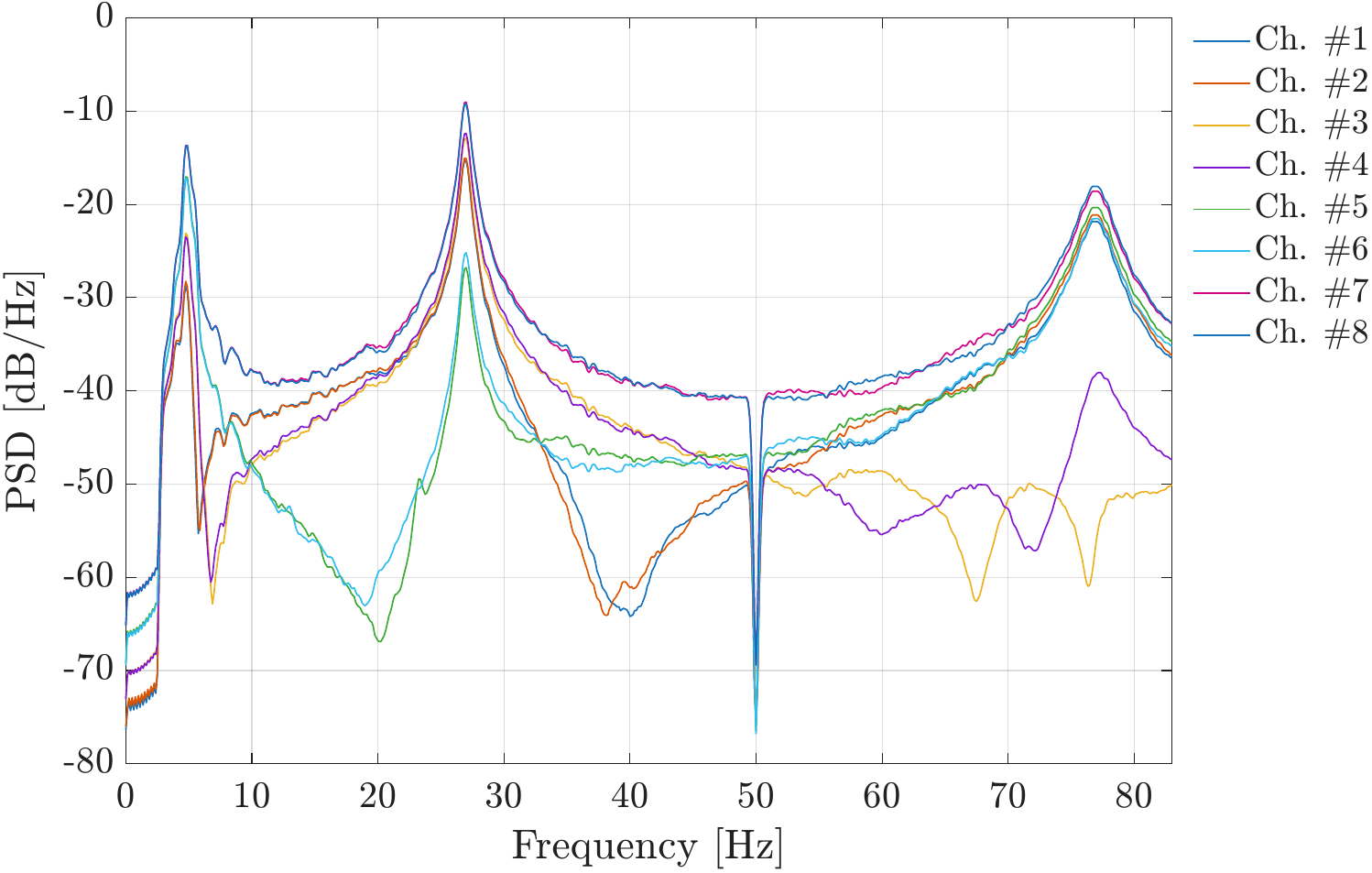}
\caption{PSDs of the eight acceleration response channels measured on the XB-2 wing main spar (retrieved from \cite{Dessena2025c}).}
\label{fig:fig3}
\end{figure}

Using the eight output-channel time histories, the response spectra can be estimated in terms of power spectral density (PSD). This is computed in MATLAB via the built-in \texttt{pwelch} function\footnote{\url{https://uk.mathworks.com/help/signal/ref/pwelch.html}}. Prior {to the} PSD estimation, the signal {is} downsampled to 256 Hz. Then, the resulting PSD estimates, as shown in \cref{fig:fig3}, are obtained by adopting 2048 discrete Fourier transform points, an overlap of 256 samples, and a Hamming window of length 1024 to reduce variance and provide a smoother spectral estimate.

As indicated by \cref{fig:fig3}, the PSD exhibits three dominant peaks, which can be attributed to the three modes of interest known to occur at approximately 
4.8, 27, and 76.8 Hz. The pronounced notch near 50 Hz arises from the applied band-stop filter. In addition, within the 60-70 Hz range, a resonance-like feature and an anti-resonance-like shape (most evident in channels \#3 and \#4, respectively) are observed. These are associated with local interactions involving the spar reinforcement plates to which the accelerometers are mounted, rather than with the global structural dynamics. Consistently, these features are absent from the PSDs of the remaining channels and do not appear in any FRF. The FRFs are omitted here for brevity, but are reported in the benchmark study \cite{Dessena2022b}.

\begin{figure}[!ht]
\begin{adjustwidth}{-\extralength}{0cm}
\centering
\subfloat[\centering]{\includegraphics[width=5cm]{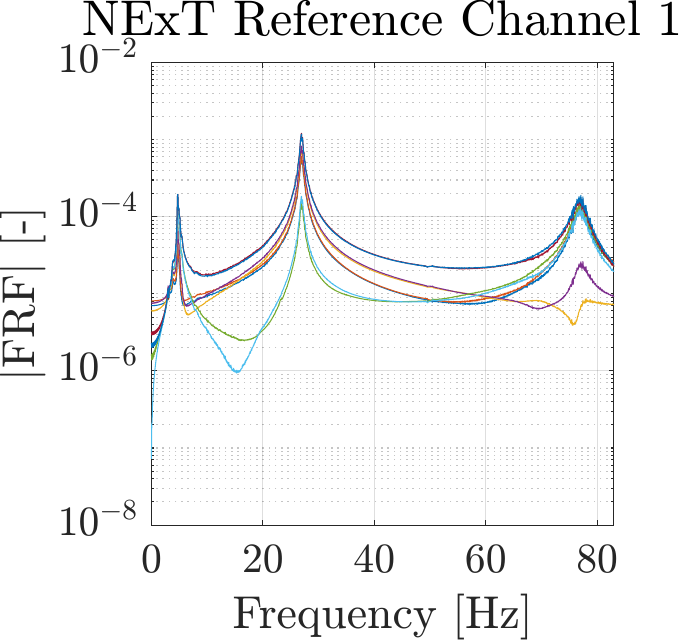}\label{fig:fig4a}}
\subfloat[\centering]{\includegraphics[width=5cm]{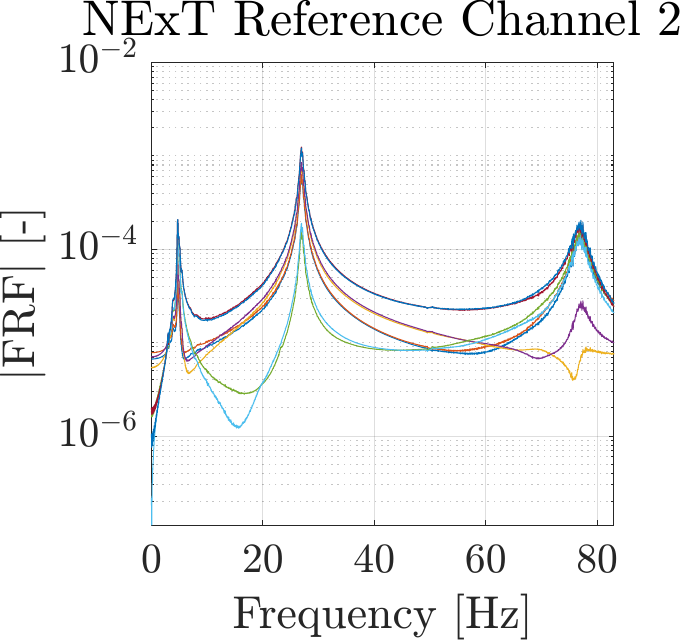}\label{fig:fig4b}}
\subfloat[\centering]{\includegraphics[width=5cm]{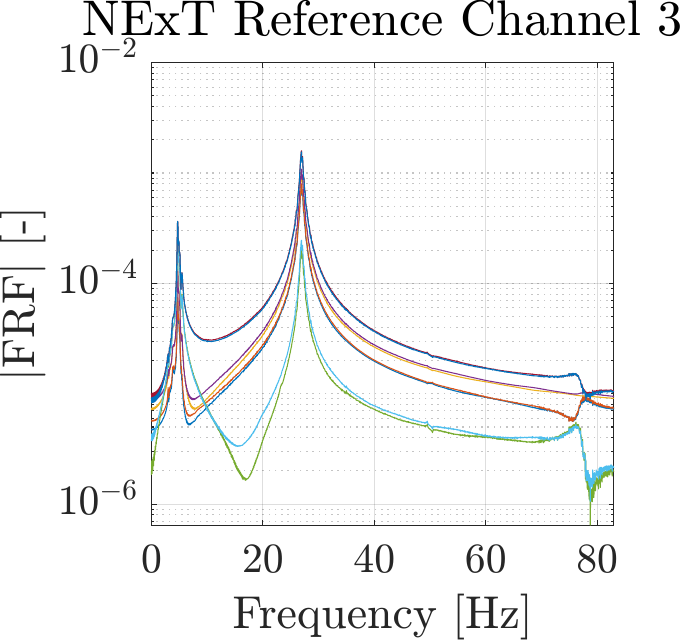}\label{fig:fig4c}}\\
\subfloat[\centering]{\includegraphics[width=5cm]{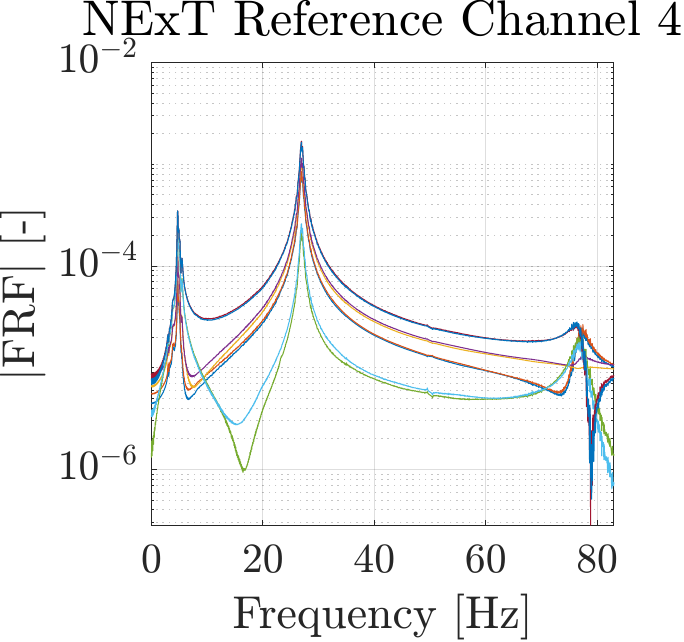}\label{fig:fig4d}}
\subfloat[\centering]{\includegraphics[width=5cm]{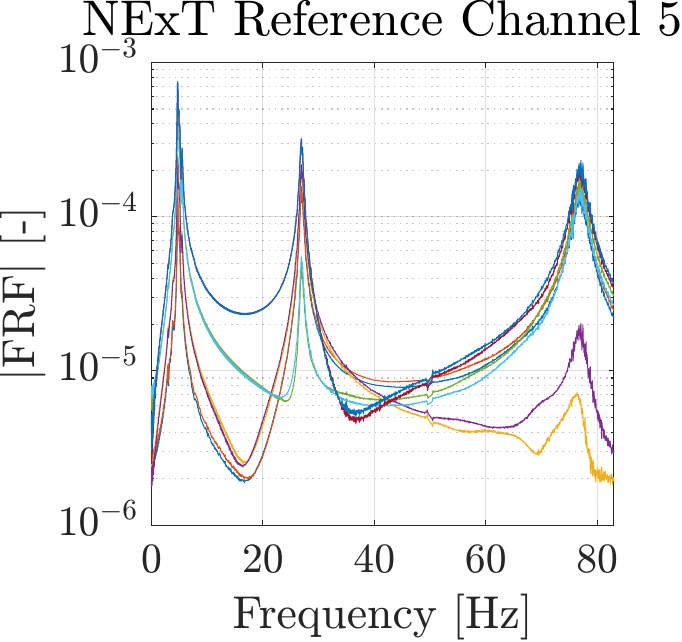}\label{fig:fig4e}}
\subfloat[\centering]{\includegraphics[width=5cm]{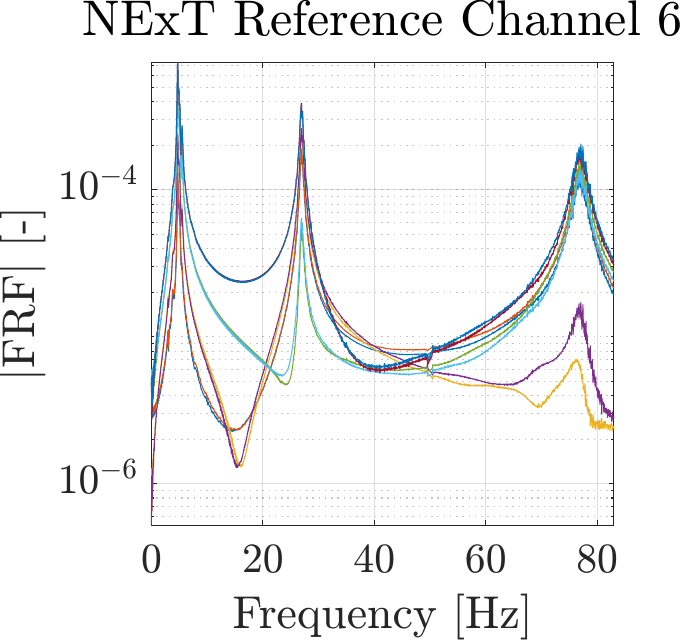}\label{fig:fig4f}}\\
\centering
\subfloat[\centering]{\includegraphics[width=5cm]{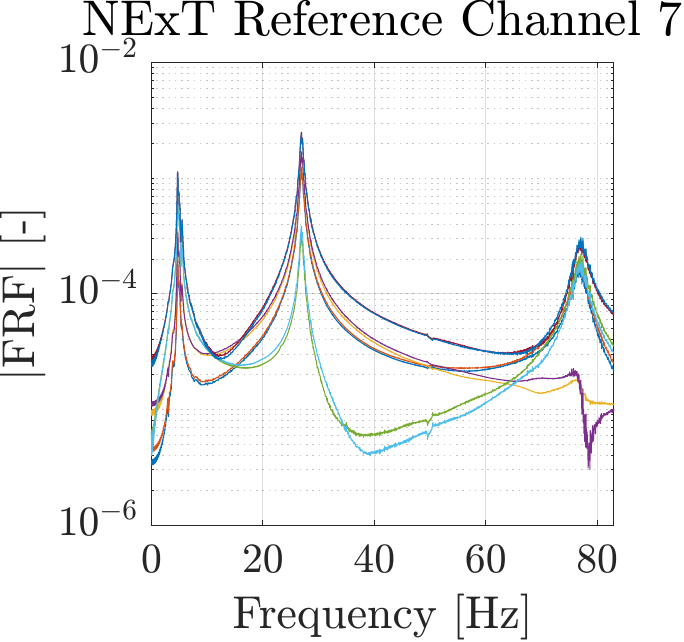}\label{fig:fig4g}}
\subfloat[\centering]{\includegraphics[width=5cm]{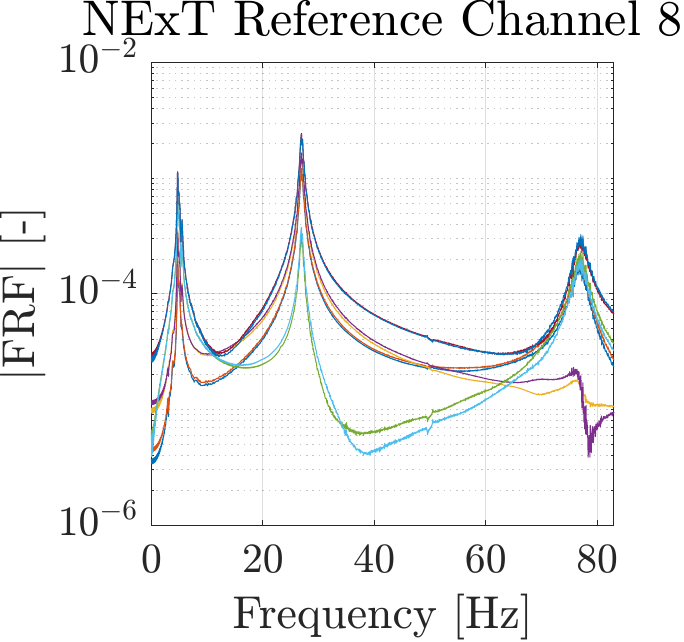}\label{fig:fig4h}}\\
\end{adjustwidth}
\caption{Equivalent FRFs obtained from the NExT-derived IRF of the XB-2 wing spar. (\textbf{a}) to (\textbf{h}) show the FRFs obtained, respectively, by setting output channel signals 1 to 8 the reference signal for NExT. \label{fig:fig4}}
\end{figure}

After this step, the identification procedure begins by providing NExT with the eight measured output acceleration time series, from which IRFs {are discretised over 2048 points.} In this work, the {acceleration signal} \#8 (see \cref{fig:fig2}) is used as the {only} reference channel for the NExT algorithm. This choice is motivated by the observation that the corresponding NExT-derived FRFs, as shown in \cref{fig:fig4}), present the three expected resonance peaks more clearly (higher amplitude) than when other reference channels are used.

These FRFs are then used as input to the LF for modal parameter estimation. In the NExT-ERA workflow, the frequency-domain step is not required, since the IRFs are provided directly to the ERA algorithm to extract the modal parameters. For SSI, modal parameters are estimated by supplying the measured acceleration time series directly to the method.

To limit the influence of outliers and non-physical modes, the identification is repeated across multiple model orders $k$, with $k\in[6,50]$, and stabilisation diagrams are used to retain only stable modes. The range of $k$ was chosen to obtain a comparable count of stable modes {between} the methods. As discussed later, this condition could not be met for ERA. The stability assessment considers $\omega_n\in[0,83]\,\mathrm{Hz}$ with a stability parameter of $0.5\%$, and $\zeta_n\in[0.01,0.03]$ with a stability parameter of $5\%$. A mode is accepted as stable only when the criteria are satisfied for five consecutive identifications.{ The consistency of }$\boldsymbol{\phi}_n$ is enforced by requiring a Modal Assurance Criterion (MAC) value not smaller than $0.95$ among the subsequent ($k$-wise) identified modes. The stabilisation diagrams for NExT-LF, NExT-ERA, and SSI-CVA are reported in \cref{fig:fig4}.

\begin{figure}[!ht]
\begin{adjustwidth}{-\extralength}{0cm}
\subfloat[\centering]{\includegraphics[width=8cm]{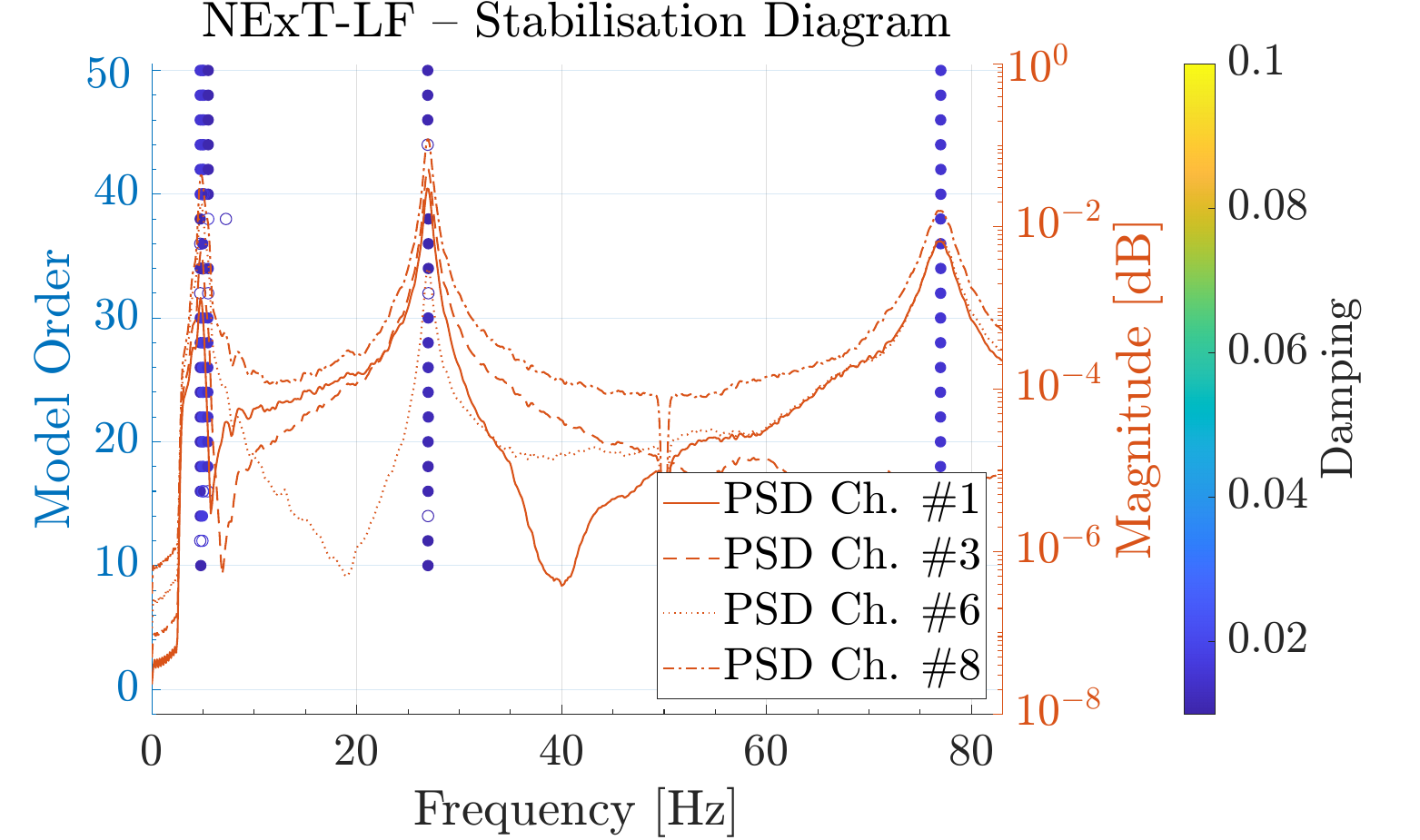}\label{fig:fig5a}}
\subfloat[\centering]{\includegraphics[width=8cm]{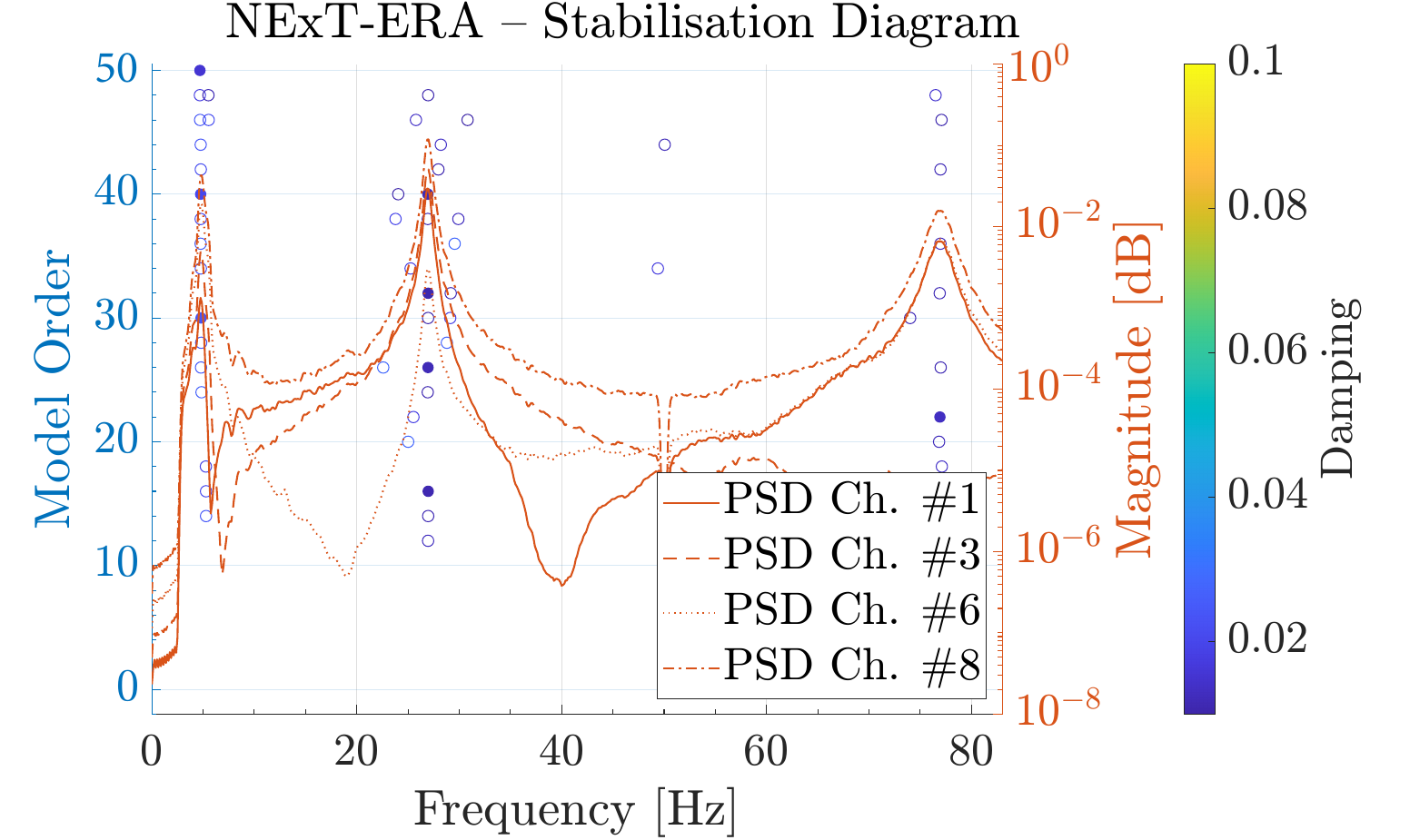}\label{fig:fig5b}}\\
\centering\subfloat[\centering]{\includegraphics[width=8cm]{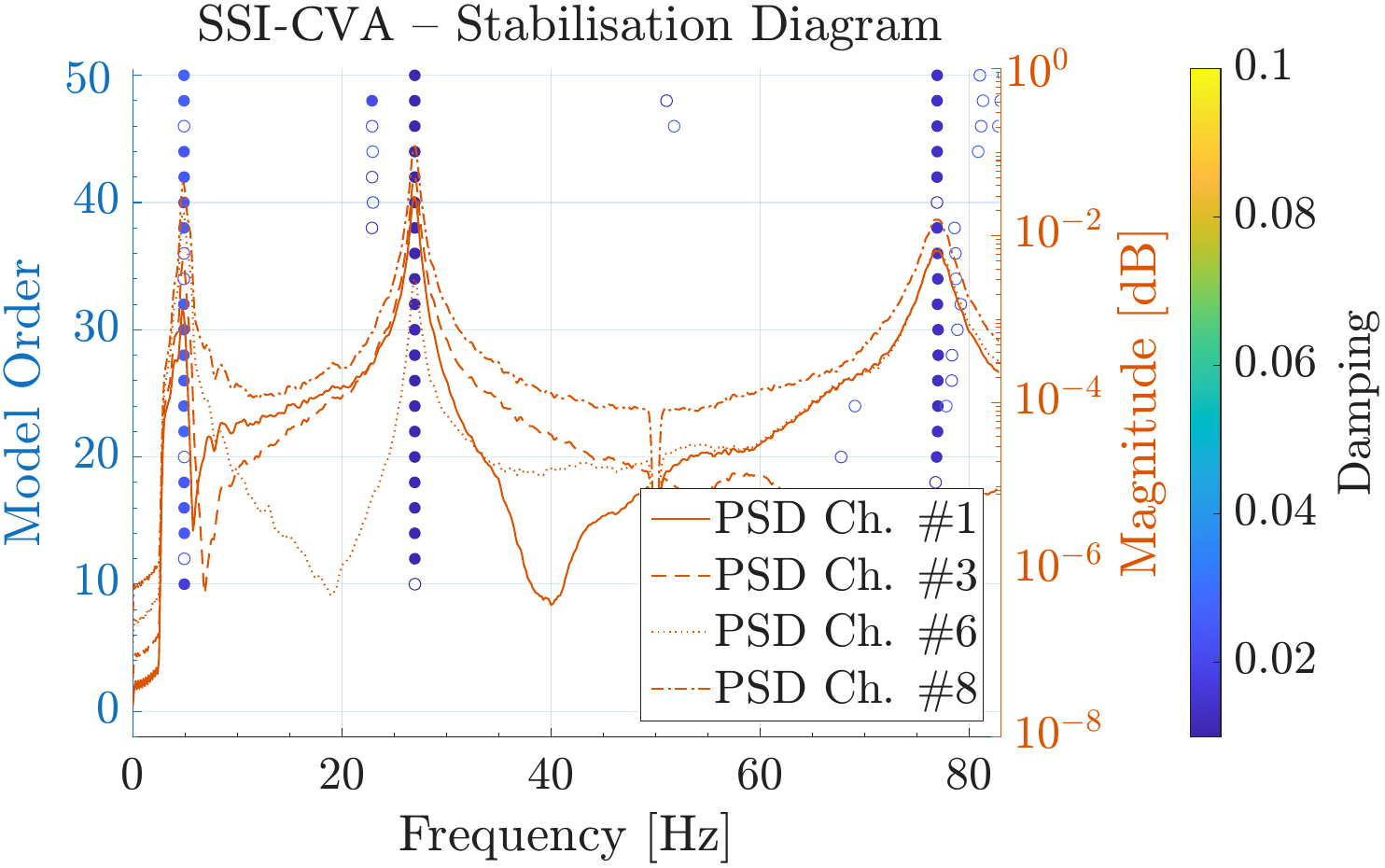}}
\end{adjustwidth}
\caption{Stabilisation diagrams obtained using NExT-LF (\textbf{a}), NExT-ERA (\textbf{b}), and SSI (\textbf{c}) for the modal identification of the XB-2 wing spar (retrieved from \cite{Dessena2025c}). \label{fig:fig5}}
\end{figure}

As shown in \cref{fig:fig5}, the {results of NExT-LF} and {SSI-CVA} exhibit markedly stronger stabilisation {than those of} the NExT-ERA case, with the difference being most evident in {damping} estimates $\zeta_n$. In particular, NExT-ERA yields virtually no modes that stabilise in damping. This behaviour is consistent with the {identified values of }$\omega_n$ and $\zeta_n$ reported in \cref{tab:tab1,tab:tab2}, respectively.

\begin{table}[h!]\small
\centering
\caption{Natural frequencies identified from the XB-2 main spar.\label{tab:tab1}}
\begin{tabular}{c c c c c}
\hline
\multicolumn{5}{c}{\textbf{Natural Frequency} [Hz] (difference w.r.t. to Benchmark [\%])}\\
\hline
Mode \# & Benchmark \cite{Dessena2022b} & NExT-LF & NExT-ERA & SSI \\
\cline{2-5}
1 & 4.855 & 4.930 & 4.689 & 4.891 \\
 & (-) & (1.53) & (-3.42) & (0.73) \\
2 & 26.966 & 26.922 & 26.942 & 26.969 \\
 & (-) & (-0.16) & (-0.09) & (0.01) \\
3 & 76.851 & 77.011 & 76.923 & 76.940 \\
 & (-) & (0.21) & (0.09) & (0.12) \\
\hline
\end{tabular}
\end{table}

\begin{table}[h!]\small
\centering
\caption{Identified modal damping ratios for the XB-2 main spar.\label{tab:tab2}}
\begin{tabular}{c c c c c}
\hline
\multicolumn{5}{c}{\textbf{Modal Damping Ratios} [-] (difference w.r.t. to Benchmark [\%])}\\
\hline
Mode \# & Benchmark \cite{Dessena2022b} & NExT-LF & NExT-ERA & SSI \\
\cline{2-5}
1 & 0.033 & 0.016 & 0.015 & 0.027 \\
 & (-) & (-51.17) & (-53.00) & (-17.01) \\
2 & 0.010 & 0.011 & 0.012 & 0.011 \\
 & (-) & (7.29) & (11.40) & (3.06) \\
3 & 0.014 & 0.015 & 0.013 & 0.014 \\
 & (-) & (7.35) & (-7.46) & (-3.17) \\
\hline
\end{tabular}
\end{table}

The $\omega_n$ identified using NExT-LF, NExT-ERA, and SSI-CVA are in close agreement with the benchmark results reported in \cite{Dessena2022b}, with the largest discrepancy being $-3.42\%$ for $\omega_1$ obtained via NExT-ERA. In contrast, the $\zeta_n$ in \cref{tab:tab2} show noticeably larger differences. Specifically, the absolute error in $\zeta_1$ exceeds $50\%$ for both NExT-LF and NExT-ERA, with NExT-LF providing a smaller deviation than NExT-ERA, {while} SSI-CVA limits the absolute error to approximately $17\%$. For $\zeta_2$ and $\zeta_3$, the NExT-LF estimates deviate by about $7.3\%$ in absolute terms, while NExT-ERA produces larger errors, namely $11.4\%$ and $7.46\%$, respectively. Overall, SSI-CVA yields damping estimates that are most consistent with the reference identification, although the absolute error for $\zeta_1$ remains above $17\%$.

\begin{figure}[h!]
\begin{adjustwidth}{-\extralength}{0cm}
\centering
\includegraphics[width=1.1\textwidth]{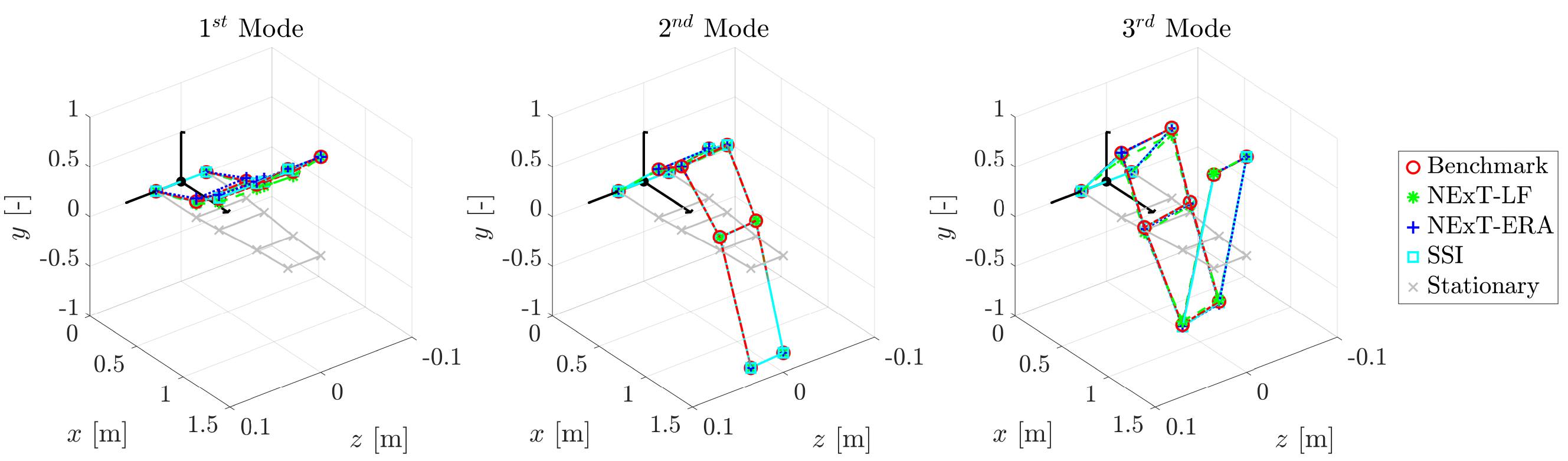}
\end{adjustwidth}
\caption{$\boldsymbol{\phi}_n$ of the XB-2 wing {spar} identified using the output-only methods, overlaid with the benchmark results (retrieved from \cite{Dessena2025c}).\label{fig:fig6}}
\end{figure}

After discussing the identifications of $\omega_n$ and $\zeta_n$, the results {for} $\boldsymbol{\phi}_n$ are now considered. Consistently with the frequency estimates, the mode shapes identified from the output-only measurements show excellent agreement with the benchmark, with the diagonal MAC values being equal, or very close, to unity. \Cref{fig:fig6} presents the $\boldsymbol{\phi}_n$ obtained via NExT-LF, NExT-ERA, and SSI from output-only data, overlaid with the reference $\boldsymbol{\phi}_n$ reported in \cite{Dessena2022b}. The identified $\boldsymbol{\phi}_n$ closely reproduce the benchmark results.

\section{Airbus Helicopters H135 Bearingless Main Rotor Blade}\label{sec:h135}
\begin{figure}[!ht]
\begin{adjustwidth}{-\extralength}{0cm}
\centering
\subfloat[\centering]{\includegraphics[height=4.5cm]{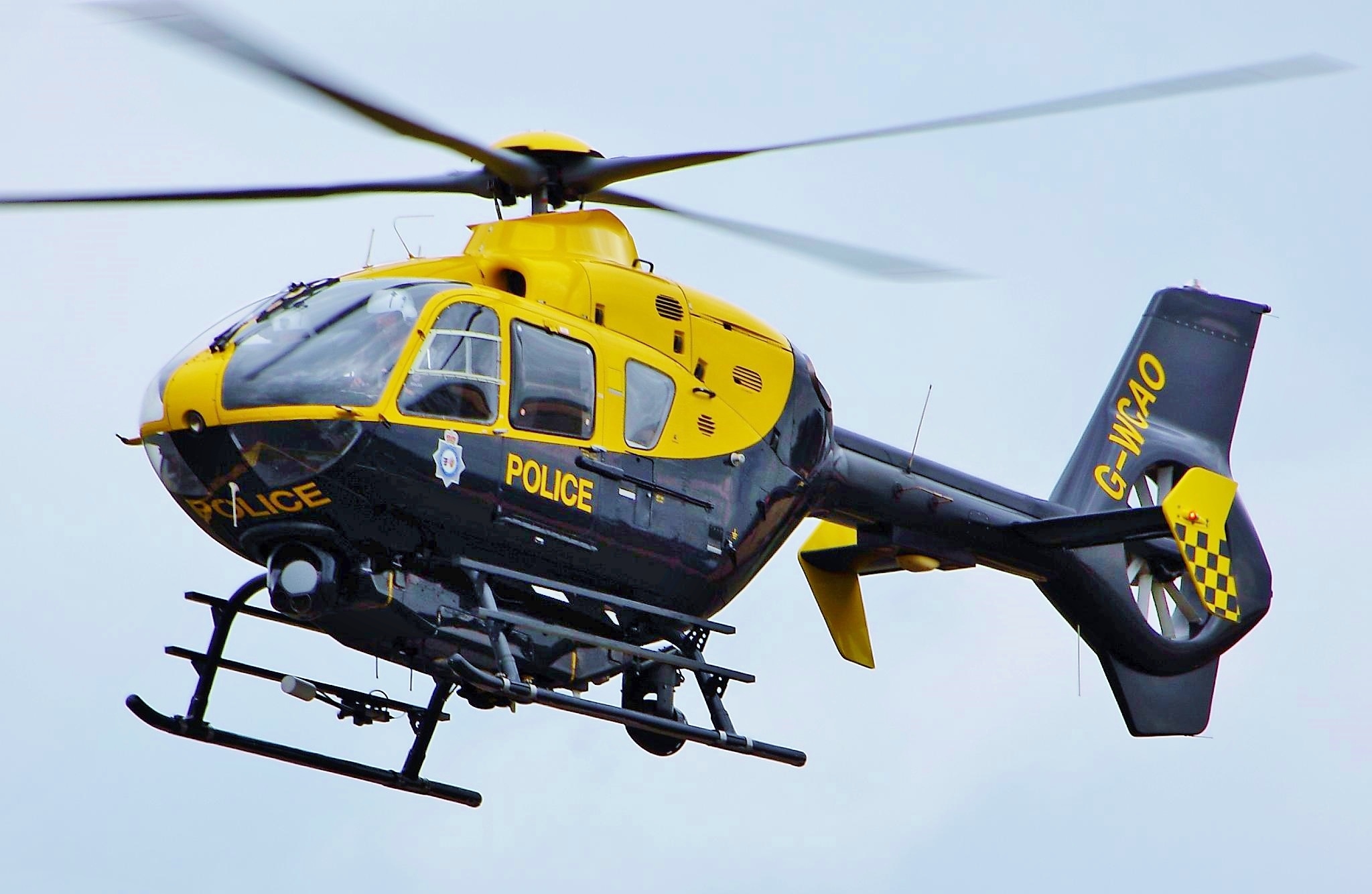}\label{fig:fig7a}}
\subfloat[\centering]{\includegraphics[height=4.5cm]{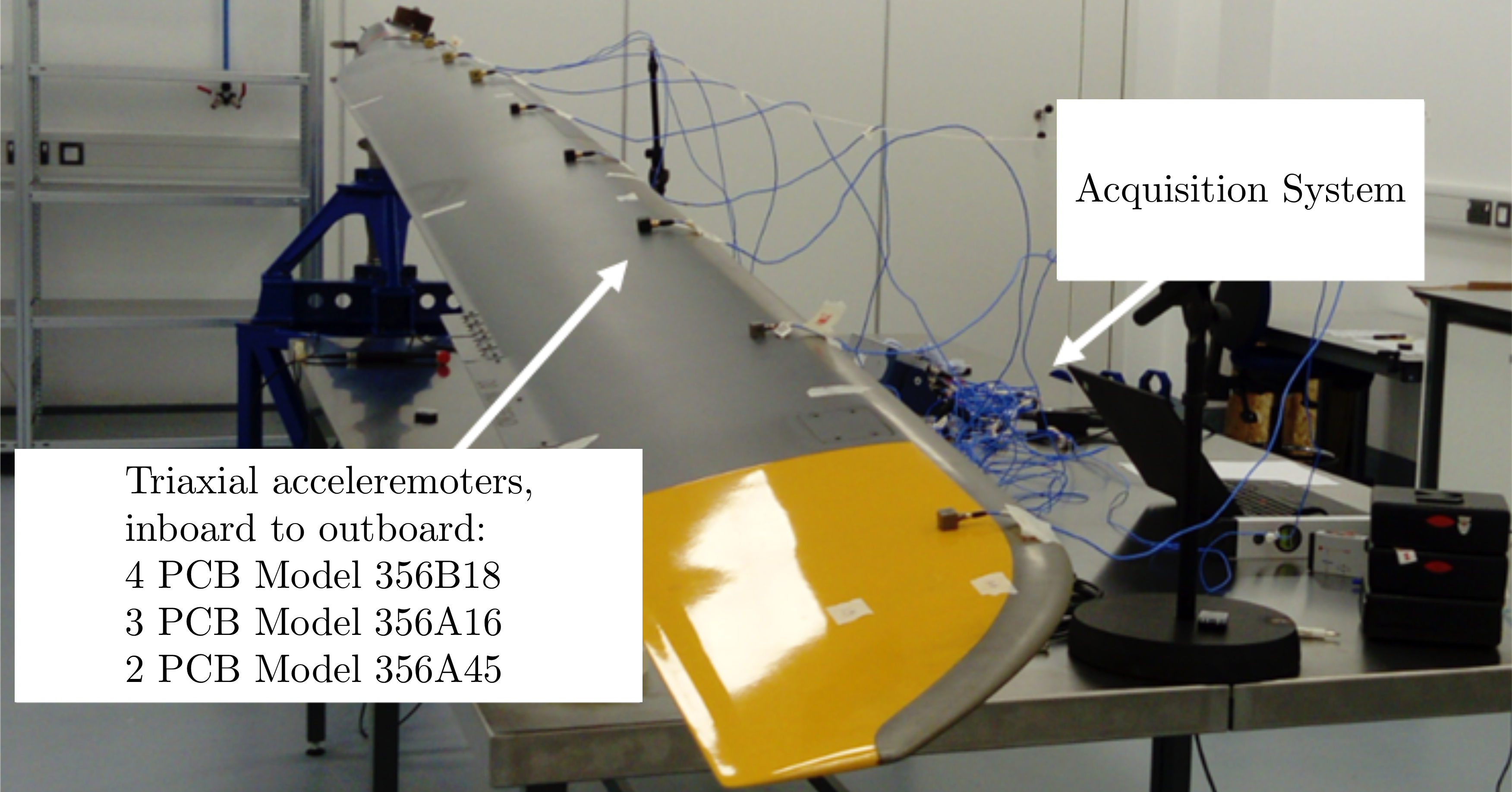}\label{fig:fig7b}}\\
\end{adjustwidth}
\caption{(\textbf{a}) Airbus Helicopters H135 helicopter in Western Counties Air Operations Unit livery (UK -- retrieved from \cite{Felce2011}) and (\textbf{b}) the test setup of the ambient vibration testing on the Airbus Helicopter H135 BMR blade (adapted from \cite{Weber2021}).\label{fig:fig7}}
\end{figure}

In order to generalise the findings of the previous sections, a second benchmark aeronautical structure should be considered. For this {aim,} an Airbus Helicopters H135 BMR blade is chosen. The Airbus Helicopters H135, shown in \cref{fig:fig7a}, is a lightweight rotorcraft capable of a maximum forward flight speed of 136 kts, a maximum range of 342 nm, and a maximum take-off mass of 3100 kg. In this work, the focus is on its BMR blade, which is made of the pitch control cuff with an elastomeric lead-lag damper, an aerofoil {section,} and a flexbeam. This arrangement allows {for incorporation of conventional} mechanical flapping, lagging, and torsional hinges {of the blade into} a single structural arrangement. {Structurally, each blade component is made of composite materials: the flexbeam is manufactured from E-glass/913 unidirectional prepregs with a cruciform cross-section whose slits reduce torsional stiffness while providing fail-safe load paths via double-lug root attachments, whereas the pitch control cuff and aerofoil skin are made of carbon fibre reinforced polymer}{. The approximate total mass of a single BMR is $40$ {kg}, of which roughly $7.5$ {kg} are tuning masses for frequency adjustment and lead-lag bending moment reduction }\cite{Bansemir1999}. More information on the Airbus Helicopters H135 BMR blade is available in \cite{Bansemir1997,Kampa1997}.

From a geometric perspective, the BMR blade has an equivalent chord of 0.288 m and spans 4.99 m from the hub attachment. {The H135 BMR blade specimen considered here, see }\cref{fig:fig7b}{, is an unserviceable}\footnote{In the aircraft maintenance sense.} {item that was retired from service in relatively good condition with over 4000 hours of remaining service life. This specimen was dynamically tested via AVT within the former Cranfield University BladeSense project} \cite{Weber2021}. {It is worth noting that the AVT considered here is not a test under the blade operating conditions, which will be in rotating conditions while attached to the rotor hub. Nevertheless, this setup can still be considered OMA, as the input data is unknown. }

\begin{figure}[h!]
\centering
\includegraphics[width=.8\textwidth]{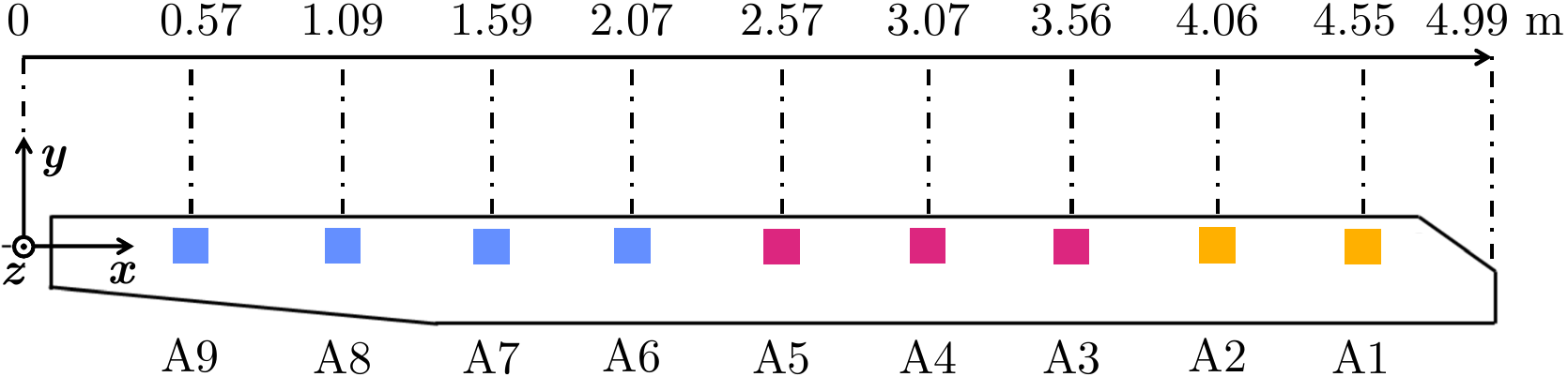}
\caption{Sensors layout of the instrumented Airbus Helicopter H135 BMR blade: A9-6 PCB Model 356B18 (\textcolor{mycolor1}{\rule{2.5mm}{2.5mm}}), A5-3 PCB Model 356A16 (\textcolor{mycolor2}{\rule{2.5mm}{2.5mm}}), and A2-1 PCB Model 356A45 (\textcolor{mycolor3}{\rule{2.5mm}{2.5mm}}). Note that the $z$-axis is positive in the out-of-page direction. \label{fig:fig8}}
\end{figure}

The AVT was carried out in the aeroelastic laboratory of
the Sir Peter Gregson Aerospace Integration Research Centre at Cranfield University, under controlled temperature and humidity conditions. The stationary, i.e. not revolving, blade is clamped at its end and instrumented with 9 nine triaxial accelerometers (A1-9), recording in-plane ($z$-axis) and out-of-plane transverse ($y$-axis) acceleration response from ambient vibration\footnote{ $x$-axis accelerations are recorded from A9, but are disregarded in this study.}.
The {types and model of accelerometer} are shown in \cref{fig:fig7b} and their layout and spanwise position in \cref{fig:fig8}. The {locations of the accelerometers along the span} in \cref{fig:fig8} are determined graphically from the Figures in \cite{Weber2021}. Additionally, note that \textcolor{mycolor1}{\rule{2.5mm}{2.5mm}} highlights the 4 PCB Model 356B18 accelerometers used, while \textcolor{mycolor2}{\rule{2.5mm}{2.5mm}} and \textcolor{mycolor3}{\rule{2.5mm}{2.5mm}} identify, respectively, the 3 PCB Model 356A16 and 2 PCB Model 356A45 accelerometers. {The use of three different accelerometer models was driven by laboratory availability rather than by a specific technical requirement; all three types provide adequate sensitivity for the low-amplitude ambient vibrations measured on this BMR blade across the frequency range of interest.}

\begin{figure}[h!]
\centering
\includegraphics[width=\textwidth]{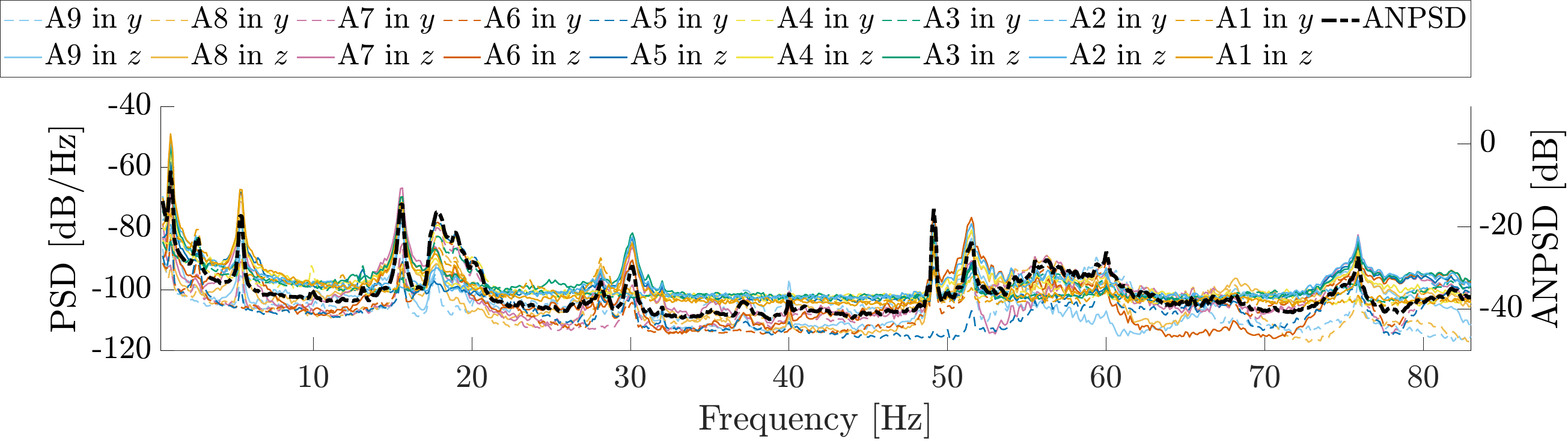}
\caption{PSDs and ANPSD from the signals recorded from accelerometers A1-9 on the Airbus Helicopter H135 BMR blade. \label{fig:fig9}}
\end{figure}

{18} acceleration time series are recorded {from the 9 accelerometers }at a sampling frequency $f_s=$ 2560 Hz for 600 s, using a similar setup to what is described for the XB-2 spar in \cref{sec:spar}, involving a National Instruments cDAQ-9178 and then {an in-house} LabVIEW programme.
Prior to identification, the signals are decimated to 256 Hz, {trimmed}, detrended, and bandpass-filtered between 0.4 and 100 Hz to exclude potential low- and high-frequency drifts with the \texttt{bandpass}\footnote{\url{https://uk.mathworks.com/help/signal/ref/bandpass.html}} built-in MATLAB function. From {this}, a PSD of the 18 channels and the ANPSD are found to qualitatively assess the quality of the {response.} The PSDs and ANPSD, shown in \cref{fig:fig9} are calculated using Welch's method over 2048 discrete points with a Hamming window of length 2048 with an overlap of 256.

{As} is clear from \cref{fig:fig9}, this case poses a tougher identification challenge than the XB-2 spar, as the AVT amplitudes are much smaller than those produced by shaker excitation, resulting in lower modal amplitudes. As for the other cases, the identification procedure starts by obtaining the IRF from NExT, discretised over 4098 points, using two reference channels, {A1 in }$y${ and A4 in} $z$ directions. This is necessary in this case, as using a reference channel for each direction increases the likelihood of obtaining accurate {estimates of the modal parameters} \cite{Caicedo2011}. The obtained IRFs are fed directly {to }ERA, while they are transformed to the frequency domain via an FFT discretised over 2048 points for use with the LF. The SSI-based identification is carried out directly from the time-domain data.

\begin{figure}[!htp]
\begin{adjustwidth}{-\extralength}{0cm}
\subfloat[\centering]{\includegraphics[width=8cm]{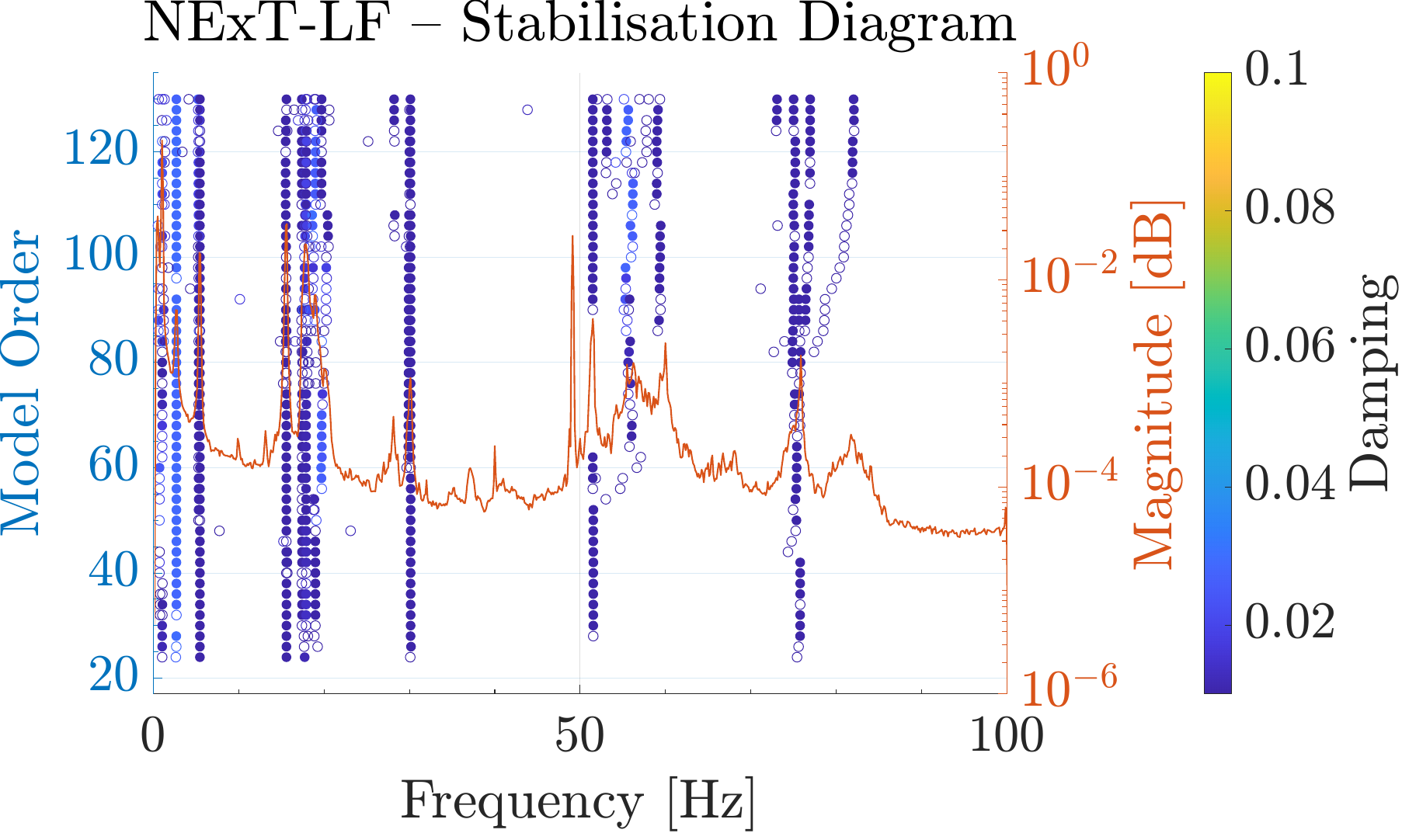}\label{fig:fig10a}}
\subfloat[\centering]{\includegraphics[width=8cm]{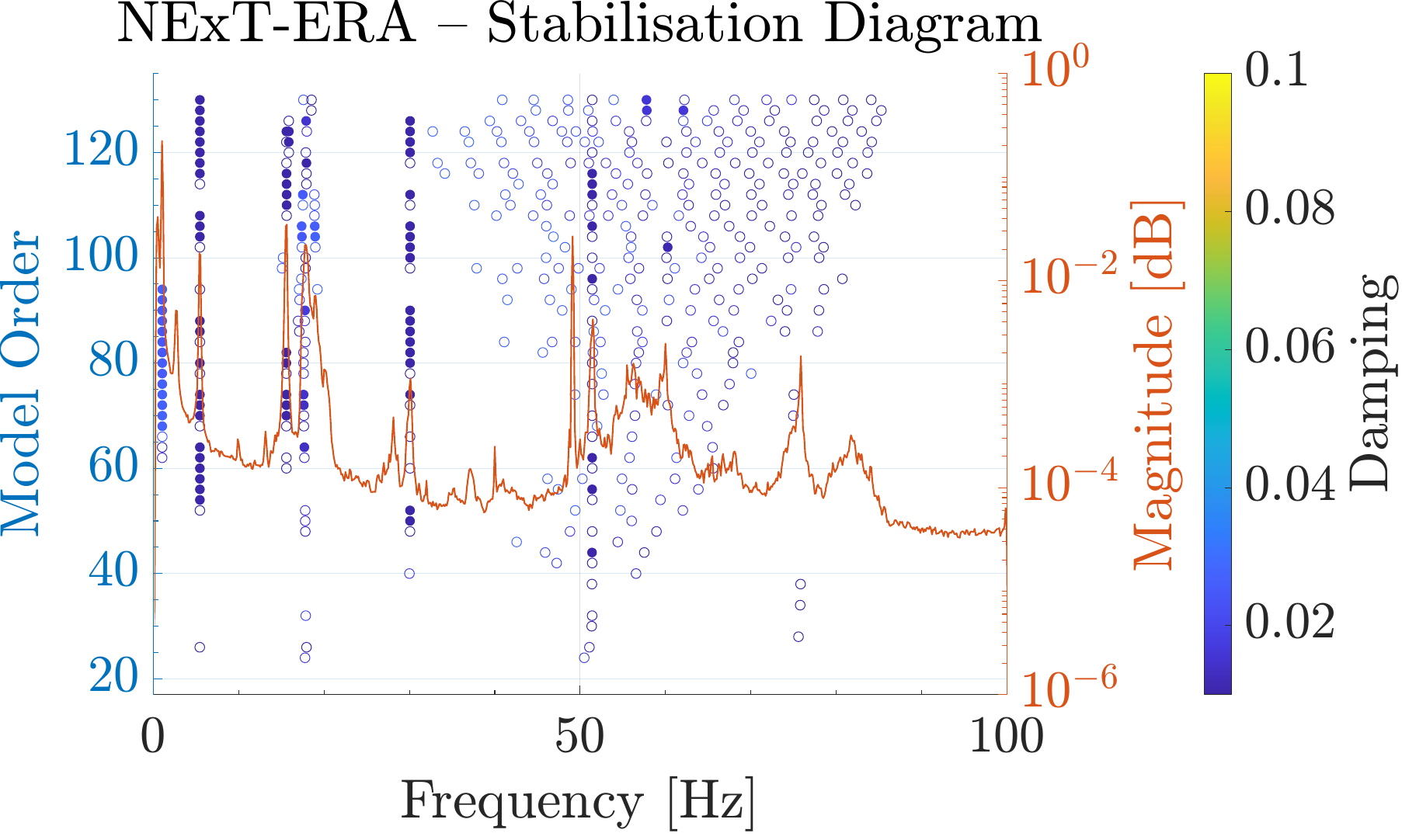}\label{fig:fig10b}}\\
\centering\subfloat[\centering]{\includegraphics[width=8cm]{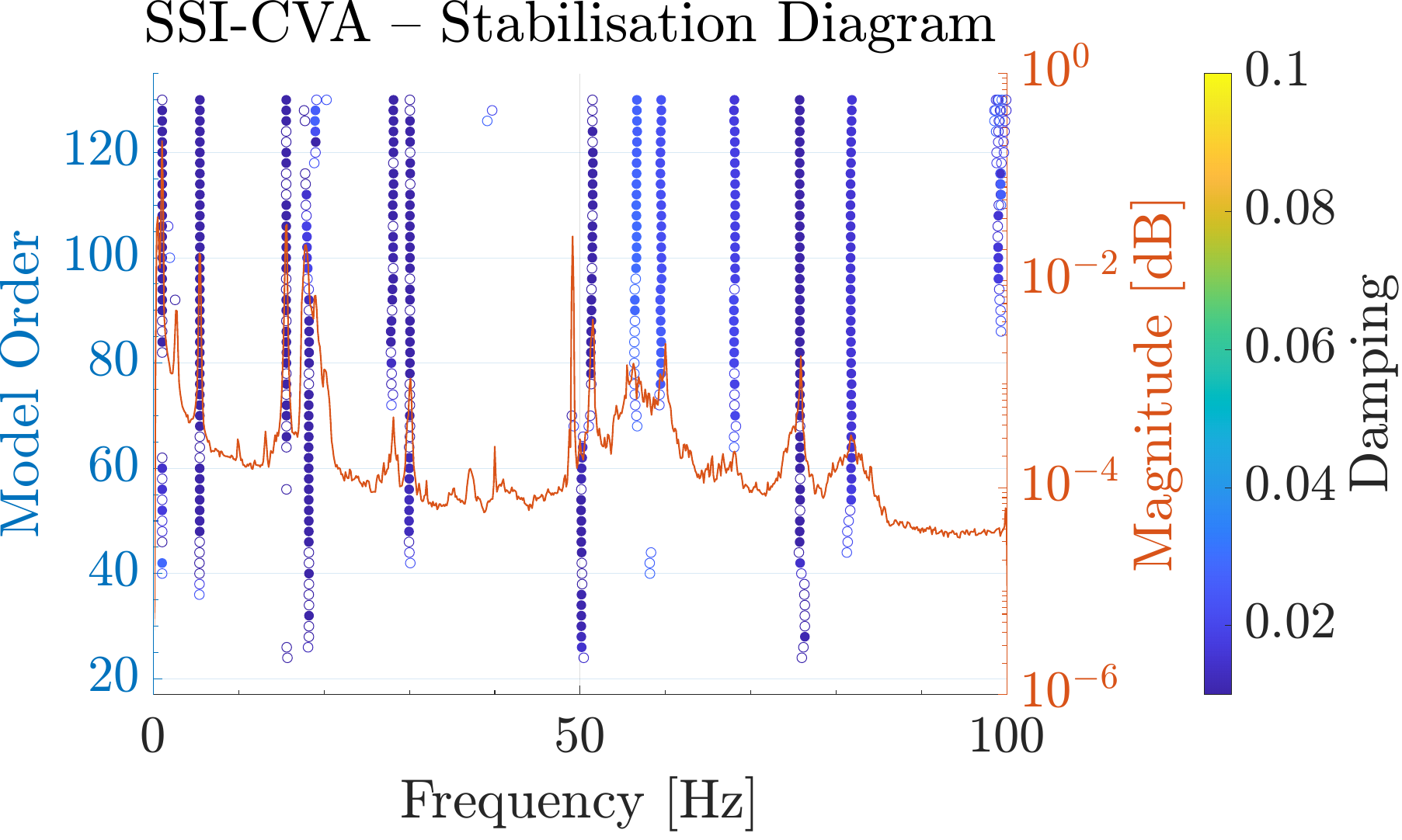}\label{fig:fig10c}}
\end{adjustwidth}
\caption{Stabilisation diagrams superimposed on the ANPSD and obtained using NExT--LF (\textbf{a}), NExT--ERA (\textbf{b}), and SSI (\textbf{c}) for the modal identification of the Airbus Helicopters H135 BMR blade. \label{fig:fig10}}
\end{figure}

The identification {is carried out in }a given order range $k\in[22,130]$ and stabilisation diagrams are used to exclude spurious and non physical modes. For this aim, stabilisation parameters{ are defined. For a fair comparison, these remain unchanged} across NExT-LF, NExT-ERA, and SSI-CVA. The stabilisation assessment is carried out {on} $f_n \in [0,100]\,$ {Hz} and $\zeta_n \in [0.003,0.03]$. Frequency stability is enforced with a relative tolerance of $0.5\%$ (i.e., $\Delta \omega_n / \omega_n \le 0.005$), while damping stability is checked with a $5\%$ tolerance (i.e., $\Delta \zeta / \zeta \le 0.05$). An additional absolute criterion {on} frequency variation is also applied, $|\Delta \omega_n| \le \epsilon$, with $\epsilon = 0.1$. A mode is accepted as stable only if these criteria are satisfied for three consecutive identifications. The consistency {of} $\boldsymbol{\phi}_n$ is enforced by requiring {a} MAC not smaller than $0.95$ among the subsequent stable identified modes. The stabilisation diagrams obtained from the NExT-LF, NExT-ERA, and SSI-CVA identification reported above are shown in \cref{fig:fig10}.

The stabilisation diagrams in \cref{fig:fig10} show that NExT--ERA, as observed for the XB-2 spar, struggles to identify most of the system modes. By contrast, the stabilisation diagrams obtained with NExT--LF and SSI--CVA are clearer and more consistent, in terms of both stability and the modes identified. It is also worth noting that some modes appear to be identified only by one of the two methods, with each approach capturing modes that are not recovered by the other. 

The extracted modal parameters, $\omega_n$ and $\zeta_n$, are {reported} in \cref{tab:tab3} with results from the literature. These include results from AOMA \cite{Mugnaini2022}, containing also $\boldsymbol{\phi}_n$, and \cite{Weber2021}, including $\omega_n$ from a calibrated FEM and $\omega_n$ and $\zeta_n$ from a separate experimental configuration, based on a series of GVTs conducted using a random-on-random excitation waveform. Please note that in \cref{tab:tab3}, \textit{flapping} refers to out-of-plane bending (within the $x$--$y$ plane), \textit{lagging} to in-plane bending (within the $x$--$y$ plane), \textit{pitching} to torsional rotations about the $x$-axis, and coupled to a mode exhibiting a combination of two or more of the dominant motions described above.

\begin{table}[htp!]\small
\caption{Natural frequencies and damping ratios obtained with different methods for the Airbus Helicopters H135 BMR blade. Values in parentheses indicate the relative difference (\%) w.r.t. \cite{Mugnaini2022}.\label{tab:tab3}}
\begin{adjustwidth}{-\extralength}{0cm}
\resizebox{1.2\textwidth}{!}{
\begin{tabular}{cccccccccccc}
\toprule
\multirow{2}{*}{\textbf{Mode \#}} &
\multicolumn{6}{c}{\textbf{Natural Frequencies [Hz]} -- (Relative difference w.r.t. \cite{Mugnaini2022} [\%])} &
\multicolumn{5}{c}{\textbf{Damping ratio [\%]} -- (Relative difference w.r.t. \cite{Mugnaini2022} [\%])} \\
\cmidrule(lr){2-7}\cmidrule(lr){8-12}
& \textbf{NExT-LF} & \textbf{NExT-ERA} & \textbf{SSI-CVA} & \textbf{AOMA \cite{Mugnaini2022}} & \textbf{Exp. \cite{Weber2021}} & \textbf{FEM \cite{Weber2021}}
& \textbf{NExT-LF} & \textbf{NExT-ERA} & \textbf{SSI-CVA} & \textbf{AOMA \cite{Mugnaini2022}} & \textbf{Exp. \cite{Weber2021}} \\
\midrule
1 - 1\textsuperscript{st} flapping  & 1.03 & 1.03 & 1.02 & 1.02 & 1.02 & 0.90 & 2.33 & 2.72 & 0.68 & 0.55 & 1.05 \\
   & (0.57) & (0.65) & (0.10) &  & (0.00) & (-11.76) & (324.22) & (394.53) & (23.07) &  & (90.91) \\
2 - 1\textsuperscript{nd} lagging  & 2.68 & - & - & 2.71 & 2.60 & 3.41 & 2.76 & - & - & 0.70 & 5.50 \\
   & (-1.01) &  &  &  & (-4.06) & (25.83) & (293.86) &  &  &  & (685.71) \\
3 - 2\textsuperscript{nd} flapping  & 5.44 & 5.44 & 5.44 & 5.43 & 5.39 & 4.98 & 0.55 & 0.60 & 0.76 & 0.50 & 1.13 \\
   & (0.18) & (0.13) & (0.15) &  & (-0.74) & (-8.29) & (10.43) & (19.06) & (51.41) &  & (126.00) \\
4 - 3\textsuperscript{rd} flapping & 15.58 & 15.57 & 15.55 & 15.55 & 15.44 & 15.28 & 0.47 & 0.34 & 0.35 & 0.30 & 0.97 \\
   & (0.18) & (0.16) & (0.03) &  & (-0.71) & (-1.74) & (55.93) & (13.85) & (17.93) &  & (223.33) \\
5 - 2\textsuperscript{nd} lagging & 17.83 & 17.90 & 17.93 & 18.11 & 18.68 & 22.72 & 1.91 & 1.49 & 1.48 & 1.12 & - \\
   & (-1.55) & (-1.18) & (-1.01) &  & (3.15) & (25.46) & (70.46) & (33.42) & (32.08) &  &  \\
6 - 1\textsuperscript{st} coupled & 19.69 & 18.90 & 18.98 & - & - & - & 0.87 & 2.60 & 2.62 & - & - \\
   &  &  &  &  &  &  &  &  &  &  &  \\
7 - 1\textsuperscript{st} pitching & 28.20 & - & 28.14 & 28.11 & 27.39 & 28.59 & 0.46 & - & 0.67 & 0.66 & 3.22 \\
   & (0.31) &  & (0.09) &  & (-2.56) & (1.71) & (-31.03) &  & (1.70) &  & (387.88) \\
8 - 4\textsuperscript{th} flapping  & 30.11 & 30.08 & 30.06 & 30.07 & 29.98 & 29.98 & 0.42 & 0.45 & 0.45 & 0.41 & 0.61 \\
   & (0.15) & (0.02) & (-0.03) &  & (-0.30) & (6.65) & (1.49) & (10.90) & (9.01) &  & (48.78) \\
9 - 5\textsuperscript{th} flapping  & 51.49 & 51.42 & 51.46 & 51.43 & 51.71 & 52.17 & 0.31 & 0.31 & 0.57 & 0.45 & 0.85 \\
   & (0.12) & (-0.03) & (0.05) &  & (0.54) & (73.50) & (-32.10) & (-31.94) & (26.72) &  & (88.89) \\
10 - 3\textsuperscript{rd} lagging & 55.65 & 57.78 & 56.65 & 56.24 & - & - & 2.64 & 1.58 & 2.68 & 2.52 & - \\
   & (-1.05) & (2.74) & (0.74) &  &  &  & (4.89) & (-37.35) & (6.20) &  &  \\
11 - 2\textsuperscript{nd} coupled& 59.14 & 60.27 & 59.50 & - & - & - & 1.08 & 1.13 & 2.34 & - & - \\
   &  &  &  &  &  &  &  &  &  &  &  \\
12 - 3\textsuperscript{rd} coupled & - & - & 68.09 & - & - & - & - & - & 1.65 & - & - \\
   &  &  &  &  &  &  &  &  &  &  &  \\
13 - 6\textsuperscript{th} flapping & 75.65 & - & 75.73 & 75.80 & 75.68 & 75.12 & 0.31 & - & 0.85 & 0.88 & 0.94 \\
   & (-0.20) &  & (-0.09) &  & (-0.16) & (-0.90) & (-65.30) &  & (-3.11) &  & (6.82) \\
14 - 2\textsuperscript{nd} pitching & 82.05 & - & 81.86 & 81.71 & 81.63 & 81.80 & 1.04 & - & 1.60 & 1.82 & 1.50 \\
   & (0.42) &  & (0.19) &  & (-0.10) & (0.11) & (-42.63) &  & (-12.16) &  & (-17.58) \\

\bottomrule
\end{tabular}}
\end{adjustwidth}
\end{table}

{It should be noted that the relative differences reported in }\cref{tab:tab3}{ are computed with respect to the AOMA results in }\cite{Mugnaini2022} {, as these represent the most directly comparable OMA-to-OMA reference for the same specimen and test configuration, and constitute the only available source from which reference mode shapes can be retrieved. The experimental values reported in }\cite{Weber2021}{ originate from a different test setup (GVT with random-on-random excitation) and are therefore retained as an external validation benchmark rather than as the primary comparison baseline. It should also be noted that a change in testing conditions (e.g. boundary conditions, excitation type) can influence the identified modal response, which further motivates the use of results obtained under the same test configuration as the primary reference.} From \cref{tab:tab3}, it is important to note that NExT-LF identifies two additional modes within the frequency range of interest, compared {to} the previous identification reported for AOMA \cite{Mugnaini2022} and the experimental results in \cite{Weber2021}. 
These modes are located at $19.69~\mathrm{Hz}$ and $59.14~\mathrm{Hz}$, respectively. In addition, SSI-CVA identifies a further mode at $68.09~\mathrm{Hz}$. As shown later, these modes are all coupled in the $z$- and $y$-axis displacements, with the suspicion that the 2\textsuperscript{nd} coupled mode may also include a torsional component; however, this cannot be fully confirmed due to the adopted instrumentation layout. Nevertheless, the coupled character of these modes provides a plausible explanation for why they are not identified by the simplified FEM reported in \cite{Weber2021}. Furthermore, it is worth noting that NExT-LF finds more modes than NExT-ERA (13 vs 9), and the same number as SSI-CVA.

In terms of the identified natural frequencies $\omega_n$, the agreement {between} the different output-only identification methods is generally very good. With respect to the reference values obtained via AOMA \cite{Mugnaini2022}, the relative differences remain below approximately $3\%$ for all coincident modes, the largest deviation being observed for NExT-ERA at the 10\textsuperscript{th} mode. A similar level of agreement is observed when {compared} with the experimental results reported in \cite{Weber2021}. The consistency of the identified $\omega_n$ across NExT-LF, NExT-ERA, and SSI-CVA confirms the robustness of the proposed framework in terms of frequency estimation, even for higher-order flapping and lagging modes.

However, the same level of consistency is not observed for {damping} ratios $\zeta_n$. In contrast to $\omega_n$, the identified $\zeta_n$ values exhibit a significantly larger dispersion {between} methods. In particular, relative differences with respect to AOMA \cite{Mugnaini2022} frequently exceed several tens of {a percentage} and, in some cases, are considerably larger, especially for {lower-frequency} modes. This behaviour is expected, as damping estimation in operational modal analysis is well known to be more sensitive to noise, limited measurement duration, and modelling assumptions than frequency estimation \cite{SrikanthaPhani2007}. Moreover, small absolute differences in $\zeta_n$ can translate into large relative deviations when the reference damping level is low. This amplification effect is well documented in the OMA literature \cite{SrikanthaPhani2007} and has been recently confirmed for aerospace-grade structures in \cite{Zhu2025}, where the sensitivity of output-only damping estimates to record length, noise floor, and spectral leakage is quantified.

Overall, the results indicate that while the identification of $\omega_n$ is highly reliable and consistent {among all} the {approaches considered}, the estimation of $\zeta_n$ remains more method-dependent. However, to fully assess the performance of NExT-LF, the identified $\boldsymbol{\phi}_n$ must also be examined. To this end, \cref{tab:tab4} reports the diagonal MAC values between the $\boldsymbol{\phi}_n$ identified via NExT-LF and those obtained in \cite{Mugnaini2022}.

\begin{table}[htp!]\small
\caption{Diagonal terms of the MAC value matrices obtained by correlating the identified (NExT-LF, NExT-ERA, and SSi-CVA) $\boldsymbol{\phi}_n$ and the reference results in \cite{Mugnaini2022}.\label{tab:tab4}}
\begin{adjustwidth}{-\extralength}{0cm}
\begin{tabular}{ccccccccccccccc}
\toprule
\multicolumn{15}{c}{Diagonal terms of the \textbf{MAC value [-]} matrix} \\
\midrule
& \multicolumn{14}{c}{\textbf{Mode} } \\
\cmidrule(lr){2-14}
\textbf{Method}& \textbf{1} & \textbf{2} & \textbf{3} & \textbf{4} & \textbf{5} & \textbf{6}\textsuperscript{1} & \textbf{7} & \textbf{8} & \textbf{9} & \textbf{10} & \textbf{11}\textsuperscript{1} & \textbf{12}\textsuperscript{1} & \textbf{13} & \textbf{14} \\
\midrule
NExT-LF  & 0.97 & 0.93 & 1.00 & 1.00 & 0.77 & - & 0.44 & 0.97 & 1.00 & 0.16 & - & - & 0.70 & 0.09 \\
NExT-ERA & 1.00 & -    & 1.00 & 1.00 & 0.99 & - & -    & 0.99 & 1.00 & 0.06 & - & - & -    & -    \\
SSI-CVA  & 1.00 & -    & 1.00 & 1.00 & 0.93 & - & 0.99 & 1.00 & 1.00 & 0.97 & - & - & 0.98 & 1.00 \\
\bottomrule
\end{tabular}
\noindent{\\\footnotesize{\textsuperscript{1} These $\boldsymbol{\phi}_n$ cannot be correlated to \cite{Mugnaini2022} as they are not reported therein.}}
\end{adjustwidth}
\end{table}

\begin{figure}[!ht]
\begin{adjustwidth}{-\extralength}{0cm}

\subfloat[\centering]{\includegraphics[width=4.8cm]{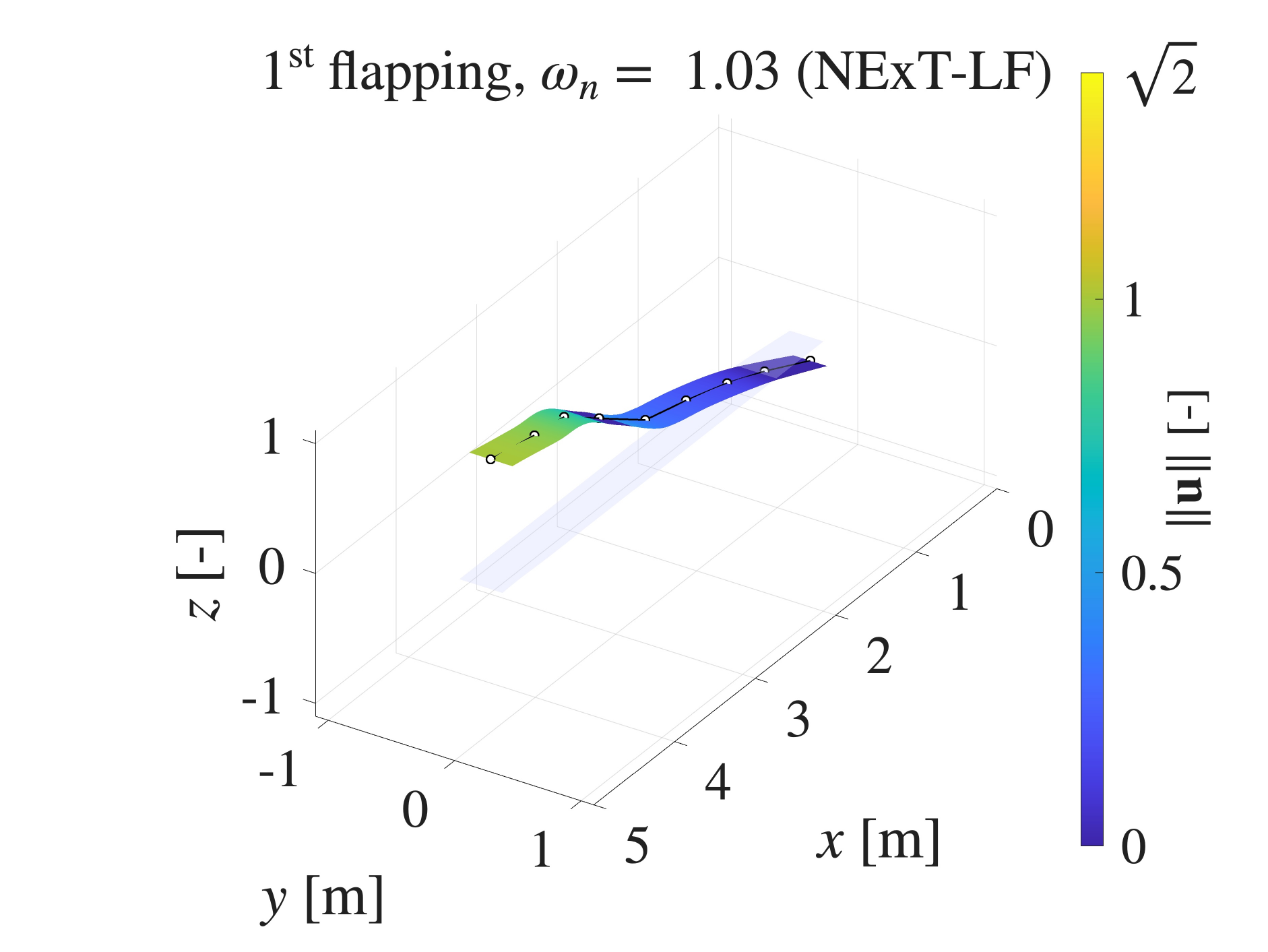}\label{fig:fig11a}}
\subfloat[\centering]{\includegraphics[width=4.8cm]{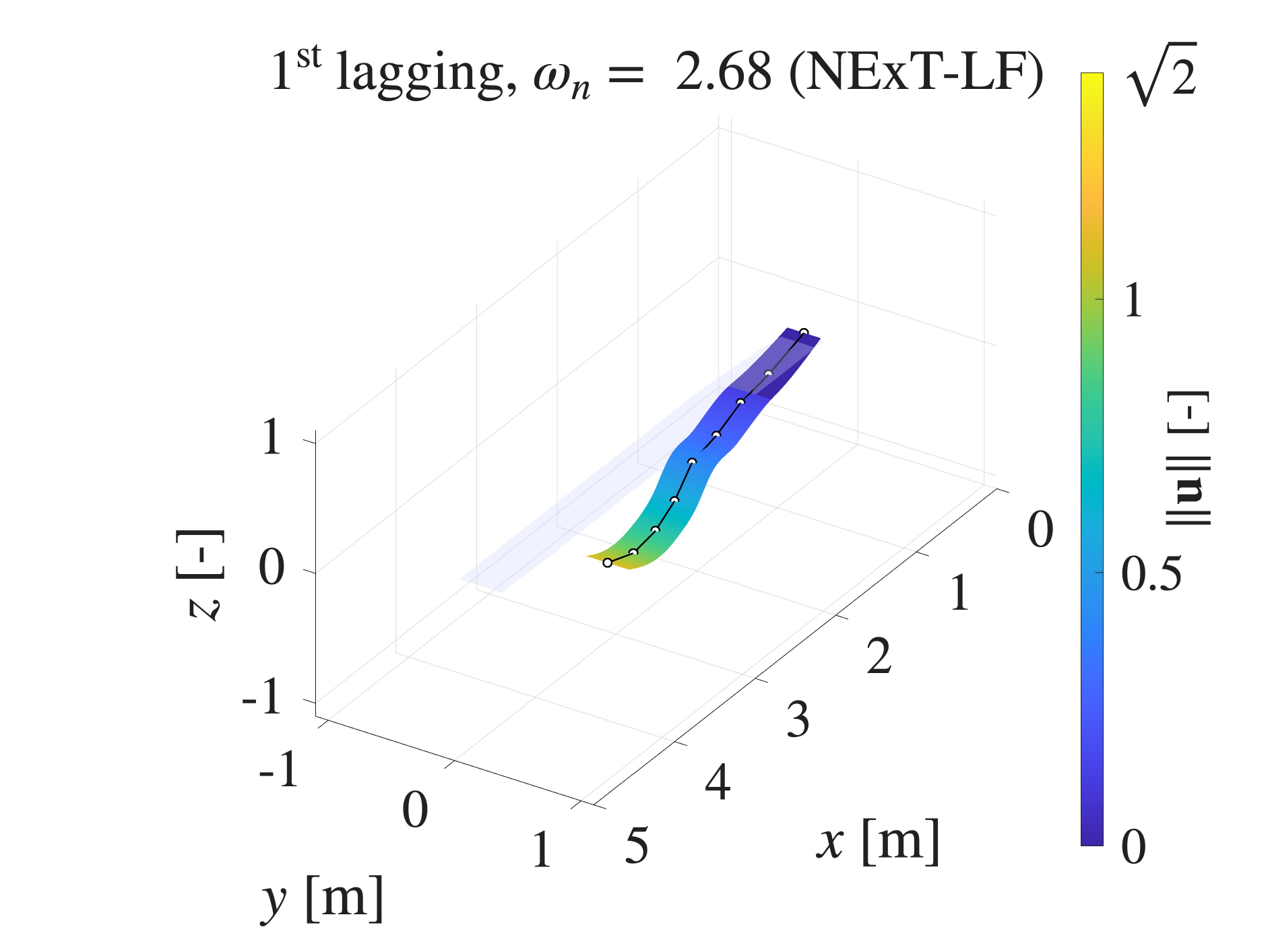}\label{fig:fig11b}}
\subfloat[\centering]{\includegraphics[width=4.8cm]{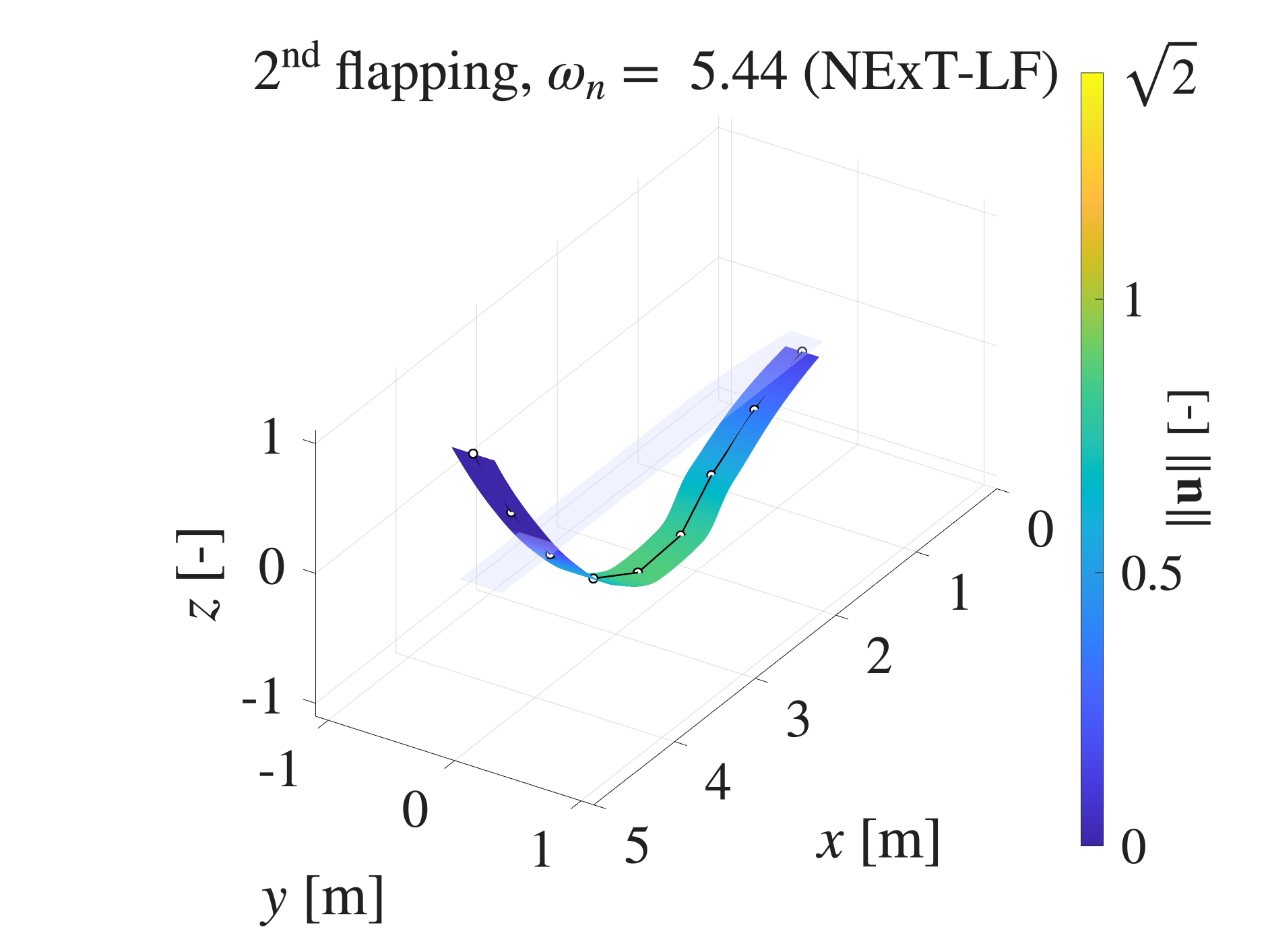}\label{fig:fig11c}}\\
\subfloat[\centering]{\includegraphics[width=4.8cm]{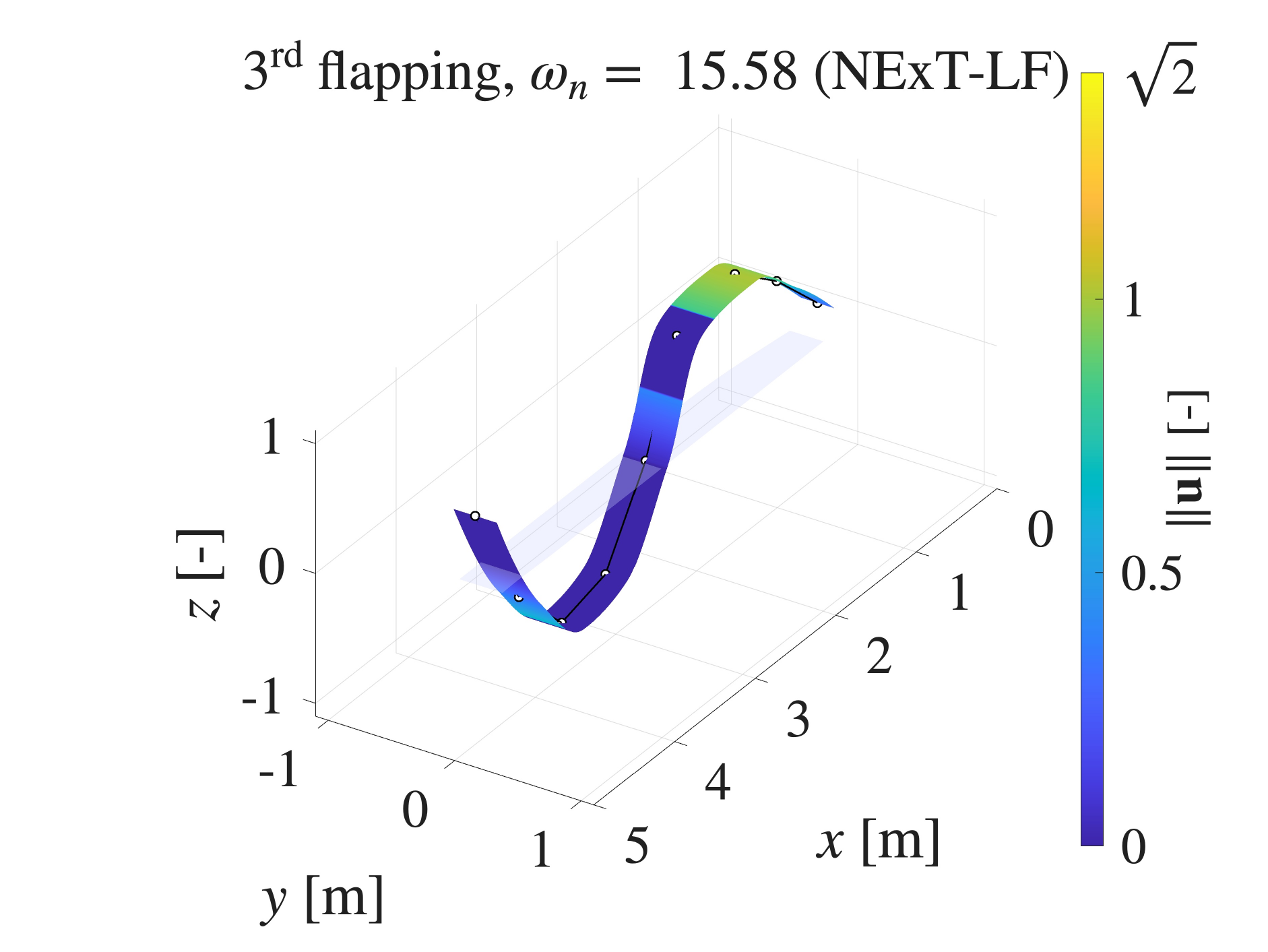}\label{fig:fig11d}}
\subfloat[\centering]{\includegraphics[width=4.8cm]{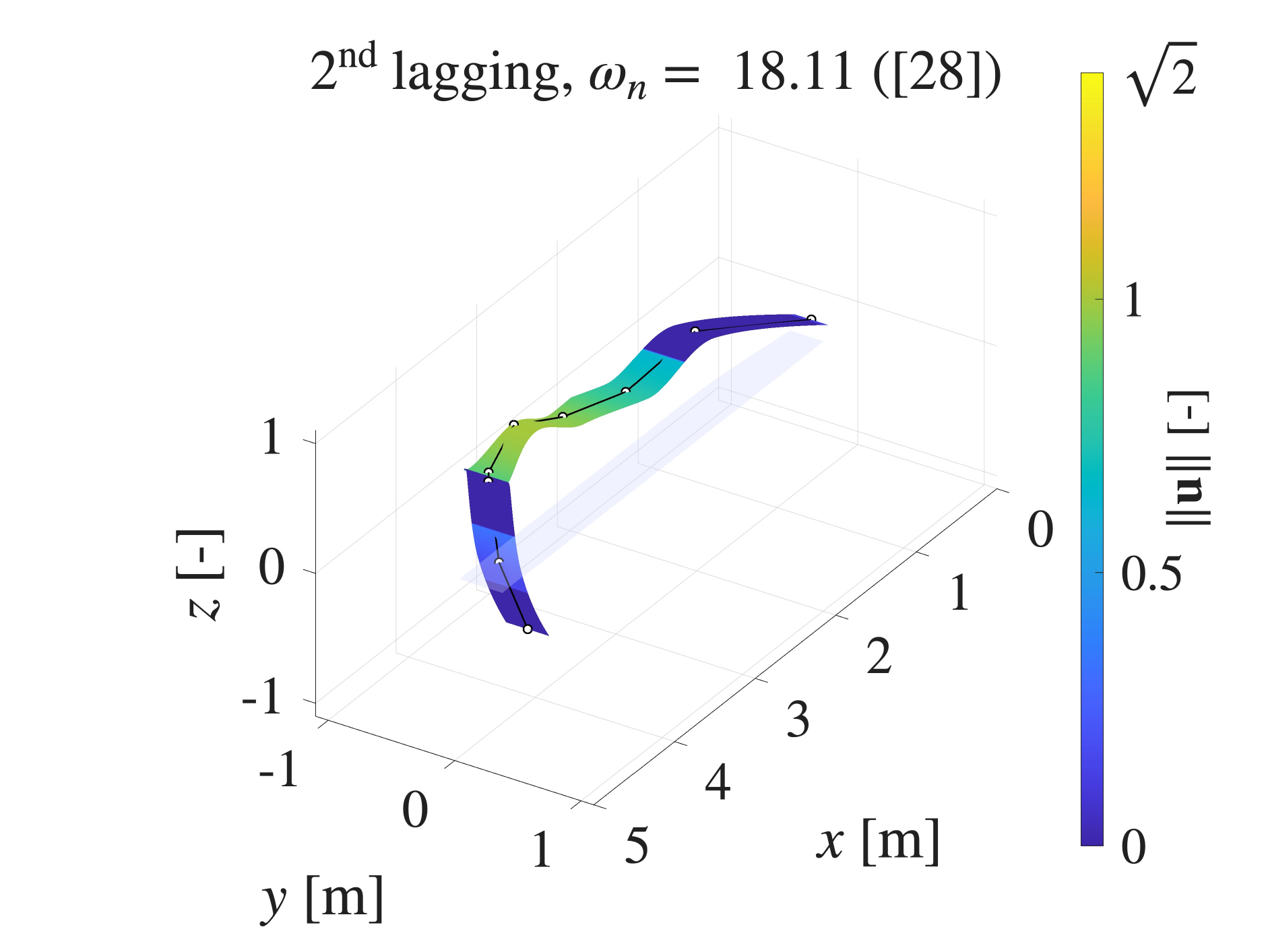}\label{fig:fig11e}}
\subfloat[\centering]{\includegraphics[width=4.8cm]{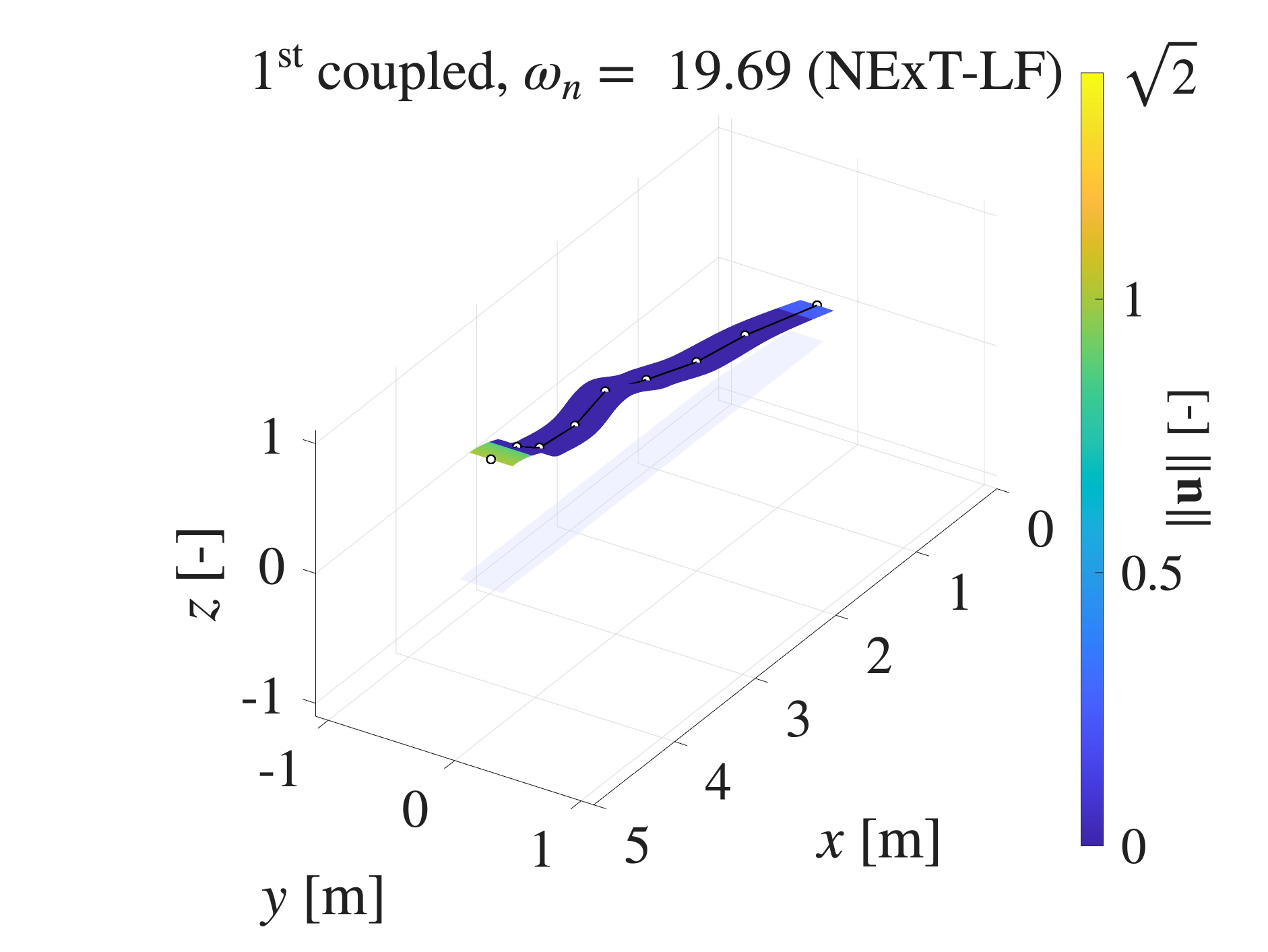}\label{fig:fig11f}}\\
\subfloat[\centering]{\includegraphics[width=4.8cm]{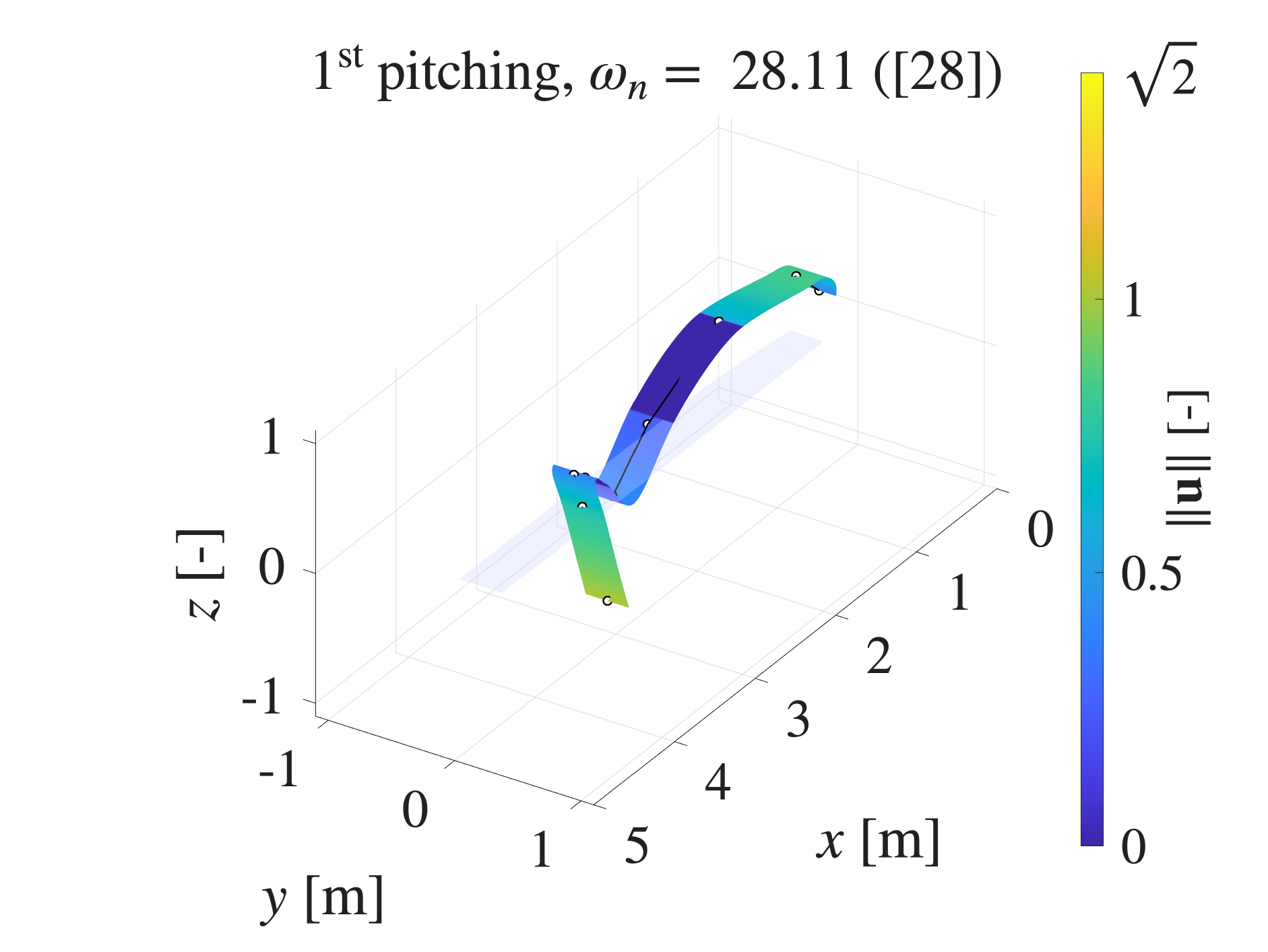}\label{fig:fig11g}}
\subfloat[\centering]{\includegraphics[width=4.8cm]{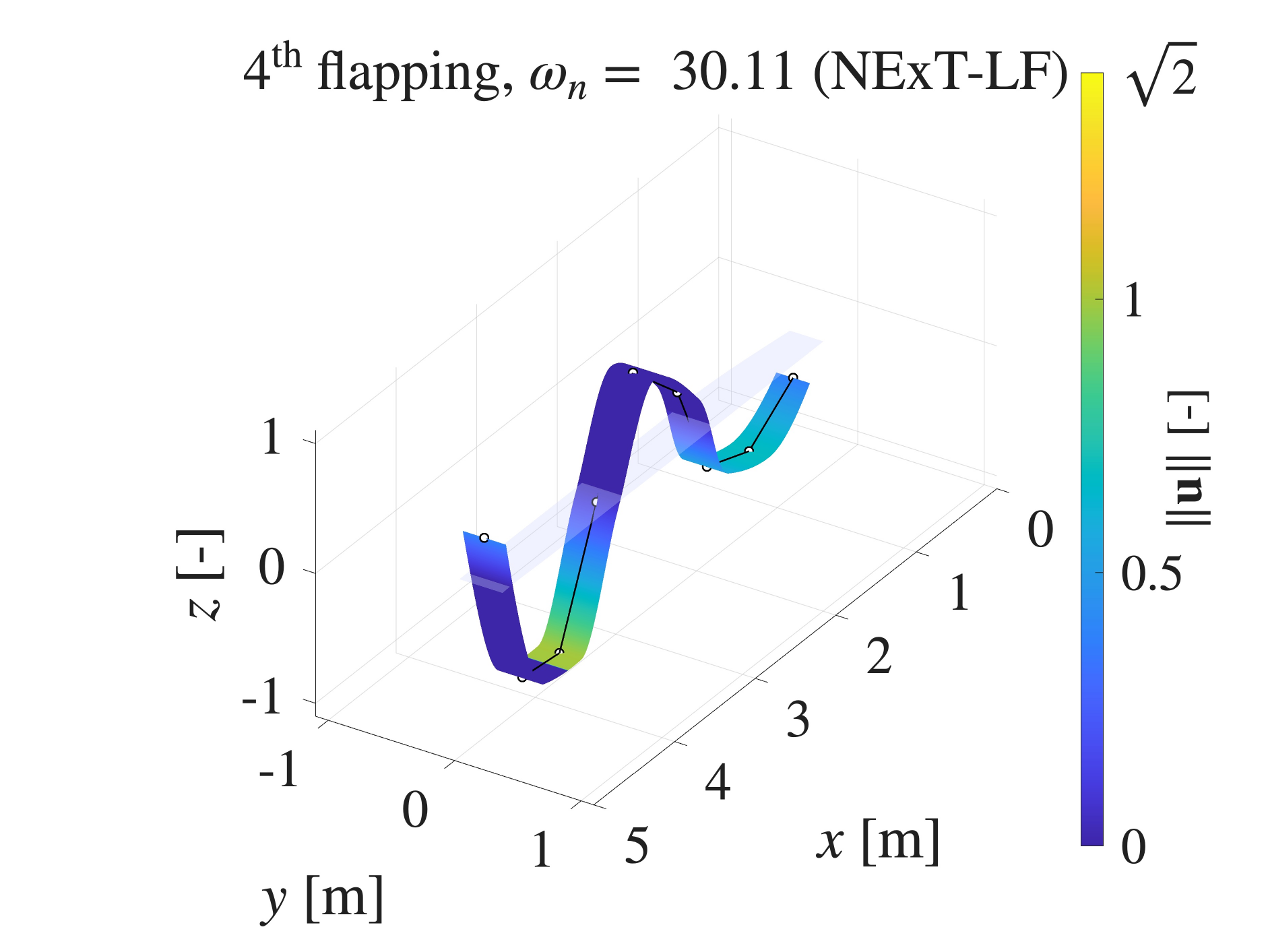}\label{fig:fig11h}}
\subfloat[\centering]{\includegraphics[width=4.8cm]{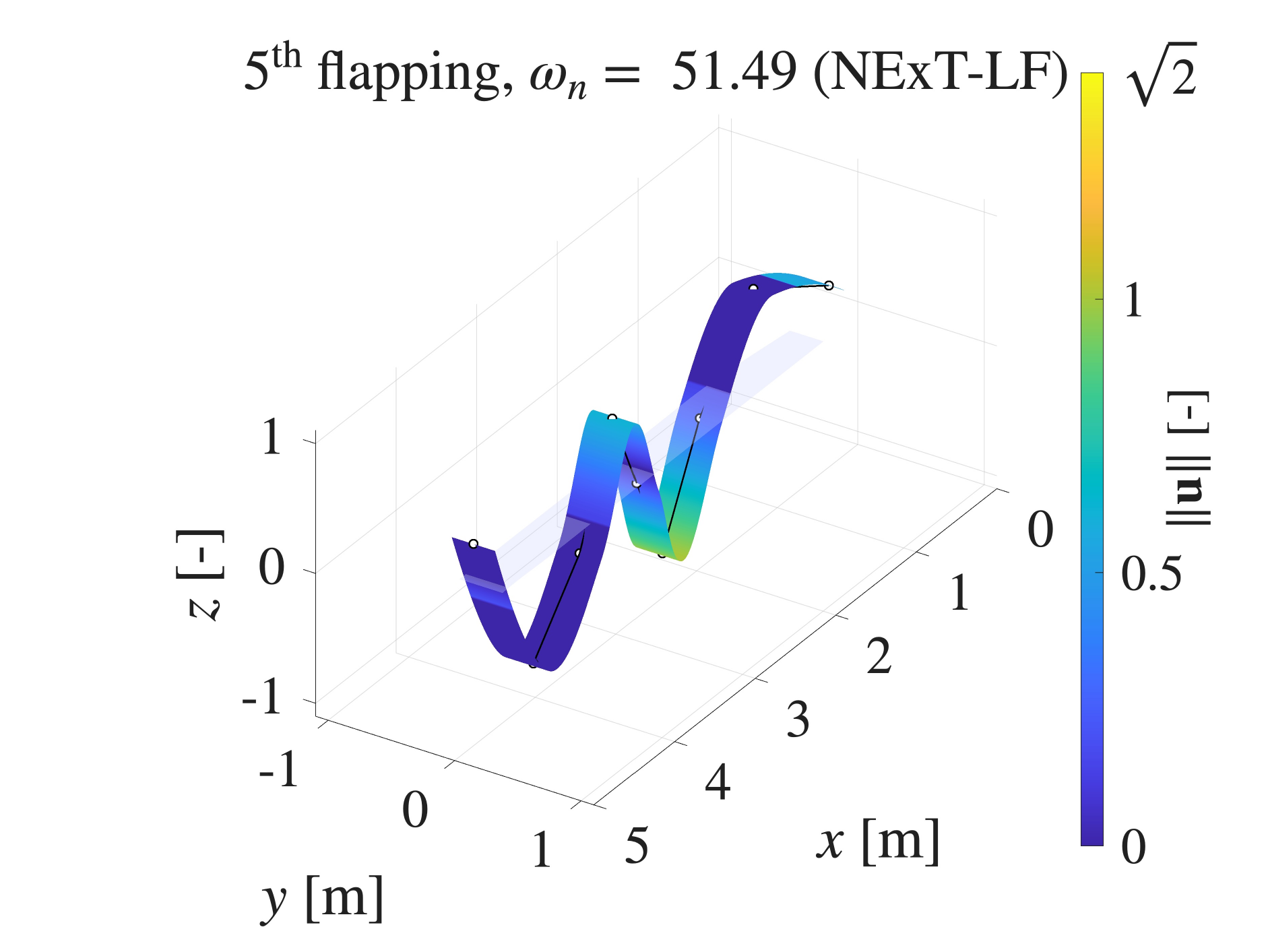}\label{fig:fig11i}}\\
\subfloat[\centering]{\includegraphics[width=4.8cm]{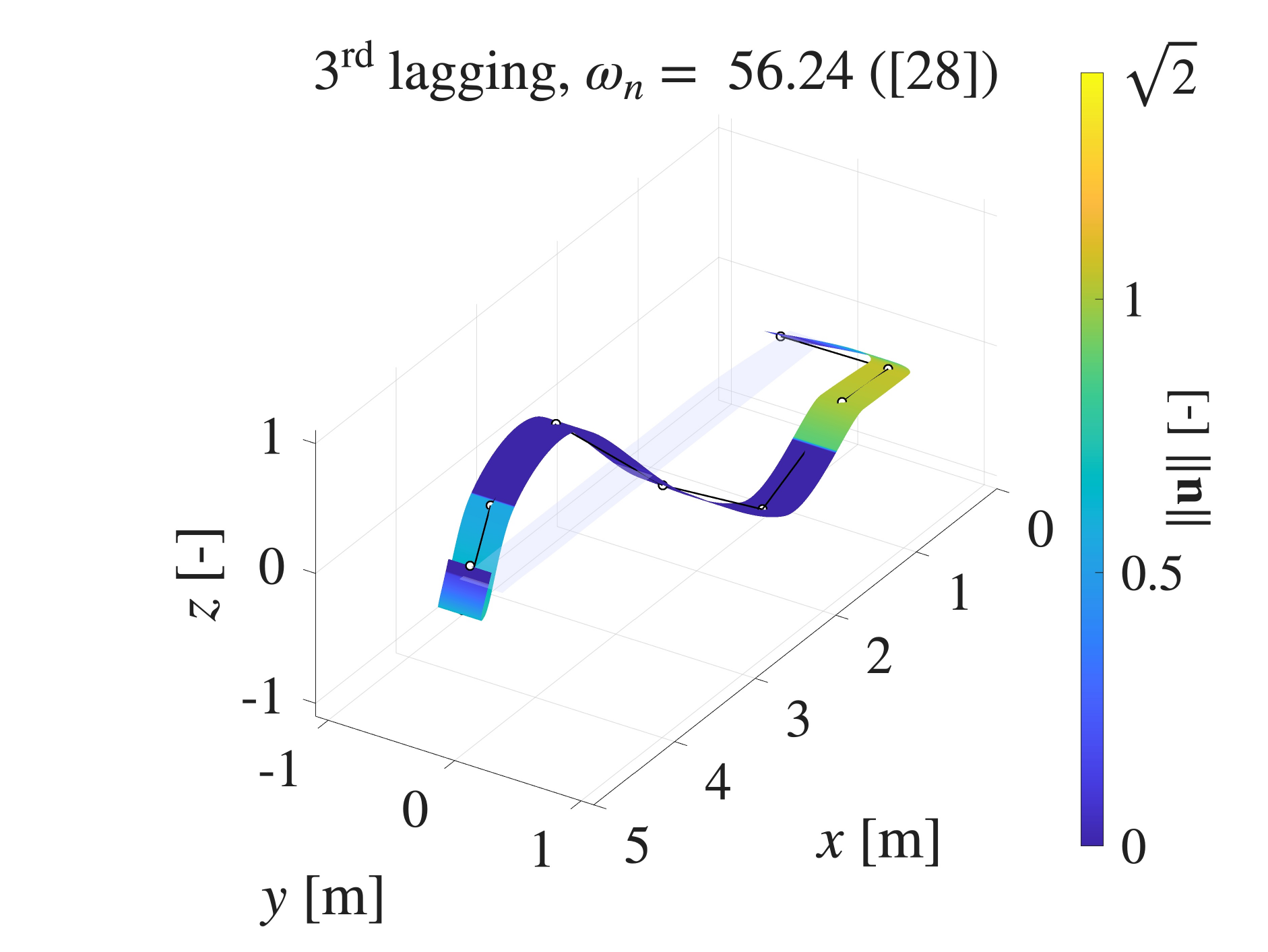}\label{fig:fig11j}}
\subfloat[\centering]{\includegraphics[width=4.8cm]{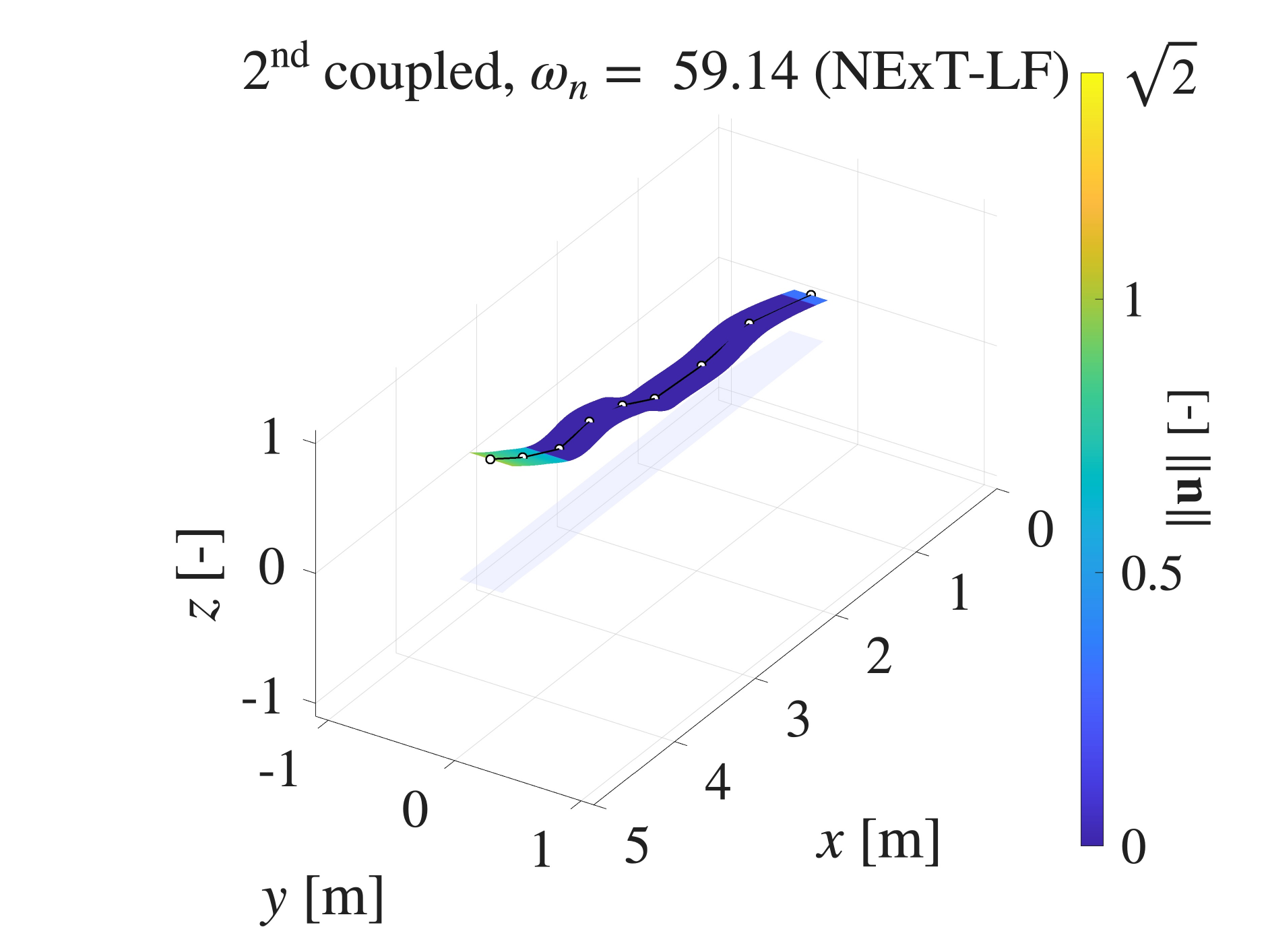}\label{fig:fig11k}}
\subfloat[\centering]{\includegraphics[width=4.8cm]{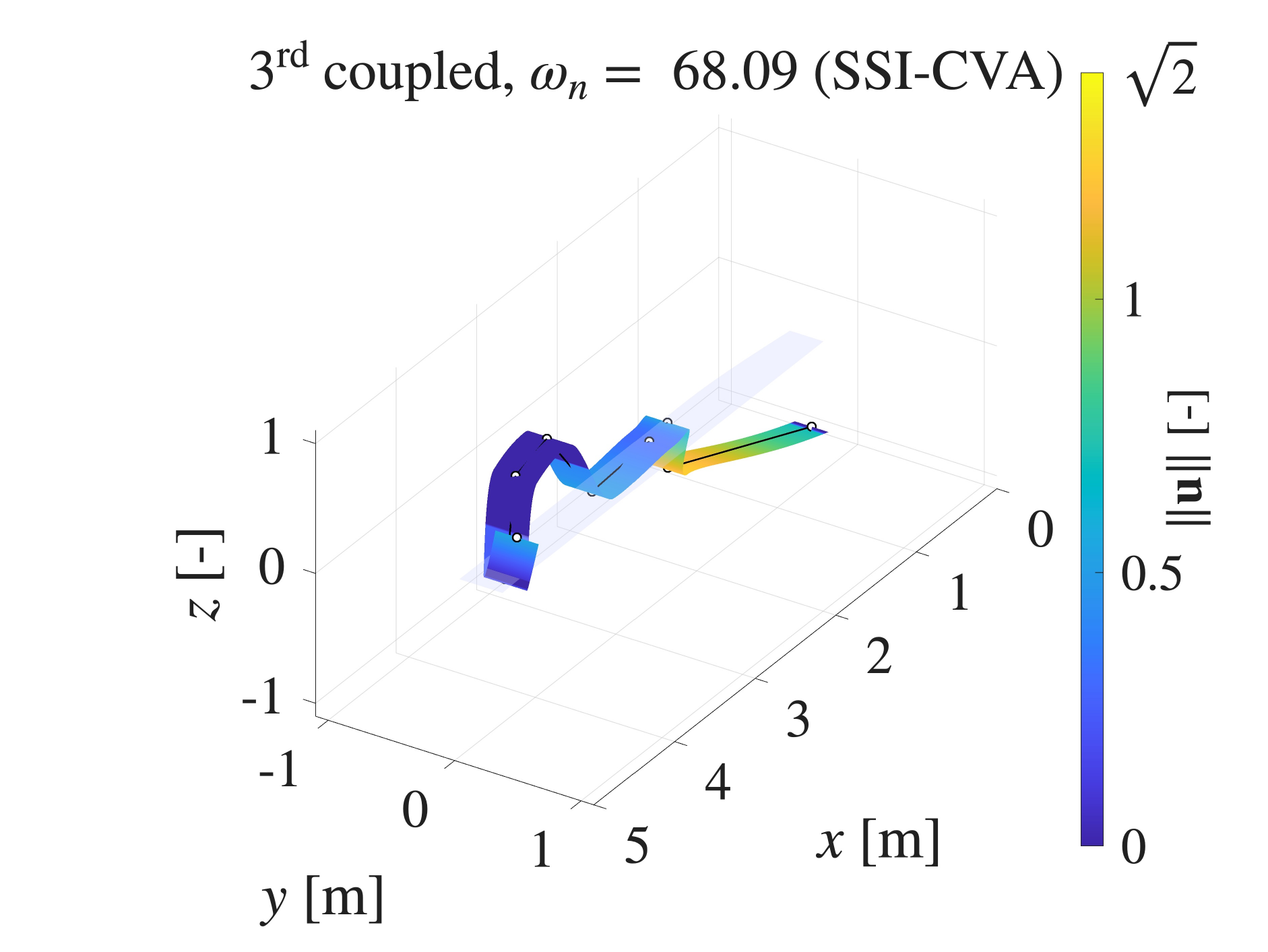}\label{fig:fig11l}}\\
\centering
\subfloat[\centering]{\includegraphics[width=4.8cm]{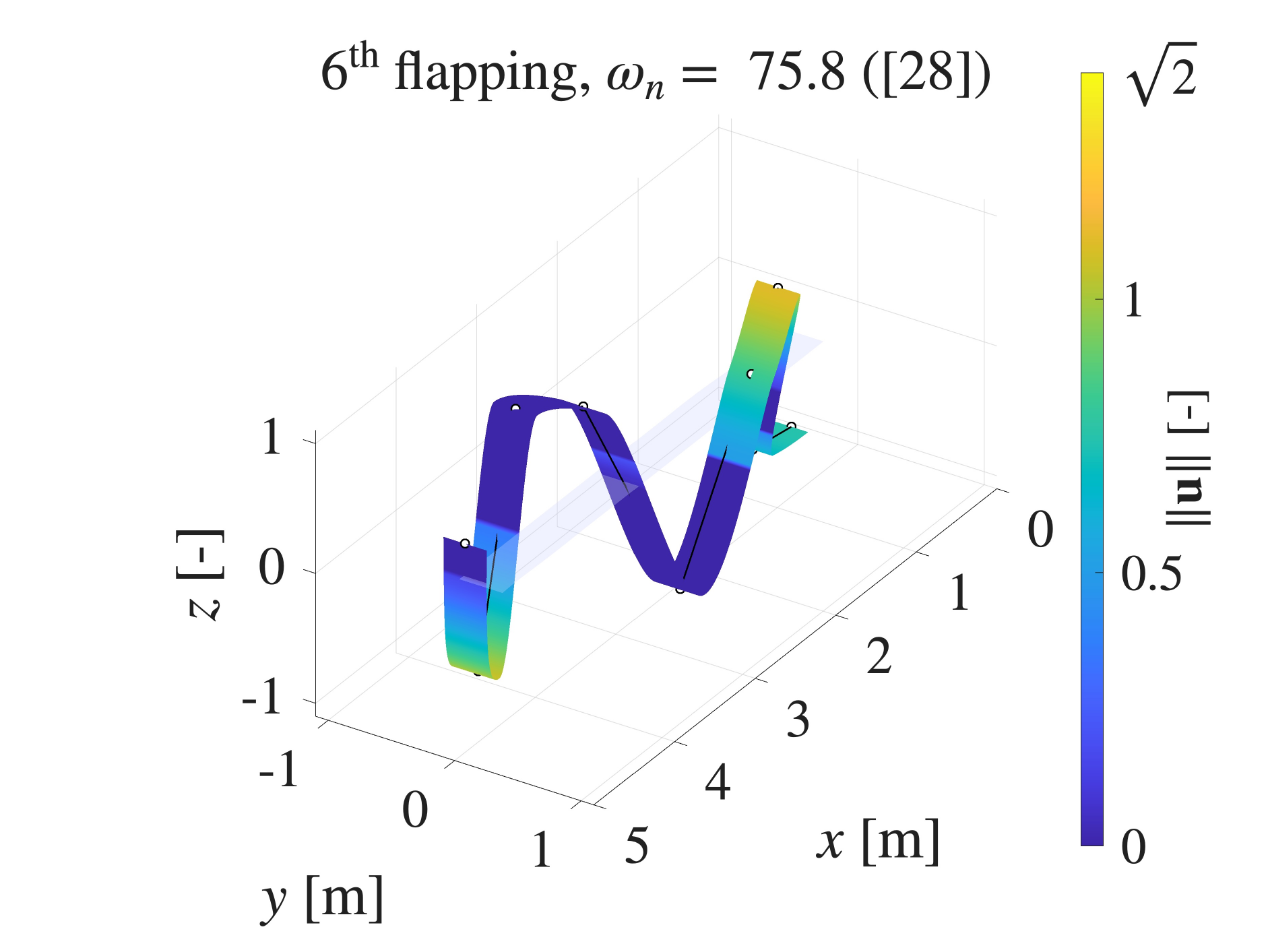}\label{fig:fig11m}}
\subfloat[\centering]{\includegraphics[width=4.8cm]{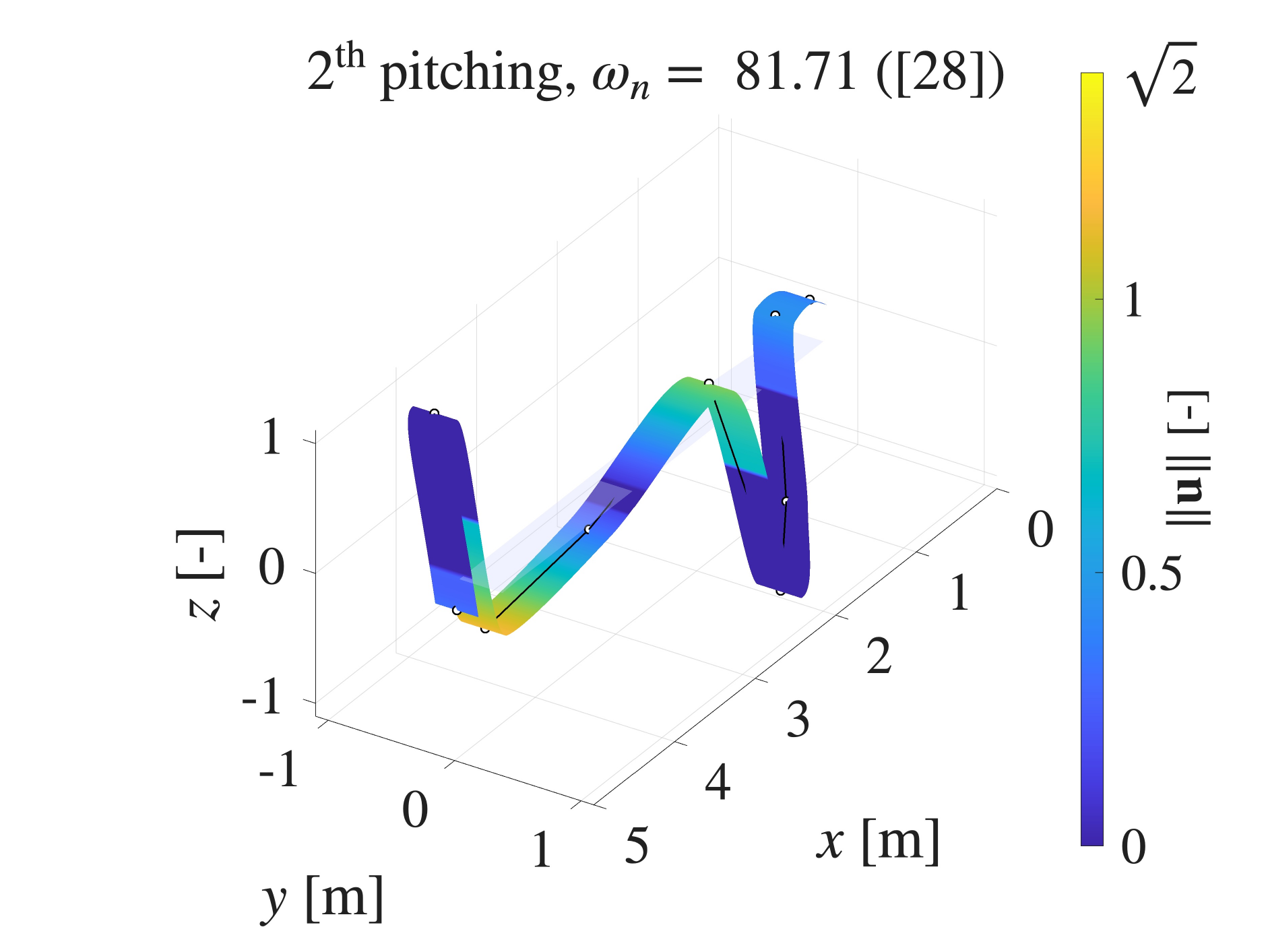}\label{fig:fig11n}}\\
\end{adjustwidth}
\caption{The 14 $\boldsymbol{\phi}_n$ (\textbf{a}-\textbf{n}) identified from the Airbus Helicopters H135 BMR blade AVT. \label{fig:fig11}}
\end{figure}

The MAC results in \cref{tab:tab4} further contextualise {the comparisons in identified} $\omega_n$ and $\zeta_n${.} In particular, NExT-LF {achieves a} very high correlation for several dominant modes (e.g{.,} Modes 1, 3, 4, 8, and 9, with MAC values close to unity), supporting the strong agreement observed for $\omega_n$ and indicating that the associated $\boldsymbol{\phi}_n$ are consistently retrieved. At the same time, the MAC values highlight that the increased modal content identified by NExT-LF comes with a non-uniform level of shape fidelity across all modes: some modes exhibit reduced correlation with respect to \cite{Mugnaini2022}, notably Mode 5 (MAC $=0.77$) and, more markedly, Mode 7 (MAC $=0.44$), as well as very low agreement for Mode 10 (MAC $=0.16$) and Mode 14 (MAC $=0.09$). This suggests that, although NExT-LF is effective in extracting a larger set of modes than NExT-ERA, a subset of the identified shapes may be more sensitive to the limited instrumentation and to the specific characteristics of the identification procedure, leading to {a weaker} correspondence with the AOMA reference in \cite{Mugnaini2022}. Conversely, SSI-CVA shows consistently high MAC values for the modes that can be compared (e.g., Mode 7 and Mode 10), indicating stronger agreement in those cases, albeit without the same increase in the number of extracted modes relative to NExT-ERA. For completeness, the 14 identified $\boldsymbol{\phi}_n$ are shown in \cref{fig:fig11}. Note that in the cases where the MAC is more than 0.95, the NExT-LF-identified $\boldsymbol{\phi}_n$ are shown; otherwise, $\boldsymbol{\phi}_n$ from \cite{Mugnaini2022} are preferred, with an exception made for mode 12 (3\textsuperscript{rd} coupled), which is retrieved from the SSI-CVA identification.

All $\boldsymbol{\phi}_n$ shown in \cref{fig:fig11} are consistent with the dominant motions reported in \cref{tab:tab3}. Nevertheless, the trajectories associated with the two pitching modes (\cref{fig:fig11g,fig:fig11n}) cannot be directly inferred from the corresponding {graphs}, as rotational degrees of freedom are not measured on the blade. {Therefore, these} modes {are} classified as pitching based on the {evidence suggested by the FE analyses} reported in \cite{Weber2021}. In addition, the three {newly identified modes} in the present {campaign -- namely}, the coupled modes (\cref{fig:fig11f,fig:fig11k,fig:fig11l}) -- do not exhibit a clear dominant displacement direction. For this reason, they are labelled as \textit{coupled}, at least in terms of combined lagging and flapping components.

\section{Discussion}\label{sec:disc}
The results obtained for the XB-2 spar (\cref{sec:spar}) and the Airbus Helicopters H135 BMR blade (\cref{sec:h135}) show that NExT-LF enables accurate identification of {modal} parameters $\omega_n$, $\zeta_n$, and $\boldsymbol{\phi}_n$. In the investigated cases, NExT-LF provides more reliable identifications than NExT-ERA, both in terms of agreement with {the benchmark values and robustness of the extracted modal set. This aspect is particularly relevant,} as both approaches rely on NExT to obtain impulse response functions from output-only data. For the helicopter blade, NExT-ERA does not retrieve the full set of modes identified by NExT-LF, whereas NExT-LF returns a richer modal content within the frequency band of interest.

Nevertheless, the helicopter blade case highlights a limitation shared by both NExT-based methods: The identification of $\boldsymbol{\phi}_n$ becomes less reliable above approximately $53~\mathrm{Hz}$. Beyond this frequency, the diagonal MAC values with respect to \cite{Mugnaini2022} drop below $0.7$, reaching $0.06$ (NExT-ERA) for Mode~10 and $0.09$ (NExT-LF) for Mode~14. This behaviour is likely driven by the distribution of excitation energy over the analysed bandwidth. The XB-2 spar is excited through broadband-limited random (white) noise using a shaker, whereas the H135 BMR blade identification is based on an AVT conducted in a laboratory environment, without externally applied broadband forcing (e.g., wind). The resulting excitation levels are markedly lower, which can prevent sufficient activation of the {modes located in the upper portion of the investigated frequency band} (here, above $\sim 50~\mathrm{Hz}$), ultimately reducing the spatial coherence of the estimated $\boldsymbol{\phi}_n$. {It should be clarified that the term \emph{higher-order} is used throughout this work to denote modes of higher rank within each modal family (e.g. 2\textsuperscript{nd}, 3\textsuperscript{rd}, etc. flapping, lagging, or pitching), not merely modes above a fixed frequency threshold. Quantitatively, the ANPSD in }\cref{fig:fig9}{ shows that the spectral energy of near peaks areas above approximately $50~\mathrm{Hz}$ is less dominant than the neighbouring area, which is consistent with the progressive degradation of the MAC values reported in }\cref{tab:tab4}{ for modes in this frequency region. This interpretation is therefore supported by the measured spectral content, although a fully decoupled verification of excitation energy at each mode would require knowledge of the input forces, which are unavailable in OMA.}

{Beyond $\boldsymbol{\phi}_n$, the damping estimates deserve separate consideration. Across both benchmarks, $\zeta_n$ exhibits a markedly larger method-to-method dispersion than $\omega_n$. For the XB-2 spar, the first-mode damping error exceeds $50$\% for both NExT-based methods, whereas for the H135 BMR blade several modes show relative deviations in excess of $30$\%. Two factors contribute to this: (i) the NExT correlation estimates are variance-limited by the available record length, which biases the exponential decay envelope from which $\zeta_n$ is inferred, and (ii) spectral leakage in the FFT step preceding the LF fit broadens the apparent resonance peaks, coupling frequency and damping estimation errors. Possible mitigation strategies include longer acquisition windows to reduce correlation variance, zero-padding or multi-taper spectral estimation to limit leakage, and dedicated regularisation of the Loewner pencil to penalise physically implausible damping values; although the latter do not constitute a real issue since they can be easily discarded via stabilisation diagram. Recent work }\cite{Zhu2025} {further confirms that output-only damping identification on lightly damped aerospace structures remains an open challenge, with reported cross-method variability of a comparable order of magnitude.}
{Therefore, importantly, the} observed degradation  appears {to be primarily} NExT-dependent rather than LF-dependent. Consistently, the effect is more pronounced for NExT-ERA, which does not yield stable modes beyond 60 Hz in this dataset. It should be mentioned that SSI-CVA does not exhibit the same limitation.

{The two benchmarks investigated in this work effectively span two distinct excitation regimes, and their comparison provides direct evidence of how excitation level and signal-to-noise ratio affect the robustness of NExT-LF. For the XB-2 spar, which is excited by a controlled broadband shaker with an RMS amplitude of 0.305 {g}, NExT-LF retrieves all three target modes with frequency errors below 1.6\% and MAC values at or near unity. For the H135 BMR blade, where the excitation is ambient and uncontrolled, the identification remains accurate for the strongly excited low-frequency modes (Modes~1--4, with the addition of 8, with frequency errors below 0.6\% and MAC values above 0.97), but degrades progressively for the modes in the upper portion of the bandwidth, where both the PSD amplitudes and the ANPSD (}\cref{fig:fig9}{) are markedly lower. In particular, the MAC values for Modes~10 and 14 fall below 0.2, and the damping dispersion across methods increases substantially. This contrast indicates that the identification quality of NExT-LF is closely tied to the effective signal-to-noise ratio at each mode: where the response is well above the noise floor, NExT-LF performs comparably to SSI-CVA; where it is not, the NExT-derived correlations become variance-limited and the subsequent LF fit inherits that uncertainty, whose effect is stronger for ERA.}

A further element is related to instrumentation and reference selection. The XB-2 spar represents a simple cantilever configuration, for which NExT can retrieve meaningful {IRFs} using a single reference channel. For the helicopter blade, two reference channels are required, as accelerations are measured along two axes. However, the absence of torsional and axial measurements may also limit the ability of NExT-based procedures to reconstruct modes with significant torsional content. In this respect, the pitching modes are not retrieved with satisfactory correlation by NExT-LF and NExT-ERA, which is consistent with the lack of measured rotational degrees of freedom. {More broadly, sensor layout affects modal observability in two respects. First, if the measurement points are sparse or lie close to the nodal lines of a given mode, the corresponding modal amplitude at those locations is small, degrading the signal-to-noise ratio and, consequently, the correlation estimate from which the mode shape is derived. This effect is particularly relevant for the higher-order flapping and lagging modes of the H135 BMR blade. Second, closely spaced or coupled modes can share similar spectral signatures, so that the identification algorithm must resolve small frequency separations under limited spectral resolution, which increases uncertainty in both damping and mode shape estimates.}

{As a limitation of this study, it is worth considering that the AVT data used here is not fully representative of the BMR blade operating conditions, which are in rotating conditions while attached to the rotor hub. Nevertheless, this is a common approach in the literature; see }\cite{Agneni2010} {for a similar test on an Agusta-Bell AB 204 blade.} {Furthermore, the individual accelerometers used in the test have masses in the range of 1-7~g. Given that the total mass of a single blade assembly is approximately 40 kg }\cite{Bansemir1999}{, the combined sensor mass (below 60 {g} for all nine accelerometers) is less than 0.15\% of the blade mass, which rules out significant mass-loading effects on the identified modal parameters. Finally, this aspect is not critical for the main scope of this work, as the H135 BMR blade dataset is employed as a benchmark for comparing NExT-LF against NExT-ERA and SSI-CVA under identical instrumentation conditions, rather than as an absolute characterisation of the blade dynamic properties.}

{Several practical strategies may improve the identification of higher-order modes in future campaigns. From an instrumentation standpoint, a denser sensor grid, particularly in the outer blade span, where higher-order characteristic displacements might be shorter, would increase modal observability. This would reduce the likelihood of measurement points coinciding with nodal regions. Multi-setup measurement schemes, in which overlapping sensor configurations are combined, offer an alternative when the available channel count is limited} \cite{Brincker2015}{, e.g. roving. On the excitation side, broader-band or higher-amplitude forcing (e.g. impact hammer sweeps or shaker excitation) would raise the signal-to-noise ratio in the upper frequency range, while longer acquisition records would reduce the variance of the NExT correlation estimates and sharpen the spectral peaks used in the LF fit. However, this would defeat the point of an OMA campaign.}

{Considering these findings, the results of this article demonstrate} that NExT-LF can be extended from prior applications on civil structures \cite{Dessena2024f} to aeronautically relevant configurations, provided that adequate attention is given to reference-channel selection and to the choice of the analysed frequency band and/or the excitation conditions adopted during testing.
\section{Conclusions}\label{sec:conc}
In this work, the Natural Excitation Technique (NExT) is paired with the Loewner Framework (LF) to extend NExT-LF from Operational Modal Analysis (OMA) of civil structures to that of aeronautically relevant configurations. 

In conclusion, the main outcomes of this study can be summarised as follows:
\begin{itemize}
    \item To the authors' knowledge, this is the first time in the literature that NExT-LF is applied for the OMA of aeronautically relevant systems;
    \item NExT-LF is applied for the OMA of the well-known eXperimental BeaRDS 2 Flexible Wing Spar Model and of an Airbus Helicopters H135 Bearingless Main Rotor Blade;
    \item The natural frequencies identified by NExT-LF are coherent to those obtained with {two} benchmark methods, NExT with the Eigensystem Realization Algorithm (NExT-ERA) and Stochastic Subspace Identification with Canonical Variate Analysis (SSI), and data;
    \item {Damping ratio estimates exhibit a significantly larger inter-method dispersion than natural frequencies and mode shapes, attributed primarily to the sensitivity of the NExT-derived correlation estimates to record length and noise level;}
    \item {The mode shapes retrieved in the lower and highly excited frequency band show a high (near 1) correlation (modal assurance criterion) with the {reference} data;}
    \item Three further modes are found w.r.t. the benchmark results of the helicopter blade. Two by NExT-LF and one by SSI.
\end{itemize} 

{Future work will focus on (i) extending the applicability of NExT-LF to poorly excited higher-frequency modes, (ii) improving damping estimation through longer acquisitions and advanced spectral preprocessing, (iii) systematically characterising the robustness of NExT-LF to measurement noise via controlled noise-injection studies, and (iv) applying NExT-LF under more representative operating conditions for helicopter blades, including in-flight or ground-run scenarios. From a practical standpoint, the computational efficiency of the LF and its compatibility with output-only data make NExT-LF a candidate tool for the modal characterisation of lightweight aerospace structures in settings where controlled excitation is impractical, such as ground vibration testing of large assemblies or in-service monitoring. In such contexts, NExT-LF could support condition-aware maintenance programmes and provide input data for finite element model updating.}



\vspace{6pt} 





\authorcontributions{Conceptualisation, G.D. and M.C.; methodology, G.D.; software, G.D.; validation, G.D. and M.C.; formal analysis, G.D.; investigation, G.D.; resources, G.D., M.C. and OE.BM.; data curation, G.D..; writing---original draft preparation, G.D.; writing---review and editing, G.D., M.C. and OE.BM. visualization, G.D.; project administration, G.D.; funding acquisition, G.D., M.C. and OE.BM..}

\funding{The first author is supported by grant JDC2024-055593-I funded by MICIU/AEI/10.13039/501100011033 and by ESF+. This work has been supported by the Madrid Government (\textit{Comunidad de Madrid}, Spain) under the Multiannual Agreement with UC3M (IA\_aCTRl-CM-UC3M).
The second author is supported by the Centro Nazionale per la Mobilità Sostenibile (MOST – Sustainable Mobility Center), Spoke 7 (Cooperative Connected and Automated Mobility and Smart Infrastructures), Work Package 4 (Resilience of Networks, Structural Health Monitoring and Asset Management).}

\institutionalreview{Not applicable.}

\informedconsent{Not applicable.}

\dataavailability{Data supporting this study (Modal identification of the XB-2 spar and H135 BMR blade modal parameters -- Data file 1) are openly available from the Zenodo Repository at \textit{(DOI to be reserved)}.
The flexible wing spar experimental data (Data file 2) used in this study are available in the CORD - Cranfield University research data repository entry \emph{Data supporting: 'Ground Vibration Testing of a Flexible Wing: A Benchmark and Case Study'} available at \url{https://doi.org/10.17862/cranfield.rd.19077023}. 
The helicopter blade experimental data (Data file 3) used in this study are not publicly available and therefore cannot be redistributed by the authors.}

\conflictsofinterest{The authors declare no conflicts of interest.} 

\isPreprints{}{
}

\reftitle{References}

\isPreprints{}{
} 

\begin{thebibliography}{999}

\bibitem[Ewins(2000)]{Ewins2000}
Ewins, D.J.
\newblock {\em Modal Testing Theory, Practice and Application}, 2nd ed.; Research Studies Press,  2000; p. 562.

\bibitem[Rainieri and Fabbrocino(2014)]{Rainieri2014}
Rainieri, C.; Fabbrocino, G.
\newblock {\em Operational Modal Analysis of Civil Engineering Structures}; Springer New York,  2014; pp. 1--131.
\newblock {\url{https://doi.org/10.1007/978-1-4939-0767-0}}.

\bibitem[Reynders(2012)]{Reynders2012}
Reynders, E.
\newblock System Identification Methods for (Operational) Modal Analysis: Review and Comparison.
\newblock {\em Archives of Computational Methods in Engineering} {\bf 2012}, {\em 19},~51--124.
\newblock {\url{https://doi.org/10.1007/s11831-012-9069-x}}.

\bibitem[Fragonara et~al.(2017)Fragonara, Boscato, Ceravolo, Russo, Ientile, Pecorelli, and Quattrone]{ZanottiFragonara2017}
Fragonara, L.Z.; Boscato, G.; Ceravolo, R.; Russo, S.; Ientile, S.; Pecorelli, M.L.; Quattrone, A.
\newblock Dynamic investigation on the Mirandola bell tower in post-earthquake scenarios.
\newblock {\em Bulletin of Earthquake Engineering} {\bf 2017}, {\em 15},~313--337.
\newblock {\url{https://doi.org/10.1007/s10518-016-9970-z}}.

\bibitem[Cadoret et~al.(2025)Cadoret, Denimal-Goy, Leroy, Pfister, and Mevel]{Cadoret2025}
Cadoret, A.; Denimal-Goy, E.; Leroy, J.M.; Pfister, J.L.; Mevel, L.
\newblock Damage detection and localization method for wind turbine rotor based on Operational Modal Analysis and anisotropy tracking.
\newblock {\em Mechanical Systems and Signal Processing} {\bf 2025}, {\em 224},~111982.
\newblock {\url{https://doi.org/10.1016/j.ymssp.2024.111982}}.

\bibitem[Jelicic et~al.(2014)Jelicic, Schwochow, Govers, Hebler, and Böswald]{Jelicic2014}
Jelicic, G.; Schwochow, J.; Govers, Y.; Hebler, A.; Böswald, M.
\newblock Real-time assessment of flutter stability based on automated output-only modal analysis.
\newblock {\em Proceedings of ISMA 2014 - International Conference on Noise and Vibration Engineering and USD 2014 - International Conference on Uncertainty in Structural Dynamics} {\bf 2014}, pp. 3633--3646.

\bibitem[Sinske et~al.(2017)Sinske, Govers, Jelicic, Buchbach, Schwochow, Handojo, Böswald, and Krüger]{Sinske2017}
Sinske, J.; Govers, Y.; Jelicic, G.; Buchbach, R.; Schwochow, J.; Handojo, V.; Böswald, M.; Krüger, W.R.
\newblock Flight testing using fast online aeroelastic identification techniques with DLR research aircraft HALO.
\newblock {\em 17th International Forum on Aeroelasticity and Structural Dynamics, IFASD 2017} {\bf 2017}, {\em 2017-June}.

\bibitem[Akers et~al.(2025)Akers, Winkel, Chin, Parks, Chandler, Stasiunas, and Allen]{Akers2025}
Akers, J.C.; Winkel, J.P.; Chin, A.W.; Parks, R.A.; Chandler, D.E.; Stasiunas, E.C.; Allen, M.S., Operational Modal Analysis of the Artemis I Dynamic Rollout Test and Wet Dress Rehearsal.
\newblock In {\em Sensors \& Instrumentation and Aircraft/Aerospace Testing Techniques Vol. 8}; Walber, C.; Stefanski, M., Eds.; River Publishers,  2025; Vol.~8, pp. 91--112.
\newblock {\url{https://doi.org/10.1007/978-3-031-68188-2_10}}.

\bibitem[Eugeni et~al.(2018)Eugeni, Coppotelli, Mastroddi, Gaudenzi, Muller, and Troclet]{Eugeni2018}
Eugeni, M.; Coppotelli, G.; Mastroddi, F.; Gaudenzi, P.; Muller, S.; Troclet, B.
\newblock OMA analysis of a launcher under operational conditions with time-varying properties.
\newblock {\em CEAS Space Journal} {\bf 2018}, {\em 10},~381--406.
\newblock {\url{https://doi.org/10.1007/s12567-018-0209-5}}.

\bibitem[James et~al.(1994)James, Carne, and Edmunds]{James1993}
James, G.; Carne, T.; Edmunds, R.
\newblock STARS missile -- Modal analysis of first-flight data using the Natural Excitation Technique, NExT.
\newblock In Proceedings of the Proc. SPIE Vol. 2251, Proceedings of the 12th International Modal Analysis Conference; DeMichele, D.J., Ed. Society for Experimental Mechanics,  1994.
\newblock {\url{https://doi.org/10.2172/10113260}}.

\bibitem[Ameri et~al.(2013)Ameri, Grappasonni, Coppotelli, and Ewins]{Ameri2013}
Ameri, N.; Grappasonni, C.; Coppotelli, G.; Ewins, D.
\newblock Ground vibration tests of a helicopter structure using OMA techniques.
\newblock {\em Mechanical Systems and Signal Processing} {\bf 2013}, {\em 35},~35--51.
\newblock {\url{https://doi.org/10.1016/j.ymssp.2012.09.013}}.

\bibitem[Sibille et~al.(2023)Sibille, Civera, Fragonara, and Ceravolo]{Sibille2023}
Sibille, L.; Civera, M.; Fragonara, L.Z.; Ceravolo, R.
\newblock Automated Operational Modal Analysis of a Helicopter Blade with a Density-Based Cluster Algorithm.
\newblock {\em AIAA Journal} {\bf 2023}, {\em 61},~1411--1427.
\newblock {\url{https://doi.org/10.2514/1.J062084}}.

\bibitem[Agneni et~al.(2010)Agneni, Crema, and Coppotelli]{Agneni2010}
Agneni, A.; Crema, L.B.; Coppotelli, G.
\newblock Output-only analysis of structures with closely spaced poles.
\newblock {\em Mechanical Systems and Signal Processing} {\bf 2010}, {\em 24},~1240--1249.
\newblock {\url{https://doi.org/10.1016/j.ymssp.2009.10.013}}.

\bibitem[III et~al.(1993)III, Carne, and Lauffer]{GeorgeIII1993}
III, G.H.J.; Carne, T.G.; Lauffer, J.P.
\newblock The Natural Excitation Technique (NExT) for Modal Parameter Extraction From Operating Wind Turbines.
\newblock Technical report, Sandia National Laboratories,  1993.

\bibitem[Overschee and Moor(1996)]{VanOverschee1996}
Overschee, P.V.; Moor, B.D.
\newblock {\em Subspace Identification for Linear Systems}; Springer US,  1996.
\newblock {\url{https://doi.org/10.1007/978-1-4613-0465-4}}.

\bibitem[Brincker et~al.(2001)Brincker, Zhang, and Andersen]{Brincker2001}
Brincker, R.; Zhang, L.; Andersen, P.
\newblock Modal identification of output-only systems using frequency domain decomposition.
\newblock {\em Smart Materials and Structures} {\bf 2001}, {\em 10},~441--445.
\newblock {\url{https://doi.org/10.1088/0964-1726/10/3/303}}.

\bibitem[O’Connell and Rogers(2024)]{OConnell2024}
O’Connell, B.J.; Rogers, T.J.
\newblock A robust probabilistic approach to stochastic subspace identification.
\newblock {\em Journal of Sound and Vibration} {\bf 2024}, {\em 581},~118381.
\newblock {\url{https://doi.org/10.1016/j.jsv.2024.118381}}.

\bibitem[O'Connell et~al.(2025)O'Connell, Champneys, and Rogers]{OConnell2025}
O'Connell, B.J.; Champneys, M.D.; Rogers, T.J.
\newblock A new perspective on Bayesian operational modal analysis.
\newblock {\em Mechanical Systems and Signal Processing} {\bf 2025}, {\em 236},~112949.
\newblock {\url{https://doi.org/10.1016/j.ymssp.2025.112949}}.

\bibitem[Amador and Brincker(2024)]{Amador2024}
Amador, S.D.R.; Brincker, R.
\newblock The new Subspace-based poly-reference Complex Frequency (S-pCF) for robust frequency-domain modal parameter estimation.
\newblock {\em Measurement: Journal of the International Measurement Confederation} {\bf 2024}, {\em 225}.
\newblock {\url{https://doi.org/10.1016/j.measurement.2023.113995}}.

\bibitem[Lv et~al.(2026)Lv, Hua, Yu, Dong, and Gong]{Lv2026}
Lv, L.; Hua, C.; Yu, H.; Dong, D.; Gong, C.
\newblock Automatic identification of operational modal parameters for train suspension systems based on an improved SSI-FCM.
\newblock {\em Engineering Structures} {\bf 2026}, {\em 349},~121868.
\newblock {\url{https://doi.org/10.1016/j.engstruct.2025.121868}}.

\bibitem[Tarinejad and Amanzad(2026)]{Tarinejad2026}
Tarinejad, R.; Amanzad, F.
\newblock A novel approach for identifying system poles using multi-reference transmissibility functions based on frequency shifts.
\newblock {\em Journal of Sound and Vibration} {\bf 2026}, {\em 626},~119638.
\newblock {\url{https://doi.org/10.1016/j.jsv.2026.119638}}.

\bibitem[Wu et~al.(2026)Wu, He, Yuan, and Yang]{Wu2026}
Wu, C.; He, S.; Yuan, B.; Yang, Z.
\newblock An improved NExT-DMD for efficient automated operational modal analysis.
\newblock {\em Applied Mathematical Modelling} {\bf 2026}, {\em 156},~116823.
\newblock {\url{https://doi.org/10.1016/j.apm.2026.116823}}.

\bibitem[Saito and Kuno(2020)]{Saito2020}
Saito, A.; Kuno, T.
\newblock Data-driven experimental modal analysis by Dynamic Mode Decomposition.
\newblock {\em Journal of Sound and Vibration} {\bf 2020}, {\em 481},~115434.
\newblock {\url{https://doi.org/10.1016/j.jsv.2020.115434}}.

\bibitem[Wu et~al.(2025)Wu, Yang, and He]{Wu2025}
Wu, C.; Yang, Z.; He, S.
\newblock Efficient modal parameter identification using DMD-DBSCAN and rank stabilization diagrams.
\newblock {\em Aerospace Science and Technology} {\bf 2025}, {\em 161},~110112.
\newblock {\url{https://doi.org/10.1016/j.ast.2025.110112}}.

\bibitem[Schimid(2010)]{Schmid2010}
Schimid, P.J.
\newblock Dynamic mode decomposition of numerical and experimental data.
\newblock {\em Journal of Fluid Mechanics} {\bf 2010}, {\em 656},~5--28.
\newblock {\url{https://doi.org/10.1017/S0022112010001217}}.

\bibitem[Mugnaini et~al.(2022)Mugnaini, Fragonara, and Civera]{Mugnaini2022}
Mugnaini, V.; Fragonara, L.Z.; Civera, M.
\newblock A machine learning approach for automatic operational modal analysis.
\newblock {\em Mechanical Systems and Signal Processing} {\bf 2022}, {\em 170},~108813.
\newblock {\url{https://doi.org/10.1016/j.ymssp.2022.108813}}.

\bibitem[Neu et~al.(2017)Neu, Janser, Khatibi, and Orifici]{Neu2017}
Neu, E.; Janser, F.; Khatibi, A.A.; Orifici, A.C.
\newblock Fully Automated Operational Modal Analysis using multi-stage clustering.
\newblock {\em Mechanical Systems and Signal Processing} {\bf 2017}, {\em 84},~308--323.
\newblock {\url{https://doi.org/10.1016/j.ymssp.2016.07.031}}.

\bibitem[de~Almeida~Cardoso et~al.(2018)de~Almeida~Cardoso, Cury, and Barbosa]{DeAlmeidaCardoso2018}
de~Almeida~Cardoso, R.; Cury, A.; Barbosa, F.
\newblock A clustering-based strategy for automated structural modal identification.
\newblock {\em Structural Health Monitoring} {\bf 2018}, {\em 17},~201--217.
\newblock {\url{https://doi.org/10.1177/1475921716689239}}.

\bibitem[Mostafaei and Ghamami(2025)]{Mostafaei2025}
Mostafaei, H.; Ghamami, M.
\newblock State of the Art in Automated Operational Modal Identification: Algorithms, Applications, and Future Perspectives.
\newblock {\em Machines} {\bf 2025}, {\em 13},~39.
\newblock {\url{https://doi.org/10.3390/machines13010039}}.

\bibitem[Dessena et~al.(2023)Dessena, Civera, Fragonara, Ignatyev, and Whidborne]{Dessena2022}
Dessena, G.; Civera, M.; Fragonara, L.Z.; Ignatyev, D.I.; Whidborne, J.F.
\newblock A Loewner-Based System Identification and Structural Health Monitoring Approach for Mechanical Systems.
\newblock {\em Structural Control and Health Monitoring} {\bf 2023}, {\em 2023},~1--22.
\newblock {\url{https://doi.org/10.1155/2023/1891062}}.

\bibitem[Dessena et~al.(2025)Dessena, Civera, Yousefi, and Surace]{Dessena2024f}
Dessena, G.; Civera, M.; Yousefi, A.; Surace, C.
\newblock NExT‐LF: A Novel Operational Modal Analysis Method via Tangential Interpolation.
\newblock {\em International Journal of Mechanical System Dynamics} {\bf 2025}, {\em 5},~401--414.
\newblock {\url{https://doi.org/10.1002/msd2.70016}}.

\bibitem[Dessena et~al.(2022)Dessena, Ignatyev, Whidborne, Pontillo, and Fragonara]{Dessena2022b}
Dessena, G.; Ignatyev, D.I.; Whidborne, J.F.; Pontillo, A.; Fragonara, L.Z.
\newblock Ground vibration testing of a flexible wing: A benchmark and case study.
\newblock {\em Aerospace} {\bf 2022}, {\em 9},~438.
\newblock {\url{https://doi.org/10.3390/aerospace9080438}}.

\bibitem[Weber et~al.(2021)Weber, Kissinger, Chehura, Staines, Barrington, Mullaney, Fragonara, Petrunin, James, Lone, and Tatam]{Weber2021}
Weber, S.; Kissinger, T.; Chehura, E.; Staines, S.; Barrington, J.; Mullaney, K.; Fragonara, L.Z.; Petrunin, I.; James, S.; Lone, M.;  et~al.
\newblock Application of fibre optic sensing systems to measure rotor blade structural dynamics.
\newblock {\em Mechanical Systems and Signal Processing} {\bf 2021}, {\em 158},~107758.
\newblock {\url{https://doi.org/10.1016/j.ymssp.2021.107758}}.

\bibitem[Lefteriu and Antoulas(2009)]{Lefteriu2009}
Lefteriu, S.; Antoulas, A.C.
\newblock Modeling multi-port systems from frequency response data via tangential interpolation.
\newblock In Proceedings of the 2009 IEEE Workshop on Signal Propagation on Interconnects,  5 2009, pp. 1--4.
\newblock {\url{https://doi.org/10.1109/SPI.2009.5089847}}.

\bibitem[Quero et~al.(2019)Quero, Vuillemin, and Poussot-Vassal]{Quero2019}
Quero, D.; Vuillemin, P.; Poussot-Vassal, C.
\newblock A generalized state-space aeroservoelastic model based on tangential interpolation.
\newblock {\em Aerospace} {\bf 2019}, {\em 6},~9.
\newblock {\url{https://doi.org/10.3390/aerospace6010009}}.

\bibitem[Vojkovic et~al.(2023)Vojkovic, Quero, Poussot-Vassal, and Vuillemin]{Vojkovic2023}
Vojkovic, T.; Quero, D.; Poussot-Vassal, C.; Vuillemin, P.
\newblock Low-Order Parametric State-Space Modeling of MIMO Systems in the Loewner Framework.
\newblock {\em SIAM Journal on Applied Dynamical Systems} {\bf 2023}, {\em 22},~3130--3164.
\newblock {\url{https://doi.org/10.1137/22M1509898}}.

\bibitem[Dessena and Civera(2025)]{Dessena2024}
Dessena, G.; Civera, M.
\newblock Improved tangential interpolation-based multi-input multi-output modal analysis of a full aircraft.
\newblock {\em European Journal of Mechanics - A/Solids} {\bf 2025}, {\em 110},~105495.
\newblock {\url{https://doi.org/10.1016/j.euromechsol.2024.105495}}.

\bibitem[Dessena et~al.(2026)Dessena, Civera, Marcos, Chiaia, and Bonilla-Manrique]{Dessena2025a}
Dessena, G.; Civera, M.; Marcos, A.; Chiaia, B.; Bonilla-Manrique, O.E.
\newblock Multiple input tangential interpolation-driven damage detection of a jet trainer aircraft.
\newblock {\em Aerospace Science and Technology} {\bf 2026}, {\em 168},~111032.
\newblock {\url{https://doi.org/10.1016/j.ast.2025.111032}}.

\bibitem[Dessena et~al.(2025)Dessena, Pontillo, Civera, Ignatyev, Whidborne, and Fragonara]{Dessena2025}
Dessena, G.; Pontillo, A.; Civera, M.; Ignatyev, D.I.; Whidborne, J.F.; Fragonara, L.Z.
\newblock Damping Identification Sensitivity in Flutter Speed Estimation.
\newblock {\em Vibration} {\bf 2025}, {\em 8},~24.
\newblock {\url{https://doi.org/10.3390/vibration8020024}}.

\bibitem[Löwner(1934)]{Lowner1934}
Löwner, K.
\newblock Über monotone matrixfunktionen.
\newblock {\em Mathematische Zeitschrift} {\bf 1934}, {\em 38},~177--216.
\newblock {\url{https://doi.org/10.1007/BF01170633}}.

\bibitem[Antoulas et~al.(2017)Antoulas, Lefteriu, and Ionita]{Antoulas2017}
Antoulas, A.C.; Lefteriu, S.; Ionita, A.C., A Tutorial Introduction to the Loewner Framework for Model Reduction.
\newblock In {\em Model Reduction and Approximation}; Society for Industrial and Applied Mathematics,  2017; pp. 335--376.
\newblock {\url{https://doi.org/10.1137/1.9781611974829.ch8}}.

\bibitem[Mayo and Antoulas(2007)]{Mayo2007}
Mayo, A.; Antoulas, A.
\newblock A framework for the solution of the generalized realization problem.
\newblock {\em Linear Algebra and its Applications} {\bf 2007}, {\em 425},~634--662.
\newblock {\url{https://doi.org/10.1016/j.laa.2007.03.008}}.

\bibitem[Kramer and Gugercin(2016)]{Kramer2016}
Kramer, B.; Gugercin, S.
\newblock Tangential interpolation-based eigensystem realization algorithm for MIMO systems.
\newblock {\em Mathematical and Computer Modelling of Dynamical Systems} {\bf 2016}, {\em 22},~282--306.
\newblock {\url{https://doi.org/10.1080/13873954.2016.1198389}}.

\bibitem[Lefteriu and Antoulas(2010)]{Lefteriu2010b}
Lefteriu, S.; Antoulas, A.C.
\newblock A new approach to modeling multiport systems from frequency-domain data.
\newblock {\em IEEE Transactions on Computer-Aided Design of Integrated Circuits and Systems} {\bf 2010}, {\em 29},~14--27.
\newblock {\url{https://doi.org/10.1109/TCAD.2009.2034500}}.

\bibitem[Pontillo et~al.(2018)Pontillo, Hayes, Dussart, Matos, Carrizales, Yusuf, and Lone]{Pontillo2018}
Pontillo, A.; Hayes, D.; Dussart, G.X.; Matos, G.E.L.; Carrizales, M.A.; Yusuf, S.Y.; Lone, M.M.
\newblock Flexible High Aspect Ratio Wing: Low Cost Experimental Model and Computational Framework.
\newblock In Proceedings of the 2018 AIAA Atmospheric Flight Mechanics Conference. American Institute of Aeronautics and Astronautics,  1 2018, pp. 1--15.
\newblock {\url{https://doi.org/10.2514/6.2018-1014}}.

\bibitem[Yusuf et~al.(2019)Yusuf, Hayes, Pontillo, Carrizales, Dussart, and Lone]{Yusuf2019}
Yusuf, S.Y.; Hayes, D.; Pontillo, A.; Carrizales, M.A.; Dussart, G.X.; Lone, M.M.
\newblock Aeroelastic Scaling for Flexible High Aspect Ratio Wings.
\newblock In Proceedings of the AIAA Scitech 2019 Forum. American Institute of Aeronautics and Astronautics,  1 2019, pp. 1--14.
\newblock {\url{https://doi.org/10.2514/6.2019-1594}}.

\bibitem[Pontillo(2020)]{Pontillo2020}
Pontillo, A.
\newblock High Aspect Ratio Wings on Commercial Aircraft: a Numerical and Experimental approach.
\newblock PhD thesis, Centre for Aeronautics, Cranfield University,  2020.

\bibitem[Hayes et~al.(2019)Hayes, Pontillo, Yusuf, Lone, and Whidborne]{Hayes2019}
Hayes, D.; Pontillo, A.; Yusuf, S.Y.; Lone, M.M.; Whidborne, J.
\newblock High aspect ratio wing design using the minimum exergy destruction principle.
\newblock In Proceedings of the AIAA Scitech 2019 Forum. American Institute of Aeronautics and Astronautics,  1 2019, p.~21.
\newblock {\url{https://doi.org/10.2514/6.2019-1592}}.

\bibitem[Dessena et~al.(2025{\natexlab{a}})Dessena, Civera, and Bonilla-Manrique]{Dessena2025c}
Dessena, G.; Civera, M.; Bonilla-Manrique, O.E.
\newblock Tangential interpolation for the operational modal analysis of aeronautical structures.
\newblock In Proceedings of the 15th EASN International Conference, 14-17 of October 2025. In press,  2025, pp. 1--8.

\bibitem[Dessena et~al.(2025{\natexlab{b}})Dessena, Pontillo, Ignatyev, Whidborne, and Fragonara]{Dessena2022h}
Dessena, G.; Pontillo, A.; Ignatyev, D.I.; Whidborne, J.F.; Fragonara, L.Z.
\newblock Identification of Nonlinearity Sources in a Flexible Wing.
\newblock {\em Journal of Aerospace Engineering} {\bf 2025}, {\em 38},~04025060.
\newblock {\url{https://doi.org/10.1061/JAEEEZ.ASENG-5508}}.

\bibitem[Bansemir and Emmerling(1999)]{Bansemir1999}
Bansemir, H.; Emmerling, S.
\newblock Fatigue Substantiation and Damage Tolerance Evaluation of Fiber Composite Helicopter Components.
\newblock In Proceedings of the RTO AVT Specialists' Meeting on "Application of Damage Tolerance Principles for Improved Airworthiness of Rotorcraft". NATO Research and Technology Organization,  1999.
\newblock {\url{https://doi.org/10.14339/RTO-MP-024-ALL-PDF}}.

\bibitem[Bansemir and Müller(1997)]{Bansemir1997}
Bansemir, H.; Müller, R.
\newblock The EC135 - Applied Advanced Technology.
\newblock In Proceedings of the 53rd American Helicopter Society International Annual Forum 1997. American Helicopter Society International,  1997, pp. 846--861.

\bibitem[Kampa et~al.(1999)Kampa, Enenkl, Polz, and Roth]{Kampa1997}
Kampa, K.; Enenkl, B.; Polz, G.; Roth, G.
\newblock Aeromechanic Aspects in the Design of the EC135.
\newblock {\em Journal of the American Helicopter Society} {\bf 1999}, {\em 44},~83--93.
\newblock {\url{https://doi.org/10.4050/JAHS.44.83}}.

\bibitem[Felce()]{Felce2011}
Felce, T.
\newblock EC-135 -- RIAT 2011.
\newblock Flickr, Inc. License: CC BY-SA 2.0 (\url{https://creativecommons.org/licenses/by-sa/2.0/}). Source listed on page: Flickr photo ID 6199185182.
\newblock {\url{https://www.flickr.com/photos/24874528@N04/6199185182/}}.

\bibitem[Caicedo(2011)]{Caicedo2011}
Caicedo, J.
\newblock Practical Guidelines for the Natural Excitation Technique (NExT) and the Eigensystem Realization Algorithm (ERA) for Modal Identification Using Ambient Vibration.
\newblock {\em Experimental Techniques} {\bf 2011}, {\em 35},~52--58.
\newblock {\url{https://doi.org/10.1111/j.1747-1567.2010.00643.x}}.

\bibitem[Phani and Woodhouse(2007)]{SrikanthaPhani2007}
Phani, A.S.; Woodhouse, J.
\newblock Viscous damping identification in linear vibration.
\newblock {\em Journal of Sound and Vibration} {\bf 2007}, {\em 303},~475--500.
\newblock {\url{https://doi.org/10.1016/j.jsv.2006.12.031}}.

\bibitem[Zhu et~al.(2025)Zhu, Wang, Liu, Lei, and Fei]{Zhu2025}
Zhu, R.; Wang, W.; Liu, Y.; Lei, M.; Fei, Q.
\newblock Novel LMD-NExT Hybrid Method for Damping Ratio Identification Under Sweep-Frequency Excitation.
\newblock {\em AIAA Journal} {\bf 2025}, pp. 1--8.
\newblock {\url{https://doi.org/10.2514/1.J066336}}.

\bibitem[Brincker and Ventura(2015)]{Brincker2015}
Brincker, R.; Ventura, C.E.
\newblock {\em Introduction to Operational Modal Analysis}; Wiley,  2015.
\newblock {\url{https://doi.org/10.1002/9781118535141}}.

\end{thebibliography}
\end{document}